\documentclass[]{msu-thesis}
\usepackage[T1]{fontenc}
\usepackage{newtxtext,newtxmath}
\usepackage{stmaryrd,latexsym,amsmath,braket}
\usepackage{todonotes}
\usepackage{graphicx}
\usepackage[numbers]{natbib}
\usepackage{hyperref}

\DeclareOldFontCommand{\rm}{\normalfont\rmfamily}{\mathrm}
\DeclareOldFontCommand{\bf}{\normalfont\bfseries}{\mathbf}

\title{Eigenstate Preparation on Quantum Computers}
\author{Joey Bonitati}
\dualmajor{Physics}{Computational Mathematics, Science and Engineering}
\date{2024}

\dedication{"How does a quantum computer greet you? With a wave function!'' - ChatGPT (When prompted for a dad joke about quantum computing)}

\begin{document}

\frontmatter
\maketitlepage
\begin{abstract}
This thesis investigates quantum algorithms for eigenstate preparation, with a primary focus on solving eigenvalue problems such as the Schrödinger equation by utilizing near-term quantum computing devices. These problems are ubiquitous in several scientific fields, but more accurate solutions are specifically needed as a prerequisite for many quantum simulation tasks. To address this, we establish three methods in detail: quantum adiabatic evolution with optimal control, the Rodeo Algorithm, and the Variational Rodeo Algorithm. 

The first method explored is adiabatic evolution, a technique that prepares quantum states by simulating a quantum system that evolves slowly over time. The adiabatic theorem can be used to ensure that the system remains in an eigenstate throughout the process, but its implementation can often be infeasible on current quantum computing hardware. We employ a unique approach using optimal control to create custom gate operations for superconducting qubits and demonstrate the algorithm on a two-qubit IBM cloud quantum computing device.

We then explore an alternative to adiabatic evolution, the Rodeo Algorithm, which offers a different approach to eigenstate preparation by using a controlled quantum evolution that selectively filters out undesired components in the wave function stored on a quantum register. We show results suggesting that this method can be effective in preparing eigenstates, but its practicality is predicated on the preparation of an initial state that has significant overlap with the desired eigenstate. To address this, we introduce the novel Variational Rodeo Algorithm, which replaces the initialization step with dynamic optimization of quantum circuit parameters to increase the success probability of the Rodeo Algorithm. The added flexibility compensates for instances in which the original algorithm can be unsuccessful, allowing for better scalability. 

This research seeks to contribute to a deeper understanding of how quantum algorithms can be employed to attain efficient and accurate solutions to eigenvalue problems. The overarching goal is to present ideas that can be used to improve understanding of nuclear physics by providing potential quantum and classical techniques that can aid in tasks such as the theoretical description of nuclear structures and the simulation of nuclear reactions.
\end{abstract}
\clearpage

\makecopyrightpage 
\clearpage

\chapter*{Acknowledgements}
\DoubleSpacing
I acknowledge the intellectual support of my advisor Dean Lee, as well as our various collaborators at MSU (Jacob Watkins, Ashe Hicks, Gabriel Given, Zhengrong Qian, Xilin Zhang, Yuanzhuo Ma, Max Bee-Lindgren, Kenneth Choi) and LLNL (Kyle Wendt, Sofia Quaglioni, Toño Coello Perez). 
I acknowledge the financial support from MSU and through grants from the DOE and NSF.
I acknowledge the professional support from my undergraduate REU advisor Filomena Nunes at MSU.
I acknowledge the emotional support from my immediate and extended family, who love to celebrate my accomplishments so that I don't have to.
Above all, I acknowledge my love, Sarah, for supporting me in many ways through the final years of my PhD program.
\clearpage

\SingleSpacing
\tableofcontents*



\mainmatter

\chapter{Introduction}
The overarching goal of the research presented in this dissertation is to discover more efficient computational methods to aid in the theoretical understanding of nuclear physics. The nucleus of an atom is so small that, in order to properly understand it, it is necessary to consider the laws of quantum mechanics. It is notoriously difficult to computationally analyze such systems, so the concept of quantum simulation, the approach of using one quantum system to emulate another, has recently gained popularity \cite{Trabesinger:2012icv}. A likely first step of a quantum simulation algorithm will be to prepare a quantum computer in a state that is analogous in some way to the initial state of the quantum system of interest. Often, this will be one of the characteristic states, called "eigenstates'', of the system. This critical step of "eigenstate preparation'' is not straightforward to perform, and it has enough significant applications in other areas to be interesting enough on its own. 

To properly build up the motivation for this particular problem, this chapter provides a brief history of the development of quantum and nuclear physics, starting with the scientific method, a topic that is often misunderstood and underappreciated. More detailed background on quantum computation and eigenvalue problems is given in Chapter \ref{cha:background}.

\section{The Scientific Method}
The scientific method is the primary way in which all kinds of scientists, including physicists, make discoveries about things that can happen in the universe. There has been a growing effort to increase awareness of the scientific method among cultures and individuals to achieve a so-called "science literacy'' \cite{national2016science}. General ignorance of science may be a factor in the proliferation of false information, an issue that has been exacerbated by social media and recent rapid developments in generative AI, which has led the World Economic Forum to declare the spread of misinformation and disinformation to be the most serious threat facing the world in 2024 \cite{WEF}. Furthermore, it is helpful to understand how the seemingly abstract field of quantum physics actually emerged from discoveries made using the scientific method, which will be explained in Section \ref{sec:historyqm}. For these reasons, it is pertinent to explore how science, particularly physics, evolved over history into what it is today.

\subsection{History of the Scientific Method}

In short, the scientific method is the way that people can use a combination of logic and observation to learn new things. The idea that we can learn about the universe by observing it from within was first established in writing by the ancient Greek philosopher Aristotle \cite{aristotle_physics}. His method of obtaining knowledge relies on two types of logical reasoning: induction and deduction. Induction (or inductive reasoning) involves deriving broad generalizations from observations, whereas deduction (or deductive reasoning) involves invoking the rules of logic to make valid conclusions based on given premises. A common type of generalization used in induction is that if you repeatedly observe a particular outcome in a certain set of circumstances, then that outcome will occur again the next time those circumstances are present; for example, if you notice that whenever you drink milk, you get an upset stomach, then you could conclude that the next time you drink milk, you will get an upset stomach again. A common type of deduction used in the scientific method is the derivation of mathematical equations from established empirical laws; for example, the ideal gas law ($PV = nRT$) can be derived by combining Boyle's law ($PV = \rm constant$), Charles's law ($\frac{V}{T} = \rm constant$), Avogadro's law ($\frac{V}{n} = \rm constant$), and Gay-Lussac's law ($\frac{P}{T} = \rm constant$) using basic algebra.

Consider the following example of how Aristotle's scientific method can be applied:

\begin{enumerate}
    \item Start by making observations of a physical phenomenon
    \item Using these observations and some intuition, design an equation that can make predictions for future observations
    \item (Deduction) If possible, use algebraic manipulation on this equation to discover more equations that can make different predictions 
    \item  (Induction) Validate the predictions from steps (2) and (3) with more observations.
\end{enumerate}

After Aristotle, the scientific method was refined further by philosophers and mathematicians such as Descartes and Isaac Newton. An important part of the scientific method that was gradually added to Aristotle's initial idea is that, in order to be considered scientifically rigorous, the predictions and hypotheses produced must be testable. In order to be considered testable, a statement must be falsifiable, meaning that it is possible for it not to be true, and it must be practically feasible to set up a reproducible experiment that can refute it. Furthermore, to avoid confirmation bias, all predictions and hypotheses must be assumed to be false until there is sufficient reason to believe otherwise. If an idea is not testable, then it cannot be considered scientific. This is not to say that such ideas cannot be true; it simply means that they are outside the scope of what is considered science.

Similarly, just because an idea is scientific does not mean that it is true. In fact, all scientific predictions are by definition not provable with complete certainty! Since the scientific method relies on induction, it is subject to the "problem of induction'' formulated by the philosopher David Hume \cite{hume}. Hume pointed out that inferences based on induction always assume the principle of induction, which requires that the past can be used to predict the future. Hume argues that this is not necessarily true, so every prediction about the future based on past observations has a chance of being wrong. When using the scientific method, one must always acknowledge the possibility that a prediction could be wrong, perhaps even for some unknown reason. Since all predictions are initially assumed to be false, this means that scientific hypotheses cannot be verified beyond all doubt. Thus, the phrase "scientific fact'', which may sometimes be used in misinformation campaigns to create a false sense of authority, is actually an oxymoron!

Despite the lack of complete certainty inherent in science, it is obvious that the scientific method is still useful for making predictions. For example, physicists at NASA were able to understand the complicated trajectories of space shuttles well enough to land humans on the moon. Thus, in order to practice science, one must ignore the problem of induction to some extent and acknowledge that, even though inductive reasoning is flawed, it is necessary. After all, the assumption that the laws of physics will be the same in the future as they were in the past, while not necessarily valid, is not a difficult assumption to believe. If we make this assumption, then we can at least infer that when we make repeated observations of a phenomenon, it is somewhat probable that we will make the same observations in the future. This is enough to justify the empirical testing of hypotheses in physics, which, along with intuition and mathematical deduction, leads to a better understanding of the universe in which we live.

\subsection{From Classical Physics to Modern Physics}
Today, physics is generally classified into two branches: classical physics and modern physics. Classical physics deals with phenomena occurring under normal conditions, whereas modern physics addresses extraordinary situations, including velocities approaching the speed of light and scales of distance similar to the size of an atom. Prominent fields within modern physics include nuclear physics, special relativity, and quantum mechanics. These new branches all deal with situations outside the realm of everyday human experiences, but there is no denying that they have had noticeable effects on society; for example, advances in these sciences have made new inventions possible, such as nuclear power plants, GPS tracking, and quantum computers.

For most of history, the study of physics has been related to the phenomena we experience directly as humans. This makes sense because empirical hypotheses must be testable and the most obvious way to test things is through direct observation with the senses. However, human sensory organs are limited in their ability to gather information; the naked eye is easily fooled by optical illusions, for example. Thus, scientific progress naturally advances alongside the development of better tools and measuring devices. For example, Isaac Newton's theory of gravity was likely based on intuition that came from his human senses; perhaps it was conceived when an apple fell on his head, as the story goes. This theory was only testable thanks to the immense amount and accuracy of astronomical data collected with meticulously calibrated tools by Tycho Brahe and Johannes Kepler on the motion of planets across the night sky. It was verified further when the telescope was invented, making it possible for humans to not only see the planets but even track the movement of the moons orbiting them.

Improvements in our ability to measure things allow us to draw stronger conclusions about the accuracy of our predictions and hypotheses. The more data a theory has supporting it, the more likely it is to be true. More accurate measurements also lead to more accurate predictions. For example, predictions about the motion of objects in space are constantly improving, thanks to more accurate estimations of the gravitational constant in Newton's law of gravity. This constant is still being tested and updated frequently, and in 2022, its relative uncertainty was only $2.2\times 10^{-5}$ \cite{NIST2024}. 

Possibly even more important than the improved predictive accuracy provided by new technology is the capability to observe phenomena that would otherwise be undetectable. For example, the discovery of the Higgs boson, a fundamental particle associated with the mechanism that gives mass to other particles, was only possible because of indirect observations provided by the Large Hadron Collider (LHC). This discovery, like many others in particle physics, would have been impossible with direct sensory observation. Even sense-enhancing tools such as microscopes are incapable of assisting with observations on such a small scale. The shift in empiricism from observation through the senses to indirect observations beyond the grasp of humans was pivotal in the transition to modern physics in the early 1900s.

Observations in modern physics, unlike those in classical physics, often conflict with human intuition. As physicists began to explore realms beyond normal experiences, they uncovered phenomena which were unexplained by classical equations. New theories were necessary to describe these phenomena and, thanks to the availability of advanced equipment, they were testable, making them distinctly different from previous hypotheses pertaining to the same subjects. For example, modern atomic theory, which evolved to include electrons, protons, and neutrons, differs greatly from ancient Greek atomism, which was based on philosophy rather than empirical evidence. It is important to note that despite our theories progressing beyond our everyday human experiences, intuition, and comprehension, their main purpose continues to be rooted in explaining observations.

\section{History of Quantum Physics}\label{sec:historyqm}
To help set the context for the rest of this thesis, it is fitting to discuss how quantum physics emerged from empirical observations. Quantum physics originated in the late 19th and early 20th centuries when researchers started exploring the properties of light and matter on atomic and subatomic scales.

\subsection{Quantization}
One of the first major departures from classical physics occurred in 1900 when Max Planck attempted to explain black-body radiation, a type of electromagnetic radiation emitted by objects, e.g. infrared light given off by humans or visible light from the Sun. Classical theories, including Newtonian mechanics and Maxwell's equations, failed to properly predict the color spectrum of light that would be emitted from an object. To explain this, Planck came up with the idea that energy is quantized, meaning that it is emitted or absorbed in small indivisible amounts, called 'quanta.' In 1905, Albert Einstein used a similar idea to come up with the photoelectric effect, which describes how materials absorb light in the form of quantized packets called photons. These were the only theories that properly explained many experimental observations that had been puzzling physicists at the time, but their consequence was that a new type of physics had to be formalized; thus, quantum physics became a new field of study.

The idea of quantization proved to be important for other theories in the following years. For example, in 1913, Niels Bohr introduced a new model of the atom, postulating that electrons occupy fixed orbits around the nucleus with quantized angular momentum. The transitions of electrons between these quantized energy levels would cause photons to be emitted at specific frequencies, so this model provided a satisfactory explanation for the observed emission spectrum from hydrogen atoms. However, this raised further questions about the nature of subatomic particles because it implied that electrons would only be observed at certain energy levels and never in-between, while somehow being able to transition between these energy levels. This mysterious instantaneous transition became known as "quantum tunneling''. 

Experiments continued to be performed to probe the nature of the newly discovered electrons and photons. Initially, theories were formulated on the idea that they were particles that occupy definite amounts of space. But in the famous double-slit experiment performed by Thomas Young on light and Davisson and Germer on beams of electrons, their behavior closely resembled that of waves. In particular, they were able to exhibit constructive and destructive interference, phenomena which only make sense for waves and not for particles. This implied that for photons and electrons, the models of "wave'' and "particle'' were appropriate descriptions in different contexts, a concept called wave-particle duality. This meant that a comprehensive description of such small physical things could not be achieved with a single label such as "particle'',  "wave'', or any known model in classical physics. This provided further evidence that an entirely new branch of science was necessary.

\subsection{Schrödinger's Equation and Cat}
In 1926, Erwin Schrödinger introduced his wave equation, marking a significant advancement in quantum mechanics by providing a way to predict the behavior of physical systems at the atomic and subatomic levels. Schrödinger based his equation on the idea that physics at the quantum level is probabilistic in nature. His wave equation models quantum mechanical systems not as particles or waves but as "wave functions'', mathematical objects that can be used to calculate the probability distribution of a particle's position and momentum. Solutions to the wave equation correspond to different states in which a system can exist. 

In its simplest form, the time-independent Schrödinger equation can be written as follows:
\begin{equation}
    \hat H \mathbf{\psi}= E\mathbf{\psi},
\end{equation}
where $\mathbf{\psi}$ (often written as $\ket\psi$) is the wave function of the system, $E$ is the energy of the system, and $\hat H$ is the Hamiltonian operator (often written as $H$), a mathematical construct analogous to the classical Hamiltonian in William Rowan Hamilton's formation of classical mechanics, which equals the total kinetic and potential energy of a system. Since $\mathbf{\psi}$ is represented by a vector and $E$ a scalar, the Schrödinger equation written in this form is an eigenvalue equation, a type of equation frequently used in linear algebra. The solutions for $\mathbf{\psi}$ and $E$ in this equation are called eigenvectors (or eigenstates) and eigenvalues, respectively. When a Hamiltonian in this equation corresponds to a particular physical system, the eigenstates and eigenvalues correspond to the different quantized states in which that system can be observed. 

An interesting consequence of this model being a linear equation is that it inherits the property from linear algebra that any sum of eigenvectors is also a valid solution to the Schrödinger equation. This suggests that new wave functions can be formed by taking sums of other wave functions. In the process, they may interfere constructively or destructively, like waves (hence the name "wave'' equation). Since such sums of wave functions are valid solutions to the equation, they also correspond to physical states in which the corresponding system can exist. These states are known as superpositions, so named because they are analogous to the classical phenomenon of wave superposition. When wave functions are combined in this way, the resulting state might not be an eigenstate of the Hamiltonian, which means that superposition states do not necessarily correspond to states that can be physically observed. This discrepancy between theoretical possible states and actual observable states is called the measurement problem, and despite the success of Schrödinger's equation in accurately predicting the outcomes of experiments, it is still the subject of much debate due to its counterintuitive nature.

In order to illustrate the measurement problem, Schrödinger came up with a thought experiment about a hypothetical cat in a closed box. A radioactive atom could be placed next to a vial of poisonous gas in the box so that if a radioactive decay occurs, the gas would be released and the cat would die. There would then be two possible observations that could be made when the box is open: either the atom will have decayed and the cat will be dead, or the atom will not have decayed and the cat will be alive. Such a setup, in which the observation of one component (i.e. the atom) reveals information about another component (i.e. the cat), is referred to as an entangled system. Since atoms exhibit quantum effects, the state of the atom can be described as a wave function, and its two observable states are "decayed'' and "not decayed''. By the superposition principle, it is possible for the atom to be in a combined state, a superposition, that looks like the sum of these two observable states. Since the state of the cat is entangled with the state of the atom, the cat could also be in a similar state that looks like the sum of its "alive'' and "dead'' states. This highlights the problem, as it seems impossible for cats to exist in such a state of superposition; they should strictly be either alive or dead.

A popular interpretation of quantum mechanics is the Copenhagen interpretation, which is based on the idea that a system in a superposition state immediately changes into one of the possible observable states for that system at the moment when a measurement (or observation) is made. Under this interpretation, there is no meaningful way to interpret the superposition state that the cat experiences in Schrödinger's thought experiment. Its state before the box is opened is often paradoxically described as "both alive and dead'', which by the definitions of "alive'' and "dead'' is not logically possible. Other interpretations exist, such as the many worlds interpretation (MWI), which explains the measurement problem by postulating that all possible outcomes of quantum measurements actually occur in alternate universes. Theories like the Copenhagen interpretation and MWI are not falsifiable and thus are philosophical in nature rather than scientific, which can also be seen in how they lead to the same empirical predictions for the outcomes of experiments.

The apparent paradox in the Copenhagen interpretation prompted many scientists, including Albert Einstein and Louis de Broglie, to search for alternative theories that consider new or hidden variables in order to avoid the measurement problem. The presence of such variables could allow the interpretation that quantum systems only physically exist in their observable states, so the apparent probabilistic nature of quantum mechanics would be explained by a lack of complete information by observers. However, it was shown by John Stewart Bell that in any case where such a hidden variable exists, a certain mathematical inequality, now called a Bell inequality, would necessarily be satisfied. Experimental evidence has shown that this inequality is not satisfied for systems exhibiting quantum entanglement, suggesting that quantum systems obey an entirely different kind of logic than classical systems.  This discovery led to the foundation of quantum information theory, and the 2022 Nobel Prize in Physics was jointly awarded to Alain Aspect, John Clauser, and Anton Zeilinger for their experiments observing entangled photons violating Bell inequalities \cite{nobel2022}.  Since logical contradictions are impossible to avoid when interpreting the meaning of entangled superposition states, the straightforward but paradoxical explanation of Schrödinger's cat being "both alive and dead simultaneously'' may be the best interpretation we can grasp.

\subsection{Heisenberg's Matrix Mechanics}
Around the same time that Schrödinger introduced wave mechanics, Werner Heisenberg, Max Born, and Pascual Jordan introduced another mathematical formalism to describe quantum mechanics, called matrix mechanics. In their formalism, they modeled measurable characteristics (called observables) of quantum systems as matrices, mathematical objects often used in linear algebra. Matrices have characteristic eigenvalues, which can be interpreted as the possible values that are measurable for an observable, and corresponding eigenvectors, which represent the state of the quantum system when that measurement is made. In matrix mechanics, two observables can be combined by multiplying them together. However, matrices have the property that they are not necessarily commutative, meaning that the order in which they are multiplied may affect the result of that multiplication (i.e. $AB \neq BA$). Some observables, such as position and momentum, do not commute with each other, which implies that the order in which they are measured matters for the outcomes of those measurements. This allowed Heisenberg to mathematically derive the uncertainty principle, which states that certain pairs of observables like position and momentum cannot be known simultaneously for a quantum system.

Schrödinger's wave mechanics and Heisenberg's matrix mechanics turned out to be equivalent descriptions of quantum physics, and they were both accurate at predicting the outcomes of experiments. These testable theories were conceived on the basis of empirical evidence and repeatedly withstand rigorous experimentation, which makes them firmly rooted in the scientific method. As more powerful tools are created to probe the extremes of nature, these theories continue to evolve to help us explain and understand the universe in which we live.

\section{Motivation for Research in Quantum Many-Body Theory}
The main objective of the research presented in this dissertation is to enhance the computational methods used to understand quantum systems, especially nuclear systems. The most advanced supercomputers today face many challenges in simulating even relatively simple systems. Therefore, to gain a better understanding of the nature at the nuclear level, it makes sense to explore alternative computing paradigms such as quantum computing. This thesis will concentrate on quantum computing algorithms for a critical aspect of this problem, namely, preparing eigenstates of quantum Hamiltonians. These algorithms will be elaborated on in subsequent sections, but first, we will examine some intriguing problems in nuclear physics and, more broadly, quantum many-body theory, as well as the limitations of current classical methods to justify using this relatively unexplored computing paradigm.

\subsection{The Strong Force}
Our observations of the universe reveal that nature is governed by four fundamental forces: gravity, electromagnetic force, weak force, and strong force. Although gravity and electromagnetism are generally the most familiar, the weak and strong forces are just as critical for the universe's operations. These latter forces dominate on the scale of the atomic nucleus; the weak force is responsible for radioactive processes such as nuclear fission and fusion, while the strong force binds together hadrons, e.g. protons and neutrons, as well as their constituent quarks.

The significance of the strong force is clear, as atoms would not exist without it (making the universe far less interesting!) However, the precise characteristics of the strong force are not yet fully understood. Experimental physicists are actively scientifically testing theories with particle accelerators, such as the heavy-ion accelerator at the Facility for Rare Isotope Beams in East Lansing, to explore the nature of matter at a nuclear level. Meanwhile, theoretical physicists are developing equations to describe the same properties, but there is still no definitive consensus between experimental results and theoretical predictions. To continue increasing our understanding of the strong force, it is likely that more accurate theoretical predictions are needed to align with the experimental results. To this end, theoretical physicists often explore and develop new computational techniques to take advantage of the vast computing resources available today.

\subsection{Quantum Chromodynamics}
One of the most well-supported and developed hypotheses to describe the strong force is the theory of quantum chromodynamics (QCD). QCD explains the observation that hadrons are made of exactly three quarks each by introducing the concept of color charge, where each quark has one of the three properties labeled red, blue, and green. Analogously to the way positive and negative electric charges attract, these colors attract each other in such a way that they tend to form triplets in the case of baryons (e.g. protons and neutrons) and pairs in the case of mesons (e.g. pions). 

QCD is theoretically capable of predicting all the properties of strongly interacting particles, but extracting information from it requires solving mathematical equations which are often very challenging or impossible to solve. The conditions of QCD are incompatible with perturbation theory, a technique that is often useful for simplifying equations describing quantum systems, so the most common technique for approximately solving these equations is the non-perturbative theory of lattice QCD. In lattice QCD, space and time are modeled as discrete rather than continuous, a simplification that allows for approximate solutions to be reached. Its equations are well defined and are therefore known to have solutions, but the process of finding such solutions is still too computationally intensive, even for the largest modern supercomputers, making lattice QCD unusable for understanding even relatively small nuclei.

\subsection{Nuclear Physics}
The further understanding of nuclear systems can lead to important progress in areas of scientific and technological interest.
In astrophysics, more accurate modeling of neutron stars, dense celestial objects composed primarily of neutrons, could lead to a better fundamental understanding of how matter behaves under extreme conditions.
In nuclear reaction theory, better understanding of the mechanisms in nuclear fusion reactions could lead to breakthroughs in clean energy production, reducing the world's dependence on fossil fuels.
There is also motivation in nuclear structure theory because a better understanding of the nuclear landscape could lead to the discovery of novel isotopes which can be used in medical imaging and radiotherapy.
These examples highlight the global scale of importance in the advancement of knowledge of nuclear physics.

Unlike the more general theory of QCD which explains forces between quarks, the objective of nuclear physics is to describe the nature of the strong force at just the level of the atomic nucleus, which allows for some simplifications. Quark interactions are prominent at higher energies ($\sim$1000 MeV), while most interactions on the scale of the whole nucleus are important at lower energies ($\sim100$ MeV or $\sim10$ MeV). This discrepancy in scale makes it possible to introduce an effective theory, allowing more accurate predictions at the expense of abstracting the theory further from its physical interpretation. By focusing on systems with lower energy, one can model only the interactions between hadrons and ignore the many smaller interactions among the quarks that make them up. 

An example of an effective theory that is useful in the low-energy regime for nuclear physics is the Chiral Effective Field Theory (cEFT), which is closely related to the underlying theory of QCD. In cEFT, an assumption is made that particles exhibit chiral symmetry, a property which implies that quarks are massless. This is not the case in QCD in general, but it is approximately true for low-energy systems involving lighter quarks, such as those found in protons and neutrons. This provides a framework for modeling interactions among hadrons in nuclear systems and has been successfully applied to few-nucleon systems \cite{Epelbaum_2009}. A relatively new approach called nuclear lattice EFT combines cEFT with a lattice approximation to take advantage of the same computational techniques that were developed to study lattice QCD \cite{Lee_2004}. This type of approach is a kind of \textit{ab initio} method, which seeks to describe nuclear systems by solving the Schrödinger equation for the nuclear Hamiltonian,. Other popular \textit{ab initio} methods in nuclear physics include the no-core shell model \cite{Nogga_2006}, Green's function Monte Carlo \cite{gfmc}, and coupled cluster \cite{Hagen_2014} methods.

Despite the simplifications introduced from the low-energy regime in nuclear physics compared to QCD, many challenges remain to fully understand nuclear phenomena. Even when considering just protons and neutrons, the interaction equations are still complex and generally unsolvable, requiring the use of numerical simulations to find approximate solutions. However, known simulation methods (some of which are described in the next chapter) are often inadequate to provide solutions at a useful level of accuracy due to the high computational complexity associated with modeling most nuclear systems.

The complexity of these numerical simulations can be elucidated by the size and number of terms in the quantum Hamiltonian for a nuclear system. The number of interaction terms in a Hamiltonian for any group of interacting particles scales, at minimum, with the number of pairs of particles in the system of interest, which increases quadratically with the number of particles. However, the strong force is known to include many-body interactions, such as the three-body interactions observed in experiments involving the Helium-3 isotope \cite{Sarty:1993zz}. The number of terms in the Hamiltonian that describe these interactions scales cubically with the number of particles. More importantly, every possible quantum state of the system must be in the domain of the Hamiltonian, so its dimensionality increases exponentially with the number of particles in the system. Consequently, the significant memory demand associated with storing quantum states often results in simulations running out of available memory.

In light of the vast scientific and technological implications, the search for more sophisticated computational techniques in nuclear physics is a meaningful endeavor. The better understanding of nuclear systems facilitated by these advancements will not only enhance our understanding of fundamental physical laws but may also lead to breakthroughs in diverse fields such as astrophysics, clean energy, and medical technology. Given the inherent complexities and computational challenges outlined in modeling nuclear interactions, it should be worthwhile to invest in innovative computational strategies, including alternative paradigms like quantum computing.

\chapter{Background}\label{cha:background}
The first part of this chapter describes quantum information and computation, a prerequisite for understanding algorithms on quantum computers. The remainder of the chapter discusses eigenvalue problems, traditional methods to solve them on both classical and quantum computers, and several interesting applications that involve solving eigenvalue problems.

\section{Quantum Information and Computation}
Quantum information and computation is a rapidly advancing field at the forefront of modern physics. It studies the computational capabilities of quantum systems, which behave fundamentally differently from classical systems due to quantum effects like superposition and entanglement. This section explores the fundamentals of quantum computation, and specific algorithms relevant to eigenvalue problems are described in more detail in Section \ref{sec:eigprobs}.

In classical computing, the basic unit of information is the bit, a theoretical unit that can exist in one of two distinct states ("binary'' states). In silicon-based computers, bits are usually stored in transistors, which act as electrical switches. The two possible states of a transistor correspond to different voltage levels, typically labeled 0 and 1. All data on computers are encoded using these bits and can be manipulated through various electrical components to execute programs. The operations performed using bits adhere to the logical principles established in classical information theory\cite{gallager2014information}.

Quantum computing explores the consequences of replacing classical bits with their quantum alternatives, called quantum bits or "qubits." A qubit is defined as a quantum system that can be observed in one of two distinct states. As a quantum system, a qubit can therefore also exist in a superposition of its two states, and multiple qubits can be entangled together so that the information stored in them becomes correlated. Even though the binary nature of qubits makes them analogous to classical bits, their quantum nature leads to behavior that has no direct analogy in classical systems. The manipulations that can be done with qubits therefore adhere to an entirely different set of logical principles established in the different, innovative framework of quantum information theory.

\subsection{Qubits}
The state of a qubit is represented in the Dirac notation as 
\begin{equation}\label{eq:qubit}
\ket{\psi} = \alpha\ket{0} + \beta\ket{1},
\end{equation}
where $\ket\psi$ is the wave function of the qubit, $\ket0$ and $\ket1$ represent its two observable states, and the coefficients $\alpha$ and $\beta$ are complex numbers that hold information about the relative probability of measuring the qubit in their respective states. The states $\ket0$ and $\ket1$ are conventionally written in vector form as $\ket0 = \begin{bmatrix}1 \\ 0\end{bmatrix}$ and $\ket1= \begin{bmatrix}0 \\ 1\end{bmatrix}$. These states together form the "computational basis'' that spans the set of possible superposition states, allowing the state of the qubit to be written as the two-dimensional vector $\ket{\psi} = \begin{bmatrix}\alpha \\ \beta\end{bmatrix}$

Information can be extracted from a qubit through a measurement process, but the measurement result by definition can only be one of the two states $\ket0$ or $\ket1$. Any information held in superposition states is necessarily destroyed when measurement occurs, a feature called wave function collapse. This type of measurement process (also called "projective'' measurement) follows Born's rule, which states that the results $\ket 0$ or $\ket 1$ occur with probabilities $|\alpha|^2$ and $|\beta|^2$, respectively, and immediately afterwards the qubit state is reset to the measured state\cite{nielsen_ma2010}.

In general, $\alpha$ and $\beta$ can be any set of complex numbers that obeys the restriction $|\alpha|^2 + |\beta|^2 = 1$ given by the law of total probability. Theoretically, if they were somehow further restricted to be strictly $0$ or $1$, each qubit would hold the same information as a single classical bit. This provides a hint that quantum computers may be able to stretch beyond the limits of classical computation. In the remainder of this section, we will examine how the capacity of qubits to be in superposed and entangled conditions results in novel and unique computational methods.

\subsubsection{The Bloch Sphere}
\begin{figure}
    \centering
    \includegraphics[width=0.5\linewidth]{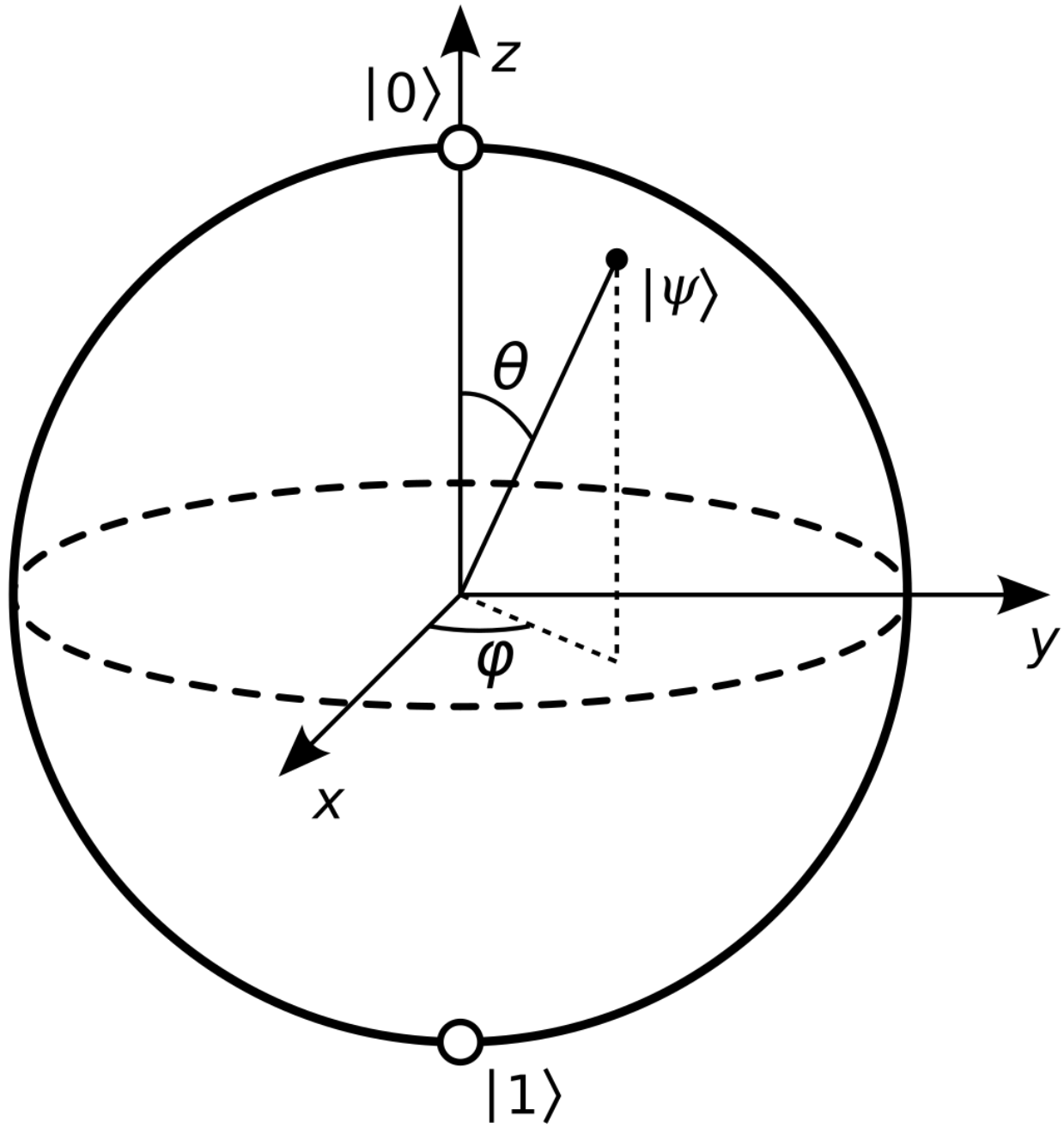}
    \caption{The Bloch sphere representing the state of a single qubit as coordinates on a sphere \cite{blochsphere}}
    \label{fig:blochsphere}
\end{figure}

The coefficients $\alpha$ and $\beta$ in Equation \ref{eq:qubit} are connected to physical meaning by the condition that the square of their magnitudes represents a probability. Since complex numbers have real-valued magnitudes, these coefficients may have imaginary components. When combined with the requirement that these probabilities add up to 1, the space defined by the possible values of $\alpha$ and $\beta$ can be represented by the points on a spherical surface, leading to a visualization aid called the Bloch sphere (shown in Figure \ref{fig:blochsphere}). 

The two observable states $\ket0$ and $\ket1$ appear at the north and south poles of the Bloch sphere, and each other point on the sphere corresponds to a distinct pure superposition of those two states. Furthermore, the points inside the Bloch sphere, rather than on its surface, represent potential mixed states of the qubit. These mixed states characterize the state of a qubit when its wave function is partially unknown, a scenario that frequently arises in practice because of unpredictable interactions of quantum systems with their surroundings.

The Bloch sphere is also useful for visualizing operations that can be performed on a qubit. Due to the conservation of probability, all operations that transform a quantum system from one state to another are necessarily unitary operations; the quantum states before and after must have the probabilities of all their possible measurements sum to 1. For a single qubit, the set of all possible transformations is therefore equivalent to the special unitary Lie group SU(2), which is homomorphic to the group of rotations around a sphere, SO(3) \cite{su2so3}. This allows for any operation on a qubit to be visualized as a rotation around the Bloch sphere.

\subsubsection{Combining Multiple Qubits}

Not many interesting computations can be done with a single qubit, so understanding the behavior of multiple qubits is crucial to unlocking the full potential of quantum computation. The combined state for a collection of qubits is represented by a tensor product of the individual states of the qubits. For example, the state of a system of two qubits, each described by an individual wave function \(\ket{\psi_1}\) and \(\ket{\psi_2}\), is written as:

\begin{equation}\label{eq:twoqubits}
\ket{\psi_1} \otimes \ket{\psi_2} \equiv \ket{\psi_1\psi_2}.
\end{equation}

The state of this two-qubit system can be expressed as a superposition of the four states: \(\ket{00} = \begin{bmatrix}
    1 \\ 0 \\ 0 \\ 0
\end{bmatrix}, \ket{01}=\begin{bmatrix}
    0 \\ 1 \\ 0 \\ 0
\end{bmatrix}, \ket{10}=\begin{bmatrix}
    0\\ 0 \\ 1 \\ 0
\end{bmatrix}, \ket{11}=\begin{bmatrix}
    0\\ 0 \\ 0 \\ 1
\end{bmatrix}\), which is referred to as a "product basis''. For example, a general two-qubit state $\ket\Psi$ can be written as:

\begin{equation}\label{eq:twoqubits_superposition}
\ket{\Psi} = \alpha_{00}\ket{00} + \alpha_{01}\ket{01} + \alpha_{10}\ket{10} + \alpha_{11}\ket{11},
\end{equation}

where \(\alpha_{00}, \alpha_{01}, \alpha_{10},\) and \(\alpha_{11}\) are complex coefficients that satisfy the normalization condition:

\begin{equation}
|\alpha_{00}|^2 + |\alpha_{01}|^2 + |\alpha_{10}|^2 + |\alpha_{11}|^2 = 1.
\end{equation}

A set of qubits taken together is called the quantum register. The set of possible pure states for a quantum register is a complex linear vector space (called a Hilbert space) spanned by its product basis states. In general, the state of an $n$-qubit system can be represented as a vector in a $2^n$-dimensional Hilbert space  \cite{nielsen_ma2010}.

\subsubsection{Entanglement}

Entanglement occurs when the state of one qubit cannot be described independently of the state of another. This phenomenon leads to correlations between measurements of the qubits that are stronger than what is possible in classical systems.

A well-known example of an entangled state is the Bell state, which for two qubits can be written as:

\begin{equation}\label{eq:bellstate}
\ket{\Phi^+} = \frac{1}{\sqrt{2}} (\ket{00} + \ket{11}).
\end{equation}

In this state, a measurement of the first qubit will yield information about the second qubit, and vice versa. For example, if the first qubit is measured as \(\ket{0}\), the second qubit will also be \(\ket{0}\), and if the first qubit is \(\ket{1}\), the second qubit will be \(\ket{1}\). This type of state is called maximally entangled because a measurement on one qubit provides complete information about the states of both qubits.

Entanglement enables quantum teleportation, a phenomenon in which information about the state of a qubit can be obtained without access to the physical qubit itself. Measurement of one qubit in a pair with maximum entanglement, for example, causes the wave function of the other qubit to collapse into a basis state $\ket0$ or $\ket1$. Since entangled qubits can theoretically remain in an entangled state even when physically separated, there are some situations in which information can be transmitted in fundamentally different ways than when it is stored classically.

Furthermore, when two qubits are in an entangled state, manipulations on one qubit can affect the wave function of the other qubit, even when they are completely isolated. For example, if two qubits are in the Bell state in Equation \ref{eq:bellstate}, and the first qubit is manipulated so that it has a 75\% chance to be consequently measured in the $\ket0$ state, then the second qubit will likewise have a 75\% chance to be measured independently in that state. The operations on one qubit in an entangled pair can therefore affect the state of the entangled qubit even before any measurements are made.

The idea that entangled states store information differently is illustrated in many pedagogical examples that have been developed in the new field of quantum game theory. One particularly striking example is the quantum prisoners' dilemma. In the well-known prisoners' dilemma problem (which we will call the classical prisoners' dilemma), two players each have to decide between options A and B without communicating. If both choose A, they both receive mild punishment; if they both choose B, they both receive slightly harsher punishment; if both players choose different options, the one who chose A receives much harsher punishment and the one who chose B receives a reward. This scenario is set up so that the optimal strategy for players acting in self-interest is to always choose option B because it leads to a better outcome than A no matter which option the other player chooses. This makes the case where both players receive slightly harsher punishment a so-called Nash equilibrium \cite{poundstone1992prisoner}. 

In the quantum prisoners' dilemma, two players are each given exactly one qubit from a pair of entangled qubits and told that the two possible measurements of the qubit correspond to options A and B. They are then allowed to perform any kind of unitary manipulations on their qubit to influence the relative probability of choosing options A and B. The players are given full information on how the entangled state was prepared, so they could theoretically undo their entanglement and choose their unitary operations to yield 100\% chance of choosing one of the options, meaning that they could choose to ignore the quantum aspect and fall back to the optimal strategy from the classical prisoners' dilemma. However, the entanglement in their qubits means that manipulations on one qubit affect the wave function of the other qubit, so the players technically share information despite not being able to communicate. It can be shown that, depending on how entangled the qubit is, the Nash equilibrium can be different for this scenario than in the classical case. In such scenarios, it can be shown that there exists a stochastic strategy for players that produces a better outcome on average than the optimal strategy for the classical case \cite{Eisert_1999}. 

This example helps to illustrate the role of quantum entanglement in information theory. In the quantum prisoners' dilemma, entanglement provides a form of shared information that can be used to inform decisions. This shows how entanglement can be used as a resource, effectively "storing'' information in the form of correlations between qubits in such a way that cannot be replicated on memory registers that store only classical information.

\subsection{Quantum Gates and Circuits}

In classical computation, information is processed through logic gates (such as the AND and OR gates) that operate on classical bits, changing their values according to predefined rules. Similarly, quantum circuits employ quantum gates to manipulate the quantum states of qubits.

An $n$-qubit quantum register can be represented as a vector in a $2^n$ dimensional Hilbert space, and the quantum gates that may act on that quantum register can be represented by unitary matrices in the Lie group SU($2^n$). Since unitary matrices are reversible, quantum gates preserve all information within the quantum register. This fundamentally distinguishes them from classical logic gates; for example, the AND gate has two input bits and one output bit, making it impossible to infer the state of the bits before the operation from the output alone.

\begin{figure}\label{fig:bellstate}
\centering
\includegraphics[width=0.5\linewidth]{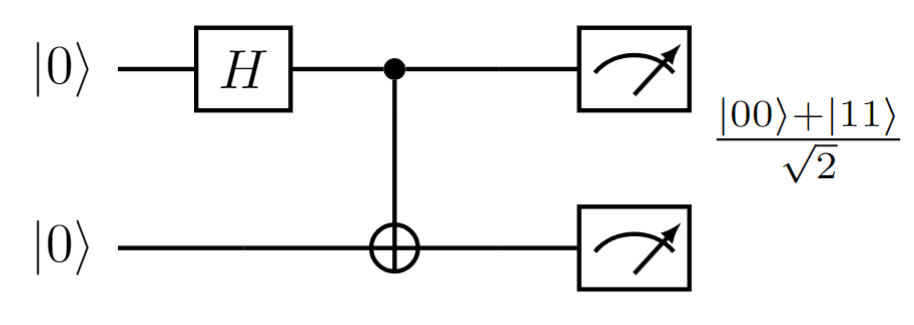}
\caption{A basic quantum circuit diagram for creating a Bell state. The initial state on the left is $\ket{00}$, the gates are applied from left to right, leading to the final state $\frac{1}{\sqrt2}(\ket{00} + \ket{11})$. The meter symbols represent measurements, which is this case will yield either both qubits in the $\ket0$ state or both qubits in the $\ket1$ state, with 50\% probability for each outcome. }
\end{figure}

Multiple quantum gates form a quantum circuit that can then be used as instructions to execute a quantum algorithm. Quantum circuits are often visually represented via quantum circuit diagrams; for example, the quantum circuit diagram to produce the Bell state in Equation \ref{eq:bellstate} is shown in Figure \ref{fig:bellstate}. The gates in these diagrams are applied sequentially from left to right on the quantum register, starting from an initial state, usually $\ket0^{\otimes_N}$ (where all qubits are in the $\ket0$ state). These operations culminate in a final quantum state which, upon measurement, provides the outcome of the quantum computation.


In a quantum circuit diagram, each qubit is depicted as a horizontal line. Gates can be represented by boxes, and some types of entangling gates can have lines connecting multiple qubits. The meter symbol, often the rightmost symbol on a qubit line, represents a measurement operation.

Because quantum operations are linear, the behavior of a quantum gate is fully characterized by its effect on a collection of basis states. Hence, the operation of each single-qubit gate can be described by its effect on the computational basis, while the operation of each multi-qubit gate can be understood through its effect on the corresponding product basis. The following are some gates that are frequently used in quantum computing algorithms.

\paragraph{The Hadamard Gate (H)} is a single-qubit gate that is typically used as the default method for creating superposition states from basis states. The Hadamard gate is represented in the quantum circuit as a boxed "H'', as in Figure \ref{fig:bellstate} and can be written in matrix form as:

\begin{equation}\label{eq:hadamardgate}
\textbf H = \frac{1}{\sqrt 2}\begin{pmatrix}
    1 & 1 \\
    1 & -1
\end{pmatrix} 
\end{equation}
This results in a rotation of $\pi$ radians about the axis $\frac{(\hat x + \hat z)}{2}$ on the Bloch sphere.  The output of a Hadamard gate acting on a qubit in either the \(|0\rangle\) or $\ket1$ state results in an even superposition state on the equator of the Bloch sphere. These states are $\textbf H\ket0 = \frac{1}{\sqrt2}(\ket{0} + \ket1) = \ket+$ and $\textbf{H}\ket1 = \frac{1}{\sqrt2}(\ket0 - \ket1) = \ket-$, which are orthonormal and form their own basis in a 2-dimensional Hilbert space. Consequently, $\textbf{H}$ can be viewed as a basis transformation that interchanges the $\hat x$ and $\hat z$ axes of the Bloch sphere. Additionally, it is involutory, meaning that it serves as its own inverse. This makes it useful for switching back and forth between the bases $\{\ket0, \ket1\}$ and $\{\ket+, \ket-\}$.

\paragraph{Pauli Gates (X, Y, Z)} perform rotations of qubit states around different axes on the Bloch sphere. They can be written in matrix form as:

$\textbf{X} = \sigma_x = \begin{pmatrix} 0 & 1 \\ 1 & 0 \end{pmatrix}$,
$\textbf{Y} = \sigma_y = \begin{pmatrix} 0 & -i \\ i & 0 \end{pmatrix}$,
$\textbf{Z} = \sigma_z = \begin{pmatrix} 1 & 0 \\ 0 & -1 \end{pmatrix}$.

All of these matrices have eigenvalues of $\pm 1$. The X gate, also known as the quantum bit-flip gate, has the effect $\ket0\rightarrow\ket1$ and $\ket1\rightarrow\ket0$, making it the quantum analogue of the classical NOT gate for the computational basis. Similarly, the Y and Z gates cause a quantum state to rotate around the $\hat y$ and $\hat z$ axes of the Bloch sphere, respectively. They are all involutory and thus serve as their own inverses. A useful property of Pauli matrices is that any $2\times2$ matrix can be represented as a linear combination of Pauli matrices and the identity matrix, $I = \begin{pmatrix}
    1 & 0 \\ 0 & 1
\end{pmatrix}$.

\paragraph{Rotation Gates ($R_x, R_y, R_z$)} are operators that can each be written as the matrix exponential of a Pauli gate, i.e. $R_j = e^{-i \sigma_j\theta/2}, j\in(x,y,z)$. These operators represent partial rotations around the $\hat x$, $\hat y$, and $\hat z$ axes of the Bloch sphere, respectively. The parameter $\theta$ dictates how many radians the state rotates, with a value of $\pi$ making the rotation equivalent to a Pauli gate operation.

\paragraph{Phase Shift (\textbf{P})} gate is written in matrix form as:

\begin{equation}\label{eq:phaseshiftgate}
\textbf P(\phi) = \begin{pmatrix}
    1 & 0 \\
    0 & e^{i\phi}
\end{pmatrix}
\end{equation}

The phase shift operation only affects the \(\ket1\) state, while leaving the \(\ket0\) state unchanged. It is equivalent to the $R_z$ gate when $\phi  = \pi$. This gate allows for precise manipulation of qubit phases, sometimes creating conditions in which wave functions destructively interfere, which is beneficial for some quantum algorithms and error correction schemes.

\paragraph{Controlled Gates } are multi-qubit gates that operate conditionally based on the state of a control qubit. The most commonly used example is the \textbf{CNOT} or "controlled-NOT'' gate, which can be represented in matrix form in the product basis as:

\begin{equation}\label{eq:cnotgate}
\textbf{CNOT} = \begin{pmatrix}
    1 & 0 & 0 & 0 \\
    0 & 1 & 0 & 0 \\
    0 & 0 & 0 & 1 \\
    0 & 0 & 1 & 0
\end{pmatrix}
\end{equation}

\begin{figure}\label{fig:cugate}
\centering
\includegraphics[width=0.25\linewidth]{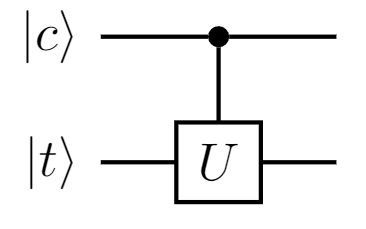}\caption{A circuit diagram for a Controlled-U gate.}
\end{figure}

The CNOT gate flips the state of the target qubit if and only if the control qubit is in the \(\ket1\) state. Specifically, it performs the transformation \(\ket{c,t} \rightarrow \ket{c,c \oplus t}\), where \(\ket{c}\) and \(\ket{t}\) are the states of the control and target qubits, respectively, and \(\oplus\) denotes the XOR operation. It is shown on circuit diagrams as a line connecting a solid circle on the control qubit line with a hollow circle on the target qubit line. It is often used as a simple way to create entangled states, as in Figure \ref{fig:bellstate}. To draw other controlled gates, the hollow circle in the diagram can be replaced with any other gate; for example, the general "controlled-U'' gate is shown in Figure \ref{fig:cugate}. For example, if $U = \begin{pmatrix}
    u_{00} & u_{01} \\
    u_{10} & u_{11}
\end{pmatrix}$, this corresponds to the matrix:

\begin{equation}\label{eq:cugate}
\textbf{CU} = \begin{pmatrix}
    1 & 0 & 0 & 0 \\
    0 & 1 & 0 & 0 \\
    0 & 0 & u_{00} & u_{01} \\
    0 & 0 & u_{10} & u_{11}
\end{pmatrix}
\end{equation}

\paragraph{Toffoli Gate (CCNOT)} is a three-qubit gate, also known as the controlled-controlled-not gate. It extends the concept of the CNOT gate by flipping the state of the target qubit only if both control qubits are in the \(\ket1\) state. The Toffoli gate can be represented in matrix form as:

\begin{equation}\label{eq:toffoligate}
\textbf{Toffoli} = \begin{pmatrix}
    1 & 0 & 0 & 0 & 0 & 0 & 0 & 0 \\
    0 & 1 & 0 & 0 & 0 & 0 & 0 & 0 \\
    0 & 0 & 1 & 0 & 0 & 0 & 0 & 0 \\
    0 & 0 & 0 & 1 & 0 & 0 & 0 & 0 \\
    0 & 0 & 0 & 0 & 1 & 0 & 0 & 0 \\
    0 & 0 & 0 & 0 & 0 & 1 & 0 & 0 \\
    0 & 0 & 0 & 0 & 0 & 0 & 0 & 1 \\
    0 & 0 & 0 & 0 & 0 & 0 & 1 & 0
\end{pmatrix}
\end{equation}

In terms of state transformation, the Toffoli gate performs \(\ket{c_1, c_2, t} \rightarrow \ket{c_1, c_2, t \oplus (c_1 \cdot c_2)}\), where \(\ket{c_1}\) and \(\ket{c_2}\) are the control qubits, and \(\ket{t}\) is the target qubit. The Toffoli gate is often used in quantum computing to create complex entangled states and implement error correction codes. It is instrumental in the construction of certain quantum circuits, such as those required for arithmetic operations and fault-tolerant quantum computation. 

\subsubsection{Universality of Quantum Gates}
In the context of quantum computation, a set of gates is considered universal if any unitary operation can be approximated to arbitrary accuracy using only gates in that set. This means that any quantum algorithm can be implemented using sequences of these gates. A commonly used set of gates are the rotation gates, the phase shift gate, and the CNOT gate \cite{williams2011}.  The Toffoli gate is notable for being sufficient by itself for universal classical computation, and the set of the Toffoli gate and the Hadamard gate is universal for quantum computation \cite{toffoliproof}. Any physical quantum computing device that can implement a set of universal quantum gates is considered a universal quantum computer, meaning that it can theoretically be used to execute any kind of quantum algorithm.

\subsubsection{Extracting information from Quantum Computers}
In quantum computers, qubits exist in superpositions which collapse into classical states when measured. Thus, the determination of the quantum state of a quantum register must be done statistically, requiring multiple circuit executions and repeated measurements. Many useful computations on a quantum computer require an iterative process of preparing a quantum register in a particular state, measuring the classical state into which it collapses, and updating a statistical analysis to estimate an expectation (average) value of an operator with respect to the quantum state. The precision of this estimation increases with the number of repetitions due to the Law of Large Numbers \cite{largenum}. 

The two states that form the computational basis, $\ket0$ and $\ket1$, are the eigenvectors of $\sigma_z$. Measurements on a quantum computer can, therefore, be used straightforwardly to estimate the expectation value of $\sigma_z$ in the state of the quantum register. To measure the expectation value of \(\sigma_x\) in the state of a qubit, one can apply a Hadamard gate prior to measurement. Similarly, the expectation value of \(\sigma_y\) can be obtained by first executing a \(\frac{\pi}{2}\) phase shift gate, followed by a Hadamard gate, before measurement. These three types of measurements are called Pauli measurements. They can be thought of as measurements in the bases $\mathbf{X}$, $\mathbf{Y}$, and $\mathbf{Z}$ of a qubit, and the gates applied before measurement can be thought of as rotations into their respective basis, starting from the $\mathbf{Z}$ basis.

Quantum state tomography, the process of approximately reconstructing a quantum state based on measurements, can be performed by using the three Pauli measurements on an ensemble of identical quantum registers. This approximation can be made more efficient by using different measurement techniques other than Pauli measurements \cite{qctomography}. Some algorithms are designed to directly estimate desired quantities without the need for tomography; for example, Grover's search algorithm uses only one measurement on each qubit at the end of a circuit \cite{grover1996fastquantummechanicalalgorithm}.

In the context of eigenvalue problems on quantum computers, it is often desirable to estimate the expectation value of a Hamiltonian with respect to the wave function stored in a quantum register. This value, written as $\braket{\psi|H|\psi}$ can be thought of as the "energy'' of the quantum state $\psi$ within the quantum system described by $H$. This energy expectation value can be obtained straightforwardly when the Hamiltonian is expressed as a sum of tensor products of single-qubit Pauli operators, i.e. $H = \sum c_j(\sigma_0\otimes \sigma_1 \otimes \dots \otimes \sigma_n)_j$ where each $\sigma_i$ is in the set $\{\sigma_x, \sigma_y, \sigma_z, I\}$ and $n$ is the number of qubits in the quantum register.  It is theoretically possible to represent any arbitrary Hamiltonian in this form because these strings of tensor products form a basis for the Hilbert space of the Hamiltonian. Due to the linearity of quantum mechanics, the energy can then be expressed as the sum of the expectation values of these Pauli strings, which can be estimated by taking the corresponding Pauli measurements of each qubit for each Pauli string and summing them together.

\subsection{Quantum Hardware and Limitations}
In the ongoing quest for quantum advantage, a diverse array of quantum hardware architectures have been developed, each with its unique attributes and inherent limitations. The types of quantum computers available in the current era can be divided into two categories: universal quantum computers and analog quantum computers. Universal quantum computers manipulate physical quantum states to perform gate operations and execute arbitrary quantum algorithms. For example, superconducting quantum computers manipulate quantum states found in Josephson junctions\cite{jjunction}, and trapped-ion quantum computers manipulate the electronic states of ions held in place by electromagnetic fields \cite{trapion}. Analog quantum computers forego the circuit model in order to leverage quantum effects of certain physical systems to perform some subset of quantum computing operations more efficiently. For example, to solve optimization problems, quantum annealers take advantage of quantum tunneling effects \cite{qannealing}, and neutral atom arrays take advantage of condensed matter physics in Rydberg atoms \cite{rydbergatoms}. 

Despite the promising capabilities of these quantum hardware architectures, they are not without limitations. Superconducting quantum computers, while scalable and compatible with existing semiconductor fabrication techniques, are highly sensitive to environmental noise. This sensitivity results in short coherence times and requires operation at cryogenic temperatures, which presents significant engineering challenges\cite{Korotkov2009}.
Trapped-ion quantum computers offer longer coherence times due to their isolation from the environment, but face challenges in scaling up while maintaining connectivity between qubits, as the interaction between ions decreases with distance, making high-fidelity operations difficult as the system size expands\cite{maksymov2021detectingqubitcouplingfaultsiontrap}.
Analog quantum computers, including quantum annealers and Rydberg atom array computers, can be advantageous in some situations, but their inability to universally execute quantum algorithms makes it unclear whether they will provide theoretical advantage over classical algorithms\cite{Albash_2018}. Still, there is much optimism that these issues will be minimized over time, as many researchers are continually developing and innovating with quantum hardware.

While researchers are working to mitigate the limitations of quantum hardware, it is still worthwhile to make use of the currently available technology by developing algorithms that can produce interesting results despite high decoherence times and environmental noise. This is the foundation of the Noisy Intermediate-Scale Quantum (NISQ) era, a term coined by John Preskill to describe the current state of quantum computing technology \cite{preskillnisq}.
NISQ devices, which may possess between 50 and a few hundred qubits, are not yet error-corrected and suffer significant operational errors, so they cannot reliably execute long sequences of gates. However, they represent the best quantum computing resources currently available; therefore, there is a pressing need for quantum algorithms that can operate effectively within the constraints of these devices.
It is challenging to devise algorithms that are robust to noise and errors, and many quantum algorithms proposed thus far require error correction or a large number of qubits to provide a significant advantage over classical computers.
For this reason, a promising alternative approach may be to design hybrid algorithms that combine techniques from currently available quantum and classical hardware to surpass the limitations on both sides.

\subsubsection{Optimal Control for Quantum Gates}
Quantum computing algorithms are generally implemented using one- and two-qubit gates, and a many-qubit algorithm can be executed precisely as a sequence of these gates~\cite{shende_vv2004, barenco_a1995}. However, the number of gates required for complex algorithms can become very large, and the short coherence times of NISQ devices impose a strict limit on the runtime of an algorithm before the output becomes dominated by noise. In order to optimize the performance of some algorithms, it is possible to create custom gates customized to the specific hardware and algorithm using optimal control theory ~\cite{holland_et2020}. Gates on physical hardware are often constructed from analog signals (e.g. microwave pulses on superconducting transmons), making this approach similar to hybrid digital-analog algorithms\cite{Parra_Rodriguez_2020}. 

\subsubsection{Qudits}
Qudits extend the concept of qubits to quantum systems with more than two possible states, allowing for more complex state representations in a higher-dimensional state space. Some physical systems used to create qubits in quantum computers are amenable to having more than two possible states when measured. For example, superconducting qubits use the energy levels of a transmon to create basis states for quantum computing, and it is possible for a transmon state to be in a superposition of more than two distinct states. This can allow more quantum information to be stored in the same amount of hardware, possibly leading to more efficient algorithms when combined with optimal control techniques to implement multilevel quantum gates \cite{qudits_xian}.

\section{Eigenvalue Problems and Methods for Solving Them}\label{sec:eigprobs}
The general form of an eigenvalue problem is as follows:

\begin{equation}
A\mathbf{v} = \lambda \mathbf{v}
\end{equation}

Where $A$ is a matrix, $\mathbf{v}$ is an unknown vector and $\lambda$ is an unknown scalar. The solutions for $\mathbf{v}$ and $\lambda$ are called eigenvectors and eigenvalues, respectively.

Eigenvalue problems are prevalent in various disciplines, including nuclear physics, chemistry, and optimization. Eigenvalues and eigenvectors can yield important insights into underlying systems, such as the energy levels of a quantum system or optimal solutions to real-world problems. However, tackling these issues, especially for large and complicated systems, can be computationally demanding. In nuclear physics and chemistry, relevant eigenvalue problems often take the form of solving the Schrödinger equation for a nuclear or molecular Hamiltonian. In combinatorial optimization, eigenvalue problems arise in graph theory, where the largest eigenvalue of a graph's adjacency matrix reveals structural information like maximum degree.

In this section, we will discuss various known computational techniques in both the classical and quantum computing regimes that aid in solving for eigenvalues and eigenvectors.

\subsection{Classical Computational Techniques}
Classical techniques for solving eigenvalue problems include direct diagonalization and iterative methods. Numerical diagonalization involves the transformation of matrices into diagonal forms to reveal eigenvalues using algorithms such as the QR or Jacobi methods \cite{bai2010}. These straightforward methods, while effective for small to medium-sized matrices, become computationally intensive for very large or dense matrices, requiring $O(N^3)$ operations for a $N \times N$ matrix. Considering that the size of the quantum Hamiltonian matrix scales exponentially with the number of particles in the quantum system, this becomes intractable for many reasonably-sized problems. For example, modeling nuclei requires $4^A$ states to fully describe the spin-isospin component, where $A$ is the number of nucleons. Thus, solving the Schrödinger equation for most nuclear Hamiltonians requires the use of iterative methods, which typically compute only a few eigenvalues and eigenvectors rather than the full spectrum. 

In this section, we will explore some iterative methods for eigenvalue problems, relating each method to how it can be used in nuclear physics. 

\subsubsection{The Power Method}
The Power Method is one of the simplest iterative techniques for finding the dominant eigenvalue and its corresponding eigenvector of a matrix. The method involves repeatedly multiplying an arbitrary non-zero vector by the matrix and normalizing the result. Mathematically, this process can be expressed as:
\[ \mathbf{v}_{k+1} = \frac{A \mathbf{v}_k}{\|A \mathbf{v}_k\|} \]
where \( \mathbf{v}_k \) is the vector at iteration \( k \), and \( A \) is the matrix whose eigenvalues are desired. After many iterations, \( \mathbf{v}_k \) converges to the eigenvector associated with the largest eigenvalue of \( A \).

In nuclear physics, the Power Method can be used to estimate the ground-state energy of a system, which corresponds to the most extremal eigenvalue of the quantum Hamiltonian. Although the Power Method is simple and easy to implement, it converges slowly, especially if the dominant eigenvalue is not well separated from the other eigenvalues. Thus, its primary use is as a pedagogical tool to explain how iterative methods can converge on a single eigenvalue more efficiently than direct diagonalization.

\subsubsection{The Lanczos Algorithm}
In many nuclear physics applications, the Hamiltonian matrix is sparse, which means that most of its elements are zero. This sparsity arises because interactions in nuclear systems are typically local and involve only a limited number of neighboring particles or basis states. As a result, each row and column of the Hamiltonian matrix contains only a few non-zero elements. This sparsity makes certain iterative algorithms more effective.

The Lanczos algorithm is an iterative method that is particularly effective in determining a few eigenvalues and eigenvectors of large sparse matrices \cite{linalgch7}. It simplifies the original matrix to a significantly smaller tridiagonal matrix, which is easier to solve. The steps of the Lanczos algorithm can be outlined as follows:

\begin{enumerate}
    \item Begin with an initial vector \( \mathbf{v}_1 \) (often chosen randomly or based on some heuristic) and normalize it. Set \(\beta_0 = 0\) and \( \mathbf{v}_0 = 0\).
    \item  For \( j = 1, 2, \ldots, k \):
    \begin{enumerate}
        \item  Compute \( \mathbf{w}_j = A \mathbf{v}_j - \beta_{j-1} \mathbf{v}_{j-1} \).
        \item Compute \( \alpha_j = \mathbf{v}_j^T \mathbf{w}_j \).
        \item Update \( \mathbf{w}_j \) to \( \mathbf{w}_j = \mathbf{w}_j - \alpha_j \mathbf{v}_j \).
        \item Orthogonalize \( \mathbf{w}_j \) with respect to all previous \( \mathbf{v}_i \) (usually done implicitly).
        \item Compute \( \beta_j = \| \mathbf{w}_j \| \).
        \item Normalize \( \mathbf{w}_j \) to get the next vector \( \mathbf{v}_{j+1} = \frac{\mathbf{w}_j}{\beta_j} \).
    \end{enumerate}
    \item Construct the tridiagonal matrix \( T \) with diagonal entries \( \alpha_j \) and off-diagonal entries \( \beta_j \).
\end{enumerate}

This process generates an orthogonal basis for the Krylov subspace \( \mathcal{K}_k(A, \mathbf{v}_1) \), and the more easily obtained solutions for the eigenvalue problem for the matrix \( T \) can be used to approximate the eigenvalues and eigenvectors of the original matrix \( A \). The algorithm is particularly useful because it significantly reduces the size of the problem compared to direct diagonalization, making it computationally efficient for large sparse matrices.

\subsubsection{Monte Carlo}
Monte Carlo algorithms (named after the city of Monte Carlo in Monaco known for its casinos) use repeated generation of random numbers to estimate quantities of interest. For example, they can be used to estimate the value of an integral of a function without requiring the knowledge of an analytical solution for that integral. For a function of one variable $f(x)$, the integral represents the area under the curve, which can be approximated for a particular interval by choosing random points within the domain and range of $f(x)$ and checking if each point is above or below $f(x)$.  In the context of eigenvalue problems, Monte Carlo techniques can be used to estimate the properties of large matrices or to solve integrals that arise in the computation of matrix elements.

Although the Monte Carlo method is capable of handling large systems, it encounters the notorious "sign problem" when applied to quantum systems. This issue arises from the necessity to sample from a distribution that includes both positive and negative values, leading to cancellations that cause the statistical error to grow exponentially with the size of the system. The sign problem is particularly an issue in simulations involving fermions due to the anti-symmetrization requirement of their wave functions, which results in probability distributions that are not always positive definite\cite{Pan_2024}.

\subsubsection{Eigenvector Continuation}
Several methods have been developed to overcome the sign problem of the Monte Carlo method when used in the context of lattice simulations for fermionic systems \cite{Cox_2000, Aarts_2008, langfeld2016formdensityofstatesmethodfinite, Fucito:1980fh}. One such method is Eigenvector continuation (EC) \cite{ec}.

EC is a variational method for finding extremal eigenvalues and eigenvectors of a Hamiltonian. It involves projecting the changing extremal eigenvector of a Hamiltonian onto a smaller subspace, reducing the dimensionality. This projection speeds up the estimation, making it versatile for various problems. By projecting the Hamiltonian onto a subspace of eigenvectors from selected training parameters and solving the generalized eigenvalue problem, we get a low-dimensional approximation of the true eigenvector. The main advantage of EC is its ability to extrapolate to regions which are inaccessible by direct calculation, similar to the technique of analytic continuation in the complex plane.

\subsection{Quantum Computational Techniques}\label{eigprep}
In this section, we will discuss various techniques that utilize quantum computing resources to solve eigenvalue problems.

\subsubsection{Variational Quantum Eigensolver}
The Variational Quantum Eigensolver (VQE) algorithm is a hybrid quantum-classical method specifically designed to mitigate the challenges of quantum computation posed by noise and other physical limitations in the NISQ era \cite{mccleanvqe}. Primarily used to find the ground-state energy of a quantum mechanical system, the algorithm combines the computational power of quantum systems with the robustness of classical optimization algorithms.

The central idea of VQE is the variational principle of quantum mechanics, which is the observation that the expectation value of the Hamiltonian, $\bra{\psi}H\ket{\psi}$, for any state $\ket\psi$ is always greater than or equal to the ground state energy $E_0$ of the system. The VQE algorithm takes advantage of this principle to approximate the ground state.

The VQE algorithm starts by preparing a quantum state $\ket{\psi(\theta)}$, also known as an ansatz, which is parameterized by a set of classical variables $\theta$. This state is typically prepared using a specialized quantum circuit on a quantum computer. The selection of ansatz can influence the algorithm's performance, and it is typically constructed to reflect some of the known physics of the problem. For example, a commonly used ansatz is the Unitary Coupled Cluster ansatz, which is notably effective for simulating molecular systems because it can include electronic correlations \cite{uccvqe}. Another example is the Hardware Efficient Ansatz, created to exploit the specific features of quantum hardware for more efficient implementation \cite{Kandala_2017}.

The algorithm proceeds by estimating the expectation value $\bra{\psi(\theta)}H\ket{\psi(\theta)}$ of the Hamiltonian using the quantum computer.
The calculated expectation value acts as an objective function for a classical optimization process that iteratively adjusts the parameters $\mathbf{\theta}$. This optimization loop continues to lower the expectation value until a predefined stopping condition is met. These procedures form a quantum-classical cycle: The quantum computer sets up the state and computes the expectation value, while the classical computer updates the parameters. 

This algorithm can be used to determine the lowest eigenvalue of a Hamiltonian, which frequently corresponds to the ground-state energy of a quantum system of interest. This is of interest for nuclear and molecular Hamiltonians because physical systems naturally gravitate towards their ground state, often making them the most important states to understand. However, higher-energy eigenstates can also correspond to physically observable states of a system, so they can also be important to compute. Since VQE is based on the variational principle, it does not readily solve for these states, making additional or alternative techniques necessary\cite{Nakanishi_2019,zhang2020variationalquantumeigensolversvariance}. 

\subsubsection{Quantum Adiabatic Evolution}\label{sec:qae}
One approach to prepare an arbitrary state on a universal quantum computer with arbitrary precision (or "fidelity'') is to simulate the process of adiabatic evolution starting from an easily accessible state. This method is based on the adiabatic theorem in quantum mechanics, which states that a system will stay in its instantaneous eigenstate as long as the perturbation influencing it changes gradually enough and there is a gap between the eigenvalue and the rest of the spectrum. For example, if a quantum system is initialized in the ground state of a basic Hamiltonian, and the time evolution of the state of the quantum register is simulated while the Hamiltonian is gradually transformed into a complex one with the desired ground state, the system will remain in the ground state throughout the evolution.

Although adiabatic evolution can theoretically generate any eigenstate of a quantum Hamiltonian, its practical implementation on NISQ devices is challenging. Achieving a desired state with high fidelity may require a large number of quantum gates, resulting in longer circuit execution times and a higher probability of qubits decohering due to undesired interactions with the environment. This can cause the evolution to deviate significantly from its intended trajectory toward the target state.

A more detailed description of the Quantum Adiabatic Evolution algorithm is given in Chapter \ref{cha:adbopt}, along with a demonstration of an implementation with custom time evolution gates created with optimal control.

\subsubsection{Quantum Approximate Optimization Algorithm}\label{sec:qaoa}

The Quantum Approximate Optimization Algorithm~\cite{Farhi:2000a} and the nearly identical Quantum Alternating Operator Ansatz~\cite{Hadfield_2019} (both called QAOA) are quantum computing algorithms closely related to VQE and Quantum Adiabatic Evolution. Like VQE, QAOA is a hybrid quantum-classical algorithm designed to take advantage of the computational advantages of quantum systems while mitigating their limitations. It uses the same variational approach as VQE but with a particular ansatz that is physically motivated by the adiabatic theorem, making it also closely related to quantum adiabatic evolution.

QAOA begins with a simple quantum state, typically a uniform superposition of all possible solutions, achieved by applying a Hadamard gate to each qubit. The algorithm then applies a sequence of unitary evolutions to this state, which perform the action of the time evolution operator $U = e^{-iHt}$, alternating $H$ between two Hamiltonians, the cost (or "problem'') Hamiltonian $H_f$ and the mixer Hamiltonian $H_i$. The values of $t$ in the time evolution operators are given by two sets of parameters, $\mathbf\beta$ for the time evolutions on $H_i$ and $\mathbf\gamma$ for the time evolution on $H_f$. These evolution operations are alternated for $p$ ``layers" (also called the "depth'') to form the ansatz. Increasing the depth theoretically leads to higher fidelity of state preparation at the expense of increased risk of decoherence due to longer circuit execution times.

The circuit of alternating unitary evolutions can then be used as an ansatz for VQE. The expectation value of $H_f$ with respect to the output state of the quantum register at the end of the circuit is typically used as the cost function in a classical minimization algorithm to update the parameters $\beta$ and $\gamma$. This forms a quantum-classical feedback loop, iteratively optimizing the parameters until the output state with the lowest energy value is found. Due to the variational principle, this state can be used to approximate the ground state of $H_f$.

Similarly to the adiabatic evolution algorithm, QAOA can be thought of as an alternative way to discretize the adiabatic process into a sequence of unitary evolution operations. The difference is that, instead of a continuous and slow evolution between two Hamiltonians, QAOA performs a series of abrupt changes between those two Hamiltonians. This approach allows QAOA to approximate the adiabatic path in a stepwise manner, potentially achieving the same result with fewer resources and in a shorter time.

\subsubsection{Quantum Phase Estimation}

The Quantum Phase Estimation (QPE) algorithm is a fundamental quantum algorithm used to estimate the eigenvalues of a unitary operator. In the context of nuclear physics, unitary operators of interest often represent the time evolution operator \( U = e^{-iHt} \) (which has the same eigenstates as the Hamiltonian $H$). Given an initial quantum state that has a non-zero overlap with a particular eigenstate of this unitary operator, the QPE algorithm estimates the phase (and thus the corresponding eigenvalue) with high precision.

The QPE algorithm works as follows:

\begin{enumerate}
    \item \textbf{Initial State Preparation}: The algorithm begins with an initial state composed of two registers, each: a register of ancilla qubits prepared in a superposition state using Hadamard gates, and a register of system qubits initialized in a state \( |\psi\rangle \) that has a nonzero overlap with an eigenstate of the unitary operator \( U \).

    \item \textbf{Controlled Unitary Operations}: Controlled applications of the unitary operator \( U \) are performed, where each qubit in the ancilla register controls the application of \( U \) on the system register, repeated $2^j$ times, where $j$ is the index of the ancilla qubit. This step encodes the phase information into the ancillary register.

    \item \textbf{Quantum Fourier Transform (QFT)}: After the controlled unitary operations, a Quantum Fourier Transform (QFT) is applied to the ancilla register. This step transforms the phase information from the time domain to the frequency domain, allowing the extraction of the phase.

    \item \textbf{Measurement}: Finally, the ancilla register is measured, yielding an estimate of the phase \( \phi \). This phase is related to the eigenvalue \( \lambda \) of the unitary operator \( U \) by \( U|\psi\rangle = e^{2\pi i \phi}|\psi\rangle \).
\end{enumerate}

In addition to the standard QFT-based QPE, there are several iterative versions of the QPE algorithm that can offer advantages in terms of circuit depth and resource requirements. These iterative methods, such as Kitaev's iterative phase estimation \cite{kitaev2002classical}, involve a sequence of measurements and classical feedback to refine the estimate of the phase \( \phi \).

Iterative QPE estimates the bits of the phase \( \phi \) sequentially. This method uses fewer qubits, reducing the overall complexity of the quantum circuit. The technique involves initially focusing on the most significant bit, where the algorithm determines each bit of the phase through a series of controlled operations and measurements. The outcome of each measurement is then used to adjust the following controlled operations, iteratively refining the phase estimate. The accuracy of the phase estimation can be tuned by the number of iterations, potentially enabling eigenvalues to be estimated with higher precision.

\subsubsection{Quantum Amplitude Estimation}

Quantum Amplitude Estimation (QAE) is a quantum algorithm that extends the principles of QPE to estimate the amplitude of a specific quantum state within a superposition. It has been shown to provide a quadratic speedup over classical Monte Carlo methods in terms of the number of samples required to achieve a given precision \cite{Brassard_2002}.

The QAE algorithm works as follows:

\begin{enumerate}
    \item \textbf{Initial State Preparation}: The algorithm begins by preparing an initial quantum state. This involves two main steps:
        \begin{enumerate}
            \item A unitary operator \( A \) is applied to the initial state \( |0\rangle \) to prepare the state \( A\ket0 = \sqrt{1-a}\ket0 + \sqrt{a}\ket1\), where \( a \) is the amplitude to be estimated.
            \item An ancilla qubit is prepared in the superposition state \( \frac{1}{\sqrt{2}}(\ket0 + \ket1) \) using a Hadamard gate.
        \end{enumerate}
    
    \item \textbf{Quantum Operator Construction}: Define the Grover operator \( Q = A S_0 A^\dagger S_\chi \), where \( S_0 \) and \( S_\chi \) are reflection operators. \( S_0 \) inverts the sign of the amplitude of the \( |0\rangle \) state, and \( S_\chi \) inverts the sign of the amplitude of the states marked by the oracle function \( \chi \).

    \item \textbf{Amplitude Amplification}: Apply the Grover operator \( Q \) iteratively. The number of applications \( Q^k \) increases the amplitude of the target state quadratically with each iteration. The amplitude of the target state becomes \( \sin((2k+1)\theta) \), where \( \theta \) is related to the amplitude \( a \) by \( \sin^2(\theta) = a \).

    \item \textbf{Phase Estimation}: Use the QPE algorithm to estimate the eigenvalues of the Grover operator \( Q \). This involves applying controlled applications of \( Q \) and performing a Quantum Fourier Transform (QFT) on the ancilla register to extract the phase \( \theta \).

    \item \textbf{Measurement and Classical Post-processing}: Measure the ancilla qubits to obtain an estimate of the phase \( \theta \). Using classical post-processing, convert the phase \( \theta \) into an estimate of the amplitude \( a \). The relationship \( a = \sin^2(\theta) \) allows for the extraction of the amplitude from the measured phase.
\end{enumerate}

The number of quantum operations required for this method scales as \( O(1/\epsilon) \), where \( \epsilon \) is the desired precision, compared to the scaling \( O(1/\epsilon^2) \) for the number of classical operations in Monte Carlo methods, representing a quadratic improvement in time complexity. This efficiency makes QAE a promising tool for problems involving probability estimation and expectation value calculations in quantum computing.

There are various modifications and enhancements to the standard QAE algorithm. One notable example is the Maximum Likelihood Amplitude Estimation (MLAE)~\cite{Suzuki_2020} which refines the amplitude estimate using maximum likelihood methods. Instead of directly using the phase estimates to determine the amplitude, the measurement data is fed into a likelihood function that models the probability of observing the data given different possible amplitudes. The likelihood function can incorporate various sources of noise and uncertainty, potentially increasing the robustness and accuracy of the estimate.

\subsubsection{Rodeo Algorithm}

The Rodeo algorithm (RA) is an innovative quantum computing approach based on the suppression of the probability amplitudes of all eigenvectors except the one of interest. Its recursive nature allows for exponential convergence with increasing cycles, and it can be used to prepare any eigenstate, including excited states.

A detailed explanation and a demonstration of the effectiveness of RA are explored as the main subject of Chapter \ref{cha:rodeo}. A novel variational method combining RA and QAOA to prepare eigenstates more efficiently is the subject of Chapter \ref{cha:vra}.

\subsection{Classical Optimization Techniques}
As noted in Section \ref{eigprep}, the preparation of eigenstates on quantum computers is frequently achieved through variational methods, as in VQE, QAOA, or the Variational Rodeo Algorithm presented in Chapter \ref{cha:vra}. These techniques require classical optimization in tandem with quantum circuit operations. Solving such optimization problems can be particularly challenging largely due to the high dimensionality of the problem space, leading to complications such as local minima and barren plateaus. 

In this section, we will explore several classical optimization methods that are commonly used as the classical component in these variational quantum algorithms. These methods are theoretically interchangeable in many applications of variational methods, although some work more efficiently than others. A short description of each approach is included for the curious reader.

\subsubsection{Gradient Descent}
Gradient Descent is a key optimization algorithm, used to find the global minimum of an objective function. It iteratively adjusts the parameters in the direction of the steepest decrease, defined by the negative gradient at the current point. The update rule is:
\[
\theta_{t+1} = \theta_t - \eta \nabla f(\theta_t)
\]
where $\eta$ is the learning rate and $\nabla f(\theta_t)$ is the gradient at $\theta_t$. Despite its simplicity, this method has limitations in high-dimensional spaces, such as in quantum variational algorithms. The convergence rate depends on selecting an appropriate learning rate. It also often becomes "trapped'' in local minima, saddle points, or areas of near-zero gradients known as barren plateaus, hindering it from converging to the true global minimum.

\subsubsection{Newton's Method}
Newton's method is a numerical root-finding algorithm adapted for optimization. It uses the second-order Taylor expansion to find minima of functions. The update rule is:
\[
\theta_{t+1} = \theta_t - H^{-1} \nabla f(\theta_t)
\]
where $H$ is the Hessian matrix of second-order partial derivatives. This method can converge faster than Gradient Descent near the optimum due to second-order information. However, its practicality in quantum optimization is limited. Computing the Hessian and its inverse is costly in high-dimensional spaces. If the Hessian isn't positive definite, convergence to a saddle point or divergence may occur. Due to these difficulties, combined with the complexity of cost functions in quantum optimization, this method is rarely used in practice, but it is still useful in the pedagogical sense to show how optimization techniques can incorporate second-order corrections to the gradient.

\subsubsection{Broyden-Fletcher-Goldfarb-Shanno (BFGS) Algorithm}\label{sec:bfgs}
The Broyden-Fletcher-Goldfarb-Shanno (BFGS) algorithm is a popular quasi-Newton method for unconstrained optimization \cite{bfgs}. Unlike Newton's method, which requires computing the Hessian matrix, BFGS approximates the inverse Hessian using gradients. The parameter update for $\theta$ is:
\[
\theta_{t+1} = \theta_t - \eta B_t \nabla f(\theta_t)
\]
where $B_t$ is iteratively updated using gradients and parameters from previous iterations. BFGS is efficient in high-dimensional spaces and avoids the computational burden of full Hessian calculations, making it suitable for variational quantum algorithms. It often converges faster than Gradient Descent, especially when the cost function is smooth \cite{guerreschi2017practicaloptimizationhybridquantumclassical}.

BFGS is effective in optimizing parameters for variational quantum algorithms such as VQE and QAOA, navigating high-dimensional landscapes more effectively than simpler gradient methods. Its adaptive Hessian refinement helps mitigate issues like barren plateaus by using curvature data to direct the optimization.

However, BFGS can be sensitive to the initial parameters and the nature of the cost function. Some hybrid methods have been proposed that combine BFGS with other strategies to improve robustness and convergence \cite{ibrahim2014, beecolony}.

\subsubsection{Nelder-Mead Method}
The Nelder-Mead method is a derivative-free optimization algorithm that works well for functions that are not smooth or lack derivatives \cite{neldermead}. This makes it a common choice for variational algorithms in which the derivative of the quantum circuit with respect to its circuit parameters may be difficult to compute. It uses a simplex, a geometric shape of \(n+1\) vertices in \(n\)-dimensional space, to iteratively search for the minimum. The algorithm performs the following operations at each step to move the simplex towards the minimum:
\begin{enumerate}
    \item \textbf{Reflection}: Reflect the worst vertex through the centroid of the remaining vertices.
    \item \textbf{Expansion}: If reflection improves the value of the function, expand further in that direction.
    \item \textbf{Contraction}: If reflection does not improve, contract the simplex around the best vertex.
    \item \textbf{Shrinkage}: If contraction fails, shrink the entire simplex towards the best vertex.
\end{enumerate}

This method is advantageous in high-dimensional optimization where gradient information is unavailable or unreliable. However, it can be slow and may not converge to the global minimum, especially in high-dimensional spaces common in variational quantum algorithms on large sets of qubits.

\subsubsection{Conjugate Gradient Method}
The Conjugate Gradient method is an optimization algorithm that is particularly effective for large-scale linear systems originally invented to minimize quadratic functions \cite{cgiterative}. It is an iterative method that improves upon Gradient Descent by considering the previous search direction, creating conjugate directions for faster convergence. The update rule is:
\[
\theta_{t+1} = \theta_t + \alpha_t p_t
\]
\[
p_{t+1} = -\nabla f(\theta_{t+1}) + \beta_t p_t
\]
where \(p_t\) is the conjugate direction, \(\alpha_t\) is the step size, and \(\beta_t\) is computed to ensure conjugacy. This method converges in at most \(n\) steps for \(n\)-dimensional quadratic functions, but is more efficient for other functions as well. Its efficiency and relatively low memory requirement make it suitable for high-dimensional problems encountered in variational quantum algorithms, although some problems tackled by these approaches are nonlinear, which may lead to a decreased efficiency of the CG method compared to quadratic functions \cite{vqenl}.

\subsubsection{Simulated Annealing}
Simulated Annealing is a probabilistic optimization technique inspired by the annealing process in metallurgy \cite{siman}. It aims to find a global minimum by allowing occasional uphill moves to escape local minima. The algorithm uses a temperature parameter that starts high and gradually cools down. The steps are:
\begin{enumerate}
    \item \textbf{Initialization}: Start with an initial solution and temperature.
    \item \textbf{Perturbation}: Modify the current solution slightly to generate a new candidate solution.
    \item \textbf{Acceptance Criterion}: Accept the new solution if it improves the objective function. If not, accept it with a probability that decreases with temperature.
    \item \textbf{Cooling Schedule}: Gradually reduce the temperature according to a predefined schedule.
\end{enumerate}

The probability of accepting worse solutions decreases over time, allowing the algorithm to explore the solution space widely at high temperatures and focus on local improvements at low temperatures. This is particularly useful for problems with many local minima, such as those in variational quantum algorithms. However, it may not be obvious how to properly choose the cooling schedule so that it leads to better convergence.

\subsubsection{Adam Optimization Algorithm}
The Adam (Adaptive Moment Estimation) optimization algorithm, an advanced version of stochastic gradient descent, combines the adaptive learning rate of AdaGrad and the momentum of RMSProp \cite{kingma2017adammethodstochasticoptimization}. The update rules are:
\[
m_t = \beta_1 m_{t-1} + (1 - \beta_1) \nabla f(\theta_t)
\]
\[
v_t = \beta_2 v_{t-1} + (1 - \beta_2) (\nabla f(\theta_t))^2
\]
\[
\hat{m}_t = \frac{m_t}{1 - \beta_1^t}
\]
\[
\hat{v}_t = \frac{v_t}{1 - \beta_2^t}
\]
\[
\theta_{t+1} = \theta_t - \eta \frac{\hat{m}_t}{\sqrt{\hat{v}_t} + \epsilon}
\]
$m_t$ and $v_t$ are the first and second moment estimates of the gradients, $\beta_1$ and $\beta_2$ are the decay rates, $\eta$ is the learning rate, and $\epsilon$ is a small constant to prevent division by zero.

Adam's adaptive learning rate is effective for training models with noisy and sparse gradients, which are common in high-dimensional optimization. It adjusts step sizes to facilitate efficient convergence compared to conventional methods.
Its momentum reduces noise and fluctuations in gradient updates, speeding up optimization and preventing getting stuck in local minima or saddle points, which is useful for cost functions with complicated landscapes.

However, while Adam is highly effective for machine learning tasks due to its robustness and adaptive nature, it is not used as often for variational methods. Currently, machine learning cost functions have much higher dimensionality than those generated from quantum circuits, meaning that methods like Adam which are optimized to work well in such regimes may not be as effective as other optimization algorithms in variational methods. Many variational circuits have cost functions with gradients that can be computed analytically, such as with the stochastic parameter shift rule \cite{Banchi_2021}, which may allow them to take advantage of simpler methods.

\section{Applications of Eigenstate Preparation}\label{sec:apps}
Though the main objective of this research is to improve computational methods to solve eigenvalue problems to better understand nuclear systems, there are many other practical applications that accompany these advances. This section will describe several application areas that may benefit from better eigenstate preparation techniques.

\subsection{Quantum Simulation}
A prime example of a case where quantum computation has the potential to efficiently address problems deemed intractable for classical computing systems is the simulation of the dynamics of large quantum systems. The resources required to solve this issue through classical computing grow exponentially in relation to the size of the system $N$. Conversely, a universal quantum computer can achieve a solution with resources that scale linearly with $N$, under the condition that local interactions drive the system's evolution~\cite{lloyd_s1996, feynman_rp1982}. Before starting the simulation of a system, it is often beneficial to prepare a quantum register in a specific eigenstate of a particular Hamiltonian that represents that system. Quantum systems of practical and theoretical interest to simulate include systems in nuclear physics, chemistry, QCD, and many other fields.

\subsection{Condensed Matter Physics}
Condensed matter physics encompasses a wide array of models that describe the collective behavior of many-body systems. Eigenstate preparation plays a crucial role in understanding these models as it allows for a detailed examination of ground states and excited states, facilitating insights into various physical phenomena.

One prominent example is the transverse field Ising model, which serves as a prototypical system for studying quantum phase transitions \cite{fieldthoeries}. Preparation of the system in its ground state enables exploration of critical behavior and scaling properties. All many-body qubit Hamiltonians can be expressed as a generalized Ising model, which makes them interesting for understanding many types of quantum systems as well \cite{verresen2023quantumisingmodel}. The Heisenberg model, another fundamental model in condensed matter physics, describes interactions in magnetic systems; its eigenstates yield insights into magnetic excitations and spin dynamics \cite{BAXTER1972323}.
In the context of strongly correlated electron systems, the Fermi-Hubbard model is a key model for understanding high-temperature superconductivity and Mott insulator transitions \cite{hubbardmodel}. Preparing eigenstates of this Hamiltonian allows for the analysis of electron correlations and pairing mechanisms.
Similarly, the Bose-Hubbard model describes bosonic particles in an optical lattice and is instrumental in studying superfluid-insulator transitions \cite{PhysRev.129.959}. Its eigenstates illuminate quantum phase transitions and coherence properties in bosonic systems.

\subsection{Quantum Chemistry}
In quantum chemistry, eigenstate problems appear in the form of solving the Schrödinger equation for electronic and vibrational structure Hamiltonians of molecules. Accurate eigenstate preparation can allow for calculation of particular molecular properties, reaction pathways, and potential energy surfaces, which are fundamental to understanding chemical reactions and designing new materials.

For electronic structure calculations, obtaining the ground state and low-lying excited states of a molecular Hamiltonian enables the determination of properties such as ionization energies, electron affinities, and dipole moments. Furthermore, the full spectrum of eigenvalues provides insight into absorption spectra and photoelectron spectroscopy \cite{szabo1996modern}.

Vibrational structure calculations, which involve the preparation of eigenstates of the vibrational Hamiltonian, are important for understanding molecular vibrations and infrared spectra. These calculations aid in the interpretation of experimental data and in the prediction of spectroscopic signatures \cite{bowman1979application}.

One of the promising applications of quantum eigensolvers in quantum chemistry is in drug discovery \cite{cao2019quantum,Santagati_2024,osti_10291929}. The process of discovering new drugs involves identifying molecules that can interact with biological targets in specific ways. This requires a detailed understanding of molecular interactions and the ability to predict how different molecules will behave. If quantum computers can be used to simulate the electronic structures of complex molecules more accurately than classical computers, they may accelerate the process of identifying potential drug candidates.

\subsection{Nuclear Physics}
Quantum eigenstate preparation holds significant promise in advancing our understanding and solving complex problems in nuclear physics \cite{osti_1631143}. This approach is particularly important when the nuclear many-body problem is addressed, where the interactions among multiple nucleons within an atomic nucleus are modeled and studied. Using quantum computing, scientists aim to achieve more accurate and efficient solutions than classical computing can provide.

\subsubsection{Nuclear Structure}
One of the primary applications of quantum eigenstate preparation in nuclear physics is the determination of nuclear structures. This involves calculating the properties of nuclei, such as binding energies, excited states, and transition probabilities. Quantum computing can enable the simulation of nuclear shell model wave functions, for example, which represent the different energy states of a nucleus \cite{P_rez_Obiol_2023}. This can be used to understand the ground and excited states of various isotopes, potentially leading to insights into nuclear stability or the discovery of new medical isotopes. 

\subsubsection{Nuclear Reactions}
Simulating nuclear reactions typically involves computing reaction cross sections and analyzing resonance phenomena by preparing the initial and final states of the interacting nuclei. Accurate simulations of nuclear reactions have applications ranging from energy production in nuclear reactors to the synthesis of elements in stellar environments.
A better understanding of the eigenstates that represent states of interacting particles before and after a nuclear reaction could lead to more efficient designs of nuclear reactors and better predictions of reaction rates in astrophysical processes.

\subsubsection{Quantum Chromodynamics and Nuclear Matter}
Quantum Chromodynamics (QCD), the theory describing the strong interaction that binds quarks and gluons in protons and neutrons, poses significant computational challenges. Better simulations of QCD could enable us to understand the behavior of nuclear matter under extreme conditions, such as those found in neutron stars or heavy-ion collisions. They can also provide insight into phase transitions in nuclear matter, such as the transition from hadronic matter to quark-gluon plasma, a state of matter believed to have existed shortly after the Big Bang~\cite{benhar2020nuclear}.

\subsection{Combinatorial Optimization}
Combinatorial optimization problems are a class of problems in which the objective is to find an optimal solution from a finite set of discrete options. These problems are ubiquitous in various fields such as logistics, scheduling, network design, and more. Many combinatorial optimization problems are classified as NP-hard, meaning that no known polynomial-time algorithm can solve all instances of these problems. Quantum algorithms offer promising approaches to address these challenges more efficiently than classical algorithms.

Importantly, many NP-hard and NP-complete problems can be reduced to the problem of finding the ground state of the transverse-field Ising model \cite{Lucas_2014}. This reduction highlights why quantum algorithms, particularly those designed to find ground states of Hamiltonians, are powerful tools for solving combinatorial optimization problems. In the following sections, we will focus on the different types of combinatorial optimization problems and how they can be formulated for quantum computation.

\subsubsection{Binary Variable Optimization}
Binary variable optimization problems involve variables that can take on one of two possible values (usually 0 or 1). These problems are fundamental in fields such as computer science, operations research, and artificial intelligence. Notable examples include MaxCut, QMaxCut, and Max-k-SAT.

\paragraph{MaxCut} is a classic NP-hard problem where the goal is to partition the vertices of a graph into two disjoint subsets such that the number of edges between the subsets is maximized~\cite{wainwright2008graphical}. The formulation of this problem on a quantum computer involves identifying a Hamiltonian whose ground state corresponds to the optimal partition. MaxCut is relevant in network design to optimize communication or transportation networks, circuit layout design to minimize the number of crossing wires, and clustering in machine learning, where the goal is to group similar items while separating dissimilar ones.

\paragraph{QMaxCut} extends the MaxCut problem to quantum graphs, where edges represent quantum interactions. The formulation involves designing a Hamiltonian that encapsulates the quantum correlations in the graph, enabling the quantum algorithm to find the optimal partition that maximizes these correlations. 
QMaxCut can be applied in quantum communication networks to optimize the quantum entanglement distribution, and in quantum error correction where optimization of the network can improve fault tolerance.

\paragraph{Max-k-SAT} is an extension of the Boolean satisfiability problem (SAT) and is also NP-hard \cite{wainwright2008graphical}. The objective is to determine the assignment of binary variables that satisfies the maximum number of clauses, where each clause has exactly $k$ literals. Formulating this problem on a quantum computer involves constructing a Hamiltonian whose ground state represents the assignment that maximizes clause satisfaction.
Max-k-SAT is relevant in artificial intelligence for constraint satisfaction problems, in software verification to check the correctness of programs, and in scheduling where various constraints must be satisfied simultaneously.

\subsubsection{Discrete Variable Optimization}
Discrete variable optimization problems involve variables that can take on a finite set of discrete values. Examples include Max-k-Cut and the Traveling Salesman Problem (TSP), both of which are fundamental in optimization theory and have numerous practical applications.

\paragraph{Max-k-Cut} generalizes the MaxCut problem to partitioning the vertices of a graph into $k$ disjoint subsets, with the aim of maximizing the number of edges between the subsets \cite{wainwright2008graphical}. Formulating this problem for quantum computation requires identifying a Hamiltonian whose ground state represents the optimal partition into $k$ different subsets \cite{fuchs2020efficientencodingweightedmax}.
Max-k-Cut is applicable in several applications, such as telecommunications to optimize the layout of cellular networks, logistics to efficiently distribute resources, and parallel computing to balance workloads between multiple processors.

\paragraph{The Traveling Salesman Problem (TSP)}~\cite{lawler1985travelling} is one of the most studied NP-hard problems, where the objective is to find the shortest possible route that visits a set of cities exactly once and returns to the origin. Formulating the TSP on a quantum computer involves constructing a Hamiltonian whose ground state corresponds to the shortest route.
TSP is critical in logistics for optimizing delivery routes, in manufacturing for minimizing the movement of robotic arms on assembly lines, and in DNA sequencing where finding the optimal path through a series of genes can significantly speed up the process.


\chapter{Adiabatic Evolution with Optimal Control}\label{cha:adbopt}

In this chapter, we discuss a noise-tolerant strategy for executing arbitrary sequences of unitary transformations with custom quantum gates. This strategy is then used to implement an adiabatic evolution algorithm designed to be run on two superconducting qubits on an IBM Quantum (IBMQ) system. These gates are created on a classical computer using optimal control theory by representing a two-qubit processor as a pair of capacitively coupled superconducting transmons driven by microwave pulses and solving the associated Lindblad master equation. This adiabatic evolution algorithm is tested by emulating specific two-qubit processors available on IBMQ. The results are then compared with those of an equivalent algorithm decomposed into native one- and two-qubit gates and executed on the same IBMQ processors. By reducing the execution time compared to the circuits using native gates, the emulations can prepare a target state with up to 95\% fidelity (measured in overlap between the target and output states), a significant improvement over IBMQ executions which had fidelities ranging from 65\% to 85\% .

\section{The Adiabatic Theorem}
This section details the mathematics behind the adiabatic theorem, the underlying mechanism behind adiabatic evolution.

Controllable quantum systems offer the capability to simulate a system that evolves from an initial Hamiltonian $H_0$ to an arbitrary "target'' Hamiltonian $H_T$ that encodes the dynamics of a system of interest. This evolution can be represented by a time-dependent Hamiltonian:

\begin{equation}
H(t) = f(t) H_0 + g(t) H_T\,,
\label{eq:Ht}
\end{equation}

where $f(t)$ and $g(t)$ are interpolation functions satisfying the conditions:

\begin{equation}
\begin{gathered}
f(0) = 1 - g(0) = 1\\
{\rm and}\\
f(T) = 1 - g(T) = 0\,.
\end{gathered}
\end{equation}

These conditions correspond to the imposing the boundary conditions $H(0)=H_0$ and $H(T)=H_T$. 

The adiabatic theorem states that a quantum system initially in the $k^{\rm th}$ eigenstate of $H_0$ will reach a state arbitrarily close to the $k^{\rm th}$ eigenstate of $H_T$ after a sufficiently long evolution time $T$, provided the $k^{\rm th}$ eigenvalue is continuous throughout evolution and does not cross other levels~\cite{born_m1928}. The time $T$ required to prepare the quantum system within a particular desired error can be estimated in terms of the parameter $s\equiv t/T$ to be of the following order \cite{albash_t2018, boixo_s2010, boixo_s2009}:

\begin{equation}
T \sim O\left(\max_s (|| \partial_s H(s) ||) / \Delta^2 \right)\,,
\end{equation}
Here, $\Delta=\min_s(|\varepsilon_k(s)-\varepsilon_{k\pm1}(s)|)$ is the minimum energy gap between the $k^{\rm th}$ eigenvalue and any other eigenvalue throughout the evolution. The error is bounded above by $\epsilon$ which is defined as follows:
\begin{equation}\label{eq:adiabaticerror}
\epsilon = || P(1) - P_k(1) ||\,,
\end{equation}

where $P(s)$ and $P_k(s)$ are the projectors onto the evolved state and the $k^{\rm th}$ eigenstate of $H$ at time $s$, respectively.

The challenge posed by the quadratic dependence of the total evolution time, $T$, on the inverse of the minimum energy gap, $1/\Delta$, is significant for the implementation of adiabatic evolution as a quantum state preparation technique on both current and near-future quantum devices. The long evolution times that can be necessitated by small energy gaps lead to prolonged implementation times that increase the chances that the physical devices will decohere due to interaction with the environment.

Despite these difficulties, there is a strong incentive to improve the performance of adiabatic evolution as even a slightly faulty implementation can be used in conjunction with alternative eigenstate preparation methods such as phase estimation~\cite{Kitaev:1995qy} and the rodeo algorithm~\cite{Choi:2020pdg, Qian:2021wya}, which have the prerequisite of substantial overlap between the initial state and the target eigenstate. Notably, even a noisy or imperfect adiabatic evolution can yield a significant enhancement in the initial-state overlap, thereby substantially improving the performance of the subsequent quantum-state preparation algorithm. For example, an increase in the initial state overlap from $0.1\%$ to $5\%$ would result in a fifty-fold improvement in the efficiency of the rodeo algorithm due to its time complexity scaling inversely with this overlap. 

\section{Adiabatic Evolution of Two-Spin Systems}
\label{sec:adb2spin}

\begin{figure}[htpb]
\centering
\includegraphics[width=0.8\linewidth]{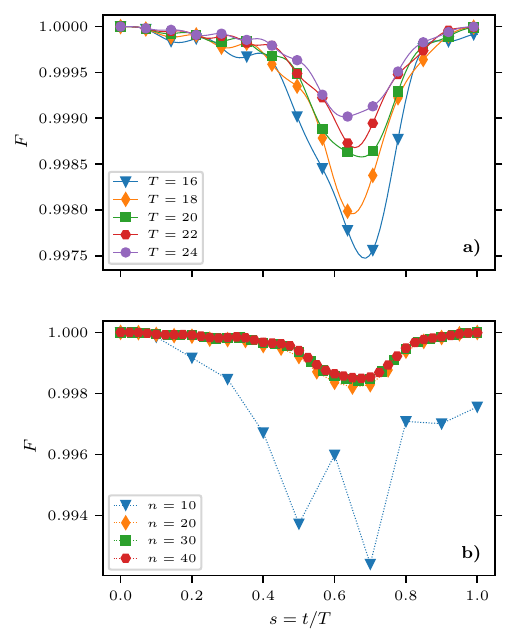}
\caption{(Figure reused from \cite{custom_gate_adiabatic}) of the fidelity between the device state and the ground state of $H(t)$ for various values of {\textbf{a)}} the evolution time $T$ and {\textbf{b)}} the number of time steps $n$ as a function of the parameter $s\equiv t/T$. The adiabatic evolution implemented on quantum devices utilizes $T=20$ and $n=20$, as this combination results in high fidelity while minimizing the number of gates required for device implementation.}
\label{fig:exactevolution}
\end{figure}

In this section, we investigate the preparation of a two-spin system in the ground state of the following Hamiltonian:
\begin{align}
H_T = - \sigma^x_1\sigma^x_2 + \sigma^y_1\sigma^y_2 +
\frac{1}{2} \sigma^z_1 \sigma^z_2 - \sum_{i=1}^2 \sigma^z_i\,,
\label{eq:HT}
\end{align}
by initializing the system in the ground state of
\begin{align}
H_0 = \sum_{i=1}^2 \sigma^x_i\,,
\label{eq:H0}
\end{align}
and carrying out the adiabatic evolution governed by the time-dependent Hamiltonian in Eq.~\eqref{eq:Ht} with interpolation functions
\begin{align}
f(t)=\cos^2(\pi t/2T) & & g(t)=1-f(t)\,.
\label{eq:interpolation}
\end{align}
We note that the ground state of $H_0$ can be represented as a linear combination of the uncoupled two-spin states.
\begin{equation}
    \ket{\phi(0)}
    =\frac{1}{2}\left(\ket{\downarrow\downarrow}-\ket{\downarrow\uparrow}-\ket{\uparrow\downarrow}+\ket{\uparrow\uparrow}\right)\,.
\end{equation}
Furthermore, the ground state of $H_T$ can be represented as follows:
\begin{equation}
    \ket{\phi(T)}
    =\mathscr{N}\left[\left(-1+\sqrt{2}\right)\ket{\downarrow\downarrow}+\ket{\uparrow\uparrow}\right]\,,
\end{equation}
where $\mathscr{N}$ is a constant to enforce the normalization of $\ket{\phi(T)}$. The eigenvalue of this state is $E_T=-2.328$. In a similar manner, the instantaneous ground state $\ket{\phi(t)}$ of $H(t)$ at any particular value of $t$ can be determined by solving the time-independent Schr\"odinger equation $H(t)\ket{\phi(t)}=E_t\ket{\phi(t)}$.  
The solution to the time-dependent Schr\"odinger equation $i\frac{d}{dt}\ket{\psi(t)}=H(t)\ket{\psi(t)}$ will approximate $\ket{\phi(t)}$ with a fidelity given by
\begin{equation}
F(t) = |\langle \phi(t) | \psi(t) \rangle|\,.
\label{eq:fid}
\end{equation}
For an adequately long evolution time $T$, the system's state at time $T$ will closely resemble the ground state of $H_T$, as shown by the exact analysis of the evolution in Figure ~\ref{fig:exactevolution}(a).
In this example, even for a relatively short time of $T=16$, the infidelity $1-F(T)$ at the end of the evolution is upper bounded by $10^{-3}$. In the following analysis, we set $T=20$. This choice is to reach an ideal balance to minimize both the run-time and the associated decoherence errors encountered during the implementation of this adiabatic evolution on current quantum hardware.

The state resulting from adiabatic evolution can be represented through the unitary-time evolution operator corresponding to a time-dependent Hamiltonian, as shown in Eq. (\ref{eq:U}):

\begin{align}
\label{eq:U}
    \ket{\psi(T)} &= \mathcal{U}(0,T)\ket{\psi(0)}\,.
\end{align}

To implement adiabatic evolution on a quantum device, the evolution must be done digitally, so the time evolution operation must be divided into discrete steps. To accomplish this, the evolution operator $\mathcal{U}(0,T)$ can be approximated by the product of $n$ short-time propagators associated with $n$ instantaneous Hamiltonians, as demonstrated in Equation \ref{eq:stp}:

\begin{equation}
\mathcal{U}(0,T) \approx \prod_{k=1}^{n} U(t_k) =
\prod_{k=1}^{n} e^{-iH(t_k)\Delta t}\,,
\label{eq:stp}
\end{equation}

where $\Delta t=T/n$. The error introduced due to the discretization of the evolution time is proportional to the time derivative of the Hamiltonian, $dH(t)/dt$, and is well-behaved as long as $H(t_k)$ provides a suitable approximation for the average value of $H(t)$ within the time interval $[t_{k-1},t_k]$ \cite{poulin_d2011}. Therefore, the discretization error can be effectively controlled by adjusting the number of steps $n$ into which the evolution time is divided, as shown in Figure \ref{fig:exactevolution} (b), where the fidelity of the final state is evaluated with respect to the solutions $\ket{\psi(t)}$ derived from the unitary time evolution of Equation \eqref{eq:U} using the approximation in Equation\eqref{eq:stp}.

Although increasing the number of steps $n$ reduces the discretization error, we must also consider the noise-induced error that occurs when the circuit is executed on a physical device. This error increases with the number of gates in the circuit implementation, which increases with the number of steps $n$. In Figure \ref{fig:exactevolution}(b), we show how the discretization error becomes negligible when $n$ is as low as 20. Therefore, we adopt $n=20$ in the subsequent analysis.

\section{Two-Qubit Quantum Adiabatic Evolution Implementation}
\label{sec:implementation}
To implement a quantum simulation of the adiabatic evolution discussed in Section \ref{sec:adb2spin}, it is necessary to map the spin degrees of freedom to the quantum processor's states and convert the short-time propagators $U(t_k)$ from Eq.~\eqref{eq:stp} into quantum gates. 
The uncoupled states of the two-spin system can be straightforwardly mapped to the computational states of a two-qubit system. Specifically, the states $|00\rangle$, $|01\rangle$, $|10\rangle$, and $|11\rangle$ are used to represent the two-spin states $\ket{\downarrow\downarrow}$, $\ket{\downarrow\uparrow}$, $\ket{\uparrow\downarrow}$, and $\ket{\uparrow\uparrow}$, respectively.
To convert short-time propagators into quantum gates, two approaches are considered for the main comparison in this chapter: $(i)$ a standard decomposition of each propagator into a circuit of elementary gates and $(ii)$ a direct implementation using a single custom two-qubit gate.

\subsection{Implementation with Elementary Gate Decomposition}\label{sec:elemdecomp}

\begin{figure}[htpb]
\centering
\includegraphics[width=0.8\linewidth]{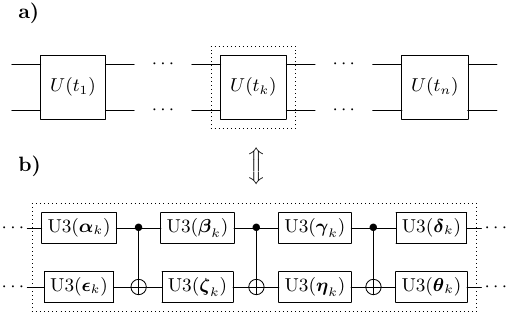}
\caption{(Figure reused from \cite{custom_gate_adiabatic}) Decomposition of the $k$-th unitary transformation in the sequence of Equation ~\eqref{eq:stp} [circuit {\textbf{a)}}] into CNOT and U3 gates [circuit {\textbf{b)}}]. Each U3 gate depends on three Euler angles. The quantum circuit resulting from this decomposition can be directly implemented on IBMQ systems.}
\label{fig:ibmqdecomp}
\end{figure}

The first approach takes advantage of the fact that any unitary operation involving two qubits, such as the short-time propagators $U(t_k)$ in Eq.~\eqref{eq:stp}, can be represented using three CNOT gates controlled by the first qubit and eight one-qubit U3 gates ~\cite{shende_vv2004,vatan_f2004,vidal_g2004}, as illustrated in Fig.~\ref{fig:ibmqdecomp}. 

The eight U3 gates can be written in terms of $x$ and $z$ one-qubit rotations as follows:

\begin{equation}
\begin{aligned}
{\rm U3}(\theta,\phi,\lambda) =& R_z(\phi) R_x\left(-\frac{\pi}{2}\right) R_z(\theta) R_x\left(\frac{\pi}{2}\right) R_z(\lambda) \\
=& \left( \begin{array}{cc}
\cos\left(\frac{\theta}{2}\right) & -e^{i\lambda}\sin\left(\frac{\theta}{2}\right) \\
e^{i\phi}\sin\left(\frac{\theta}{2}\right) & e^{i(\phi+\lambda)}\cos\left(\frac{\theta}{2}\right)
\end{array} \right).
\end{aligned}
\end{equation}

\begin{figure}[htpb]
\centering
\includegraphics[width=0.8\linewidth]{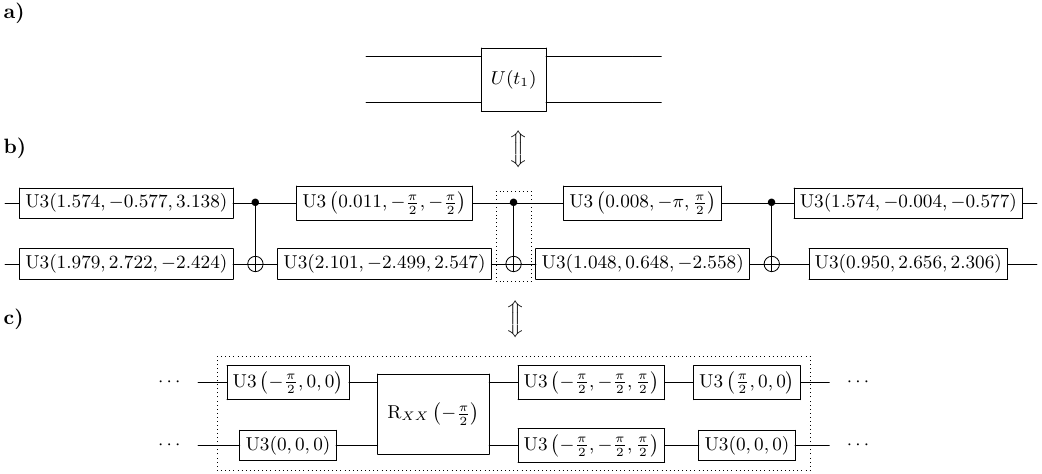}
\caption{(Figure reused from \cite{custom_gate_adiabatic}) Decomposition of the first short-time propagator in Eq.~\eqref{eq:stp} [circuit {\textbf{a)}}] into CNOT and U3 gates [circuit {\textbf{b)}}]. Circuit {\textbf{c)}} shows how the same type of decomposition can be used to decompose the CNOT gate into elementary gates.}
\label{fig:firststpdecomp}
\end{figure}

The combination of these techniques for decomposing an arbitrary two-qubit unitary operation into elementary gates is built in to Qiskit in the function \textit{quantum\_info.two\_qubit\_cnot\_decompose}. As a specific example, the decomposition of the first short-time propagator in Equation \eqref{eq:stp} is shown in Figure \ref{fig:firststpdecomp}

This decomposition is readily implementable on cloud-based quantum computing platforms. The efficacy of this technique was assessed by performing the adiabatic evolution on various IBMQ systems~\cite{ibmq_2021}.
The quantum circuit embodying the corresponding algorithm was constructed utilizing the open-source quantum information software kit (Qiskit)~\cite{qiskit_2021}, which facilitated the decomposition of each short-time propagator into elementary gates using the built-in function \textit{quantum\_info.two\_qubit\_cnot\_decompose}. 
The results of these digital quantum simulations on IBMQ processors are discussed in Section ~\ref{sec:strategy1}.

\subsubsection{Error Mitigation with Confusion Matrix}\label{sec:confusionmatrix}
Simulations of NISQ devices include several types of error, including coherence error and measurement error, some of which can be easily mitigated by basic techniques. In this work, measurement errors are mitigated by inverting the confusion matrix (also known as the "error matrix'') \cite{roggero_a2020}. The occupation probabilities measured in an experiment, arranged in the vector $\ket{c_{\rm exp}^2}$, are related to the true ones, $\ket{c_{\rm true}^2}$, through the confusion matrix $P$ in the following way:

\begin{equation}
\begin{gathered}
\ket{|c|_{\rm exp}^2} = P \ket{|c|_{\rm true}^2}\\
{\rm or}\\
\ket{|c|_{\rm true}^2} = P^{-1} \ket{|c|_{\rm exp}^2},
\end{gathered}
\end{equation}
Assuming that the measurement errors for qubits 1 and 2 are independent, the confusion matrix can be expressed using the single qubit measurement probabilities $p_{ij}$ as follows:
\begin{equation}
\begin{aligned}
\left( \begin{array}{cccc}
(1-p_{10})_1 (1-p_{10})_2 &
(1-p_{10})_1 (p_{01})_2 &
(p_{01})_1 (1-p_{10})_2 &
(p_{01})_1 (p_{01})_2 \\
(1-p_{10})_1 (p_{10})_2 &
(1-p_{10})_1 (1-p_{01})_2 &
(p_{01})_1 (p_{10})_2 &
(p_{01})_1 (1-p_{01})_2 \\
(p_{10})_1 (1-p_{10})_2 &
(p_{10})_1 (p_{01})_2 &
(1-p_{01})_1 (1-p_{10})_2 &
(1-p_{01})_1 (p_{01})_2 \\
(p_{10})_1 (p_{10})_2 &
(p_{10})_1 (1-p_{01})_2 &
(1-p_{01})_1 (p_{10})_2 &
(1-p_{01})_1 (1-p_{01})_2
\end{array}\right)
\end{aligned}
\end{equation}

Prior to the execution of each circuit, we carry out two calibration experiments by initializing the device in the states $ket{00}$ and $\ket{11}$ states. The results of these experiments correspond to the first and fourth columns of the confusion matrix. These results allow us to determine the single qubit measurement probabilities $p_{ij}$ which are then used to form the complete confusion matrix.

\subsection{Modeling the Device Hamiltonian}\label{sec:devham}

\begin{figure}[htpb]
\centering
\includegraphics[width=0.8\linewidth]{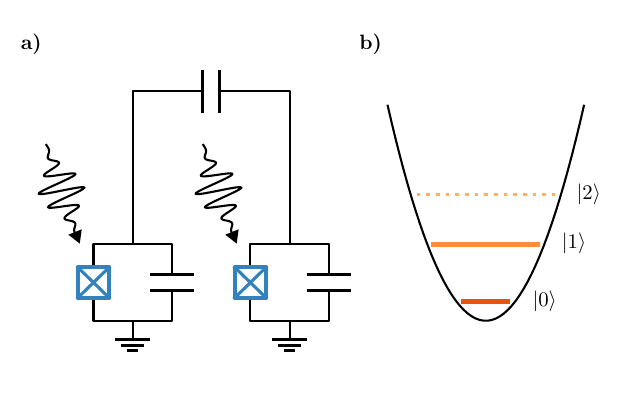}
\caption{(Figure reused from \cite{custom_gate_adiabatic}) Schematic depiction of {\textbf{a)}} the model two-qubit processor, consisting of two capacitively coupled superconducting transmons controlled with microwave pulses, and {\textbf{b)}} the energy spectrum of a superconducting transmon in which the lowest two levels, shown as solid lines, are the computational qubit.}
\label{fig:modelprocessor}
\end{figure}

The alternative strategy involves a realistic representation of a physical quantum device to actualize each short-time propagator in Eq.~\eqref{eq:stp} with a single custom gate. We emulate a two-qubit processor as two capacitively connected superconducting transmons regulated by microwave pulses, as illustrated schematically in Fig.~\ref{fig:modelprocessor} {\textbf{a)}}.

The Hamiltonian for a superconducting transmon, when expressed with respect to the quantity of Cooper pairs ($n$) and the magnetic flux ($\phi$), can be formulated up to the fourth order in $\phi$. This representation is given by~\cite{koch_j2007}:

\begin{equation}
\begin{aligned}
H =& 4E_C n^2 - E_J\cos(\phi)\\
\approx& 4E_C n^2 - E_J + \frac{E_J}{2}\phi^2 - \frac{E_J}{24}\phi^4,
\end{aligned}
\end{equation}

where the energies stored in the capacitor and the Josephson junction are denoted as $E_C$ and $E_J$, respectively. To proceed further, we introduce the creation and annihilation operators for the transmon, with which $n$ and $\phi$ can be defined as follows: 

\begin{equation}
n = i\left(\frac{E_J}{32E_C}\right)^{\tfrac{1}{4}}(a^\dagger-a)
\qquad
\phi = \left(\frac{2E_C}{E_J}\right)^{\tfrac{1}{4}}(a^\dagger+a),
\end{equation}

In terms of these operators, the Hamiltonian takes the form,

\begin{equation}
\begin{aligned}
H \approx& 
\omega a^\dagger a - \frac{\alpha}{6}(a^\dagger+a)^4,
\end{aligned}
\end{equation}

where we defined $\omega\equiv(8E_CE_J)^{1/2}$ and
$\alpha\equiv E_C/2$. According to the Baker-Campbell-Hausdorff formula \cite{hall2003lie}, it can be shown that for a transformation
$U=\exp\left(-i\Omega ta^\dagger a\right)$, the following is true:
\begin{equation}
Ua_1\cdots a_lU^\dagger=e^{\left(-i(m-n)\Omega t\right)}a_1\cdots a_l,
\end{equation}

where $a_1\cdots a_l$ is a chain of creation and annihilation operators ($a_i\in\{a^\dagger,a\}$), and $m$ and $n$ are the number of creation and annihilation operators in the chain, respectively.

Under this transformation, ignoring constant and fast-rotating terms with $|m-n|>1$, the Hamiltonian of the transmon takes the form
\begin{equation}
\begin{aligned}
H\rightarrow H' =& UHU^\dagger + i\dot{U}U^\dagger\\
\approx& (\omega+\alpha+\Omega)a^\dagger a - \alpha a^\dagger aa^\dagger a.
\end{aligned}
\end{equation}

The Hamiltonian of two capacitively coupled transmons controlled by microwave pulses can be approximately written as follows ~\cite{jones_t2021, krantz_p2019}
\begin{equation}
\begin{aligned}
H \approx& \sum_{i=1}^2\left(4E_{C_i}n_i^2 + \frac{E_{J_i}}{2}\phi_i^2 - \frac{E_{J_i}}{24}\phi_i^4\right)
+ \frac{8E_{C_1}E_{C_2}}{E_{C_g}}n_1n_2\\
&+ 2\sum_{i=1}^2 \eta_i \left[\epsilon_{\rm I}^i(t)\sin(\Omega_it) - \epsilon_{\rm Q}^i(t)\cos(\Omega_it)\right] n_i\\
=& \sum_{i=1}^2\left[\omega_i a_i^\dagger a_i - \frac{\alpha_i}{6}(a_i^\dagger + a_i)^4\right] + g(a_1^\dagger - a_1)(a_2^\dagger - a_2)\\
&+ 2i\sum_{i=1}^2 \left[\epsilon_{\rm I}^i(t)\sin(\Omega_it) - \epsilon_{\rm Q}^i(t)\cos(\Omega_it)\right] (a_i^\dagger-a_i),
\end{aligned}
\end{equation}

where the term proportional to $g\equiv 8E_{C_1}E_{C_2}/E_{C_g}\eta_1\eta_2$ with $\eta_i\equiv(32E_{C_i}/E_{J_i})^{1/4}$ describes the crosstalk interaction between the transmons due to their capacitive coupling.
Under the transformation

\begin{equation}
U=\exp\left(-i\Omega_1t a_1^\dagger a_1 - i\Omega_2t a_2^\dagger a_2\right),
\end{equation}
By substituting $\Omega_i=-\omega_i-\alpha_i$, assuming that $\Omega_1\approx\Omega_2$ and dropping constant and fast-rotating terms, the above Hamiltonian takes the form
\begin{equation}
\begin{aligned}
H \approx& -\sum_{i=1}^2 \alpha_i a_i^\dagger a_i a_i^\dagger a_i - g\left(a_1^\dagger a_2 + a_1 a_2^\dagger \right)\\
&+ \sum_{i=1}^2 \left[\epsilon_{\rm I}^i(t)(a_i^\dagger+a_i) - i\epsilon_{\rm Q}^i(t)(a_i^\dagger-a_i)\right],
\end{aligned}
\end{equation}

For simplicity, we rewrite this Hamiltonian as follows.

\begin{equation}
H_{\rm QPU}(t) = H_{\rm d} + H_{\rm c}(t)\,,
\label{eq:HQPU}
\end{equation}

where

\begin{equation}
H_{\rm d} \approx
-\sum_{i=1}^2 \alpha a_i^{\dagger}a_i a_i^{\dagger}a_i -
g \left( a_1^\dagger a_2 + a_2^\dagger a_1 \right)
\end{equation}

constitutes the drift Hamiltonian of the undisturbed device, and 

\begin{equation}
H_{\rm c}(t) = \sum_{i=1}^2 \left[
\epsilon_{\rm I}^i(t) (a_i^\dagger + a_i) -
i \epsilon_{\rm Q}^i(t) (a_i^\dagger - a_i) \right]
\end{equation}

represents the time-variant Hamiltonian delineating the governance of the quantum processor through irradiation with resonant microwave pulses.

\subsection{Implementation with Custom Gates}

We can now use the Hamiltonians from Section \ref{sec:devham} to aid in an optimal control scheme to devise custom gates for transmon qubits. For the anharmonicity of both transmons, we use $\alpha=200$~MHz, and for the strength of the crosstalk interaction strength between the transmons due to capacitive coupling, we use $g=3$~ MHz. The time-dependent amplitudes $\epsilon_{\rm I}^i(t)$ and $\epsilon_{\rm Q}^i(t)$ correspond respectively to the in-phase and quadrature tunable pulse sequences that control transmon $i$. We consider the first three energy levels of each transmon. The computational two-qubit states are defined by the subspace of states with zero and one quanta per transmon. By explicitly including states with two quanta in at least one of the transmons, we account for higher-energy states that can be populated due to gate error and decoherence. This inclusion also enables our control pulses to reduce leakage by blocking transitions to these states, analogously to the Derivative Removal by Adiabatic Gate (DRAG) algorithm ~\cite{motzoi_f2009}.

To compute the custom two-qubit gates that realize each of the short-time propagators in Eq.~\eqref{eq:stp}, we determine the pulse sequences $\epsilon_{\rm I}^i(t)$ and $\epsilon_{\rm Q}^i(t)$ that solve the optimization problem as follows:

\begin{align}
\label{eq:QOC}
U_{\rm QPU}(t_k) &\simeq {\mathcal U}_{\rm QPU}(0,\tau)\\
& = \mathcal{T} \exp\left( -\frac{i}{\hbar} \int_0^\tau
\left[H_{\rm d} + H_{\rm c}(\tau') \right] d\tau' \right),\nonumber
\end{align}

Here, $U_{\rm QPU}(t_k)$ represents the short-time propagator $U(t_k)$ embedded in the Hilbert space spanned by the two-transmon states, and $\mathcal{T}\exp$ represents the time-ordered exponential. Using the gradient ascent pulse engineering algorithm (GRAPE)~\cite{khaneja_n2005}, we can find the solution to Equation~\eqref{eq:QOC} with acceptable accuracy by minimizing the objective function.

\begin{equation}
\Phi = 1 - \frac{F_{\rm gate}^2}{2} + \chi \frac{\exp(\bar{\epsilon}^{2n})-1}{\exp(1)-1}\,,
\label{eq:of}
\end{equation}

where

\begin{equation}
F_{\rm gate} = \left|\frac{{\rm tr}(U_{\rm QPU}^\dagger\mathcal{U}_{\rm QPU})}
{{\rm dim}_{\rm QPU}}\right|\,,
\end{equation}

with ${\rm dim}_{\rm QPU}$ representing the dimension of the considered Hilbert space. In the objective function, $\bar\epsilon$ represents the root-mean-squared amplitude of the control pulse normalized to $\epsilon_{\rm cut}$ and can be written as follows:

\begin{equation}
\bar{\epsilon} = \frac{1}{\epsilon_{\rm cut}}
\sqrt{
\frac{1}{\tau} \sum_{i=1}^2\sum_{j\in\{{\rm I, Q}\}} \int_0^\tau \epsilon^i_j(\tau')^2 d\tau' }
\end{equation}

The gate fidelity, $F_{\rm gate}$, appearing in the second term on the right-hand side of Equation~\eqref{eq:of}, serves as a measure of how accurately the pulse-controlled gate reproduces the desired unitary operation. The final term imposes a penalty on large amplitudes by means of the parameters $\epsilon_{\rm cut}$ (cutoff amplitude) and $n$ (cutoff harshness), with their significance determined by the parameter $\chi$. This penalty term prevents high-amplitude solutions where the Hamiltonian approximation used in the optimization of Equation ~\eqref{eq:QOC} becomes invalid and where the hardware implementing the control pulses may not function as intended.

\begin{figure}[htpb]
\centering
\includegraphics[width=0.5\linewidth]{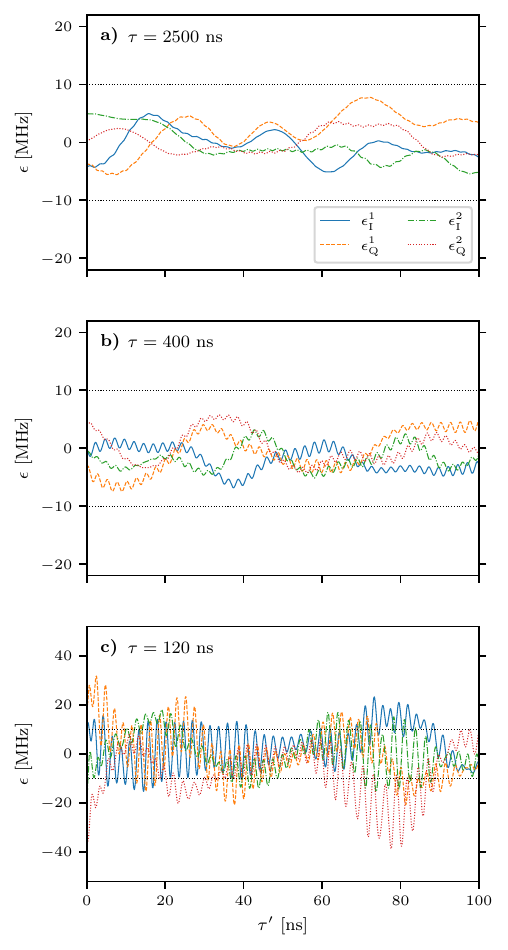}
\caption{(Figure reused from \cite{custom_gate_adiabatic}) First 100~ns of control pulses with lengths {\textbf{a)}} $\tau=2500$~ns, {\textbf{b)}} $\tau=400$~ns, and {\textbf{c)}} $\tau=120$~ns, realizing the first unitary in the sequence of Equation ~\eqref{eq:stp}. The correlation between the amplitude of the control pulse with its length establishes a lower bound for the implementation time of two-qubit gates on current quantum devices.}
\label{fig:controlpulses}
\end{figure}

The loss term is structured such that a zero-amplitude pulse corresponds to a loss of zero, and a pulse with a root mean square amplitude equal to $\epsilon_{\textrm{cut}}$ contributes $\chi$ to the total loss. Figure~\ref{fig:controlpulses} shows the first 100~ns of separately optimized control pulses, each realizing the first short-time propagator in the adiabatic evolution algorithm, using three different total pulse lengths $\tau$ (2500~ns, 400~ns, and 120~ns) as per Equation ~\eqref{eq:QOC}, with a sampling rate of 8 samples per ns.

The pulse length $\tau=2500$~ns approximately matches the implementation time $\tau_U$ of the short-time propagators on the \textit{ibmq\_belem} system, enabling a comparison between our emulated output and runs on that system. The pulse length $\tau=400$~ns corresponds to the standard implementation time of CNOT gates for IBMQ systems. The pulse length $\tau=120$~ns was the shortest pulse of our investigation that kept the gate infidelity $1-F_{\rm gate}$, below $10^{-4}$. The parameters in the objective function of Equation ~\eqref{eq:of} were $\epsilon_{\text{cut}}=30$~MHz, $n=3$ and $\chi=10^{-3}$ for all minimization attempts.

As the control pulse length decreases, the root-mean-squared amplitude and the relative influence of its high-frequency components increase. This relationship between the control pulse amplitude and its length sets a lower bound for two-qubit gate implementation times, and, consequently, the overall implementation time for the entire evolution on the types of quantum devices that follow this qubit model. This is because the quantum computer's Hamiltonian approximations (used to solve Equation ~\eqref{eq:QOC}) and the hardware controlling it both operate optimally within a specific energy regime.

In order to investigate the performance of adiabatic evolution with custom gates as a state preparation method, we used classical emulations to predict the output of a two-transmon processor by solving the Lindblad master equations for its density matrix. The analysis of this method is given in Section ~\ref{sec:strategy2}.

\section{Results and Discussion}
In this section, we present and discuss the results of our investigation of the two possible implementations of quantum adiabatic evolution to prepare the ground state of a target Hamiltonian.

\subsection{Adiabatic Evolution with Elementary Gates on IBMQ}
\label{sec:strategy1}

In this section, we evaluate the feasibility of implementing adiabatic evolution, as described in Equation ~\eqref{eq:stp}, on IBMQ systems by performing experiments using circuits composed only of elementary gates. The calibration data for the IBMQ systems at the time of this study can be found in Table~\ref{table:IBMQspecs}.

\begin{table}[b]
\centering
\caption{Calibration data for \textit{ibmq\_belem}, \textit{casablanca}, \textit{ibmq\_lima}, and \textit{ibmq\_manila}. The relaxation and dephasing times, $T_1$ and $T_2$, are provided for qubits 0 and 1 of each system throughout the simulation. The implementation times for the CNOT gate and the decomposition of the short-time propagators are denoted as $\tau_{\rm CNOT}$ and $\tau_U$, respectively.}
\begin{tabular}{c|c|c|c|c|c|c}
\hline\hline
IBMQ system & \multicolumn{2}{c|}{$T_1$ [$\mu$s]} & \multicolumn{2}{c|}{$T_2$ [$\mu$s]} &
$\tau_{\rm CNOT}$ [ns] & $\tau_{U}$ [ns] \\
\hline
\textit{ibmq\_belem} & 102.6 & 70.4 & 127.3 & 104.5 & 810.7 & $\approx$2500 \\
\textit{ibmq\_casablanca} & 111.7 & 130.1 & 40.7 & 102.2 & 760.9 & $\approx$2400 \\
\textit{ibmq\_lima} & 101.6 & 113.0 & 180.0 & 106.9 & 305.8 & $\approx$1000 \\
\textit{ibmq\_manila} & 136.0 & 244.2 & 112.8 & 46.7 & 277.3 & $\approx$900 \\
\hline\hline
\end{tabular}
\label{table:IBMQspecs}
\end{table}

The quantum circuits were initialized by preparing the IBMQ processors in the ground state of the initial Hamiltonian $H_0$ [refer to Equation ~\eqref{eq:H0}] through the application of the two-qubit Pauli $X^{(2)}= \sigma^x_1\sigma^x_2$ and Hadamard $H^{(2)}= H_1H_2$ gates to the default initial state $\ket{00}$, resulting in the following initial wave function for the quantum register:
\begin{equation}
\ket{\psi(0)}=H^{(2)}X^{(2)}\ket{00}
=\frac{1}{2}\left(\ket{00}-\ket{01}-\ket{10}+\ket{11}\right),
\label{eq:psi0}
\end{equation}
The rest of the adiabatic evolution circuit was constructed using elementary quantum gates derived from the decomposition of the $n$ short-time propagators, as discussed in Section ~\ref{sec:implementation}.

By truncating this adiabatic evolution circuit after $k$ short-time propagator operations, the final state corresponds to the instantaneous wave function $\ket{\psi(t_k)}$. We used Pauli measurements on these truncated circuits to estimate the expectation value of the target Hamiltonian $\braket{H_T}(t) = \braket{\psi(t)|H_T|\psi(t)}$ at each value of $t_k$ to provide insight into how the state of the quantum register evolves during the execution of the circuit.


We can also easily estimate the instantaneous fidelity $F(t_k)$ using the definitions in Equations \eqref{eq:stp} and \eqref{eq:psi0}. We start with the following observation:

\begin{align}
\bra{\phi(t_k)} \approx \bra{\psi(t_k)} =& \bra{00} X^{(2)} H^{(2)} {\mathcal U}^\dagger(0,t_k)\nonumber\\
\approx& \bra{00} X^{(2)} H^{(2)} \prod_{i=k}^1 U^\dagger(t_i), 
\end{align}

Thus, the overall unitary operator for the truncated circuit is:

\begin{equation}
\widetilde U(t_k)=X^{(2)} H^{(2)} \prod_{i=k}^1 U^\dagger(t_i),
\end{equation}

This is enough to derive an expression for $F(t_k)$ as follows:

\begin{equation}
\begin{aligned}
F(t_k)=|\braket{\phi(t_k)|\psi(t_k)}| \approx& |\bra{00}\widetilde U(t_k)\ket{\psi(t_k)}|\\
=& |\braket{00|\widetilde\psi(t_k)}|\\
=& |\widetilde c_{00}|.
\end{aligned}
\label{eq:figexpect}
\end{equation}

Thus, each instantaneous fidelity $F(t_k)$ is equivalent to $|c_{00}|$, which is the square root of the probability of measuring $\ket{\psi_k}$ in the $\ket{00}$ state, which can be estimated from repeated measurements of $\ket{\psi_k}$ in the default $\mathbf{Z}$ basis.

\begin{figure}[htpb]
\centering
\includegraphics[width=0.8\linewidth]{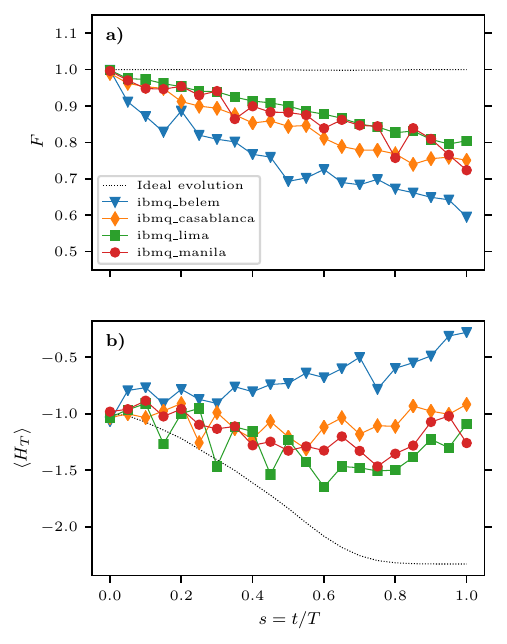}
\caption{(Figure reused from \cite{custom_gate_adiabatic}) Adiabatic evolution of {\textbf{a)}} instantaneous fidelity $F(t)$ and {\textbf{b)}} expectation value $\braket{\psi(t)|H_T|\psi(t)}$, obtained through the application of the unitary sequence from Equation ~\eqref{eq:stp} decomposed into elementary gates on four IBM quantum processors: \textit{ibmq\_belem} (blue triangles), \textit{ibmq\_casablanca} (orange diamonds), \textit{ibmq\_lima} (green squares), and \textit{ibmq\_manila} (red circles). The significant discrepancies between the experimental results and the ideal quantum processor simulations (dotted lines), conducted using Qiskit's Aer simulator, can be attributed to decoherence due to long implementation times.}
\label{fig:elementaryevo}
\end{figure}

This process is repeated to obtain estimations for $\braket{H_T}$ and $F(t)$ for each IBMQ system in question. In Fig.~\ref{fig:elementaryevo}, we compare the results we obtained from the IBMQ systems with those produced by the same process on an ideal quantum processor simulation using Qiskit's Aer simulator, which is not affected by environmental interactions.

The ideal output, depicted by a dotted line, reveals that the gate error included in our calculations for $F(t)$ (from Equation \ref{eq:fid}) does not considerably interfere with the ability to prepare a target state in two-qubit systems. However, experiments on \textit{ibmq\_belem}, \textit{ibmq\_casablanca}, \textit{ibmq\_lima}, and \textit{ibmq\_manila}, represented by color markers, diverge from the ideal evolution, achieving target state fidelities of 60\%, 75\%, 80\%, and 72\%, respectively. This loss of fidelity during evolution leads to significant discrepancies between the properties extracted from the reached state and those of the ground state of the target Hamiltonian, as illustrated by the evolution of $\braket{H_T}(t) = \braket{\psi(t)|H_T|\psi(t)}$ in Fig.~\ref{fig:elementaryevo} {\textbf{b)}}. The mean values for this expectation value obtained from the IBMQ system range from $-1.25$ to $-0.25$, while the ideal simulation attains the target value of $E_T=-2.328$.

Interestingly, the fidelity loss rate remains approximately constant throughout the evolution. This observation is unexpected since the interpolation functions defining $H(t)$ (defined in Equation~\eqref{eq:interpolation}) cause the evolution to be ``slowest" at its initial and final points (see Fig.~\ref{fig:exactevolution}), where minimal fidelity losses were anticipated. The constant fidelity loss rate implies that this effect is predominantly influenced by either the gate error accumulated from the elementary gates realizing the short-time propagator (with the most significant contributions arising from CNOT gate errors) or the coherence loss during their combined implementation time.

These findings highlight the primary challenge of preparing eigenstates with NISQ devices using adiabatic evolution. The implementation time of the quantum circuit that expresses the adiabatic evolution of Equation~\eqref{eq:stp} in terms of elementary gates is comparable to the decoherence times of current quantum devices. In this study, the implementation times of adiabatic evolution on \textit{ibmq\_belem}, \textit{ibmq\_casablanca}, \textit{ibmq\_lima}, and \textit{ibmq\_manila} are approximately 52.5~$\mu$s, 50.4~$\mu$s, 21~$\mu$s, and 18.9~$\mu$s, respectively, which are similar to the relaxation and dephasing decoherence times ($T_1$ and $T_2$) of these systems, as listed in Table~\ref{table:IBMQspecs}.

\subsection{Customized Gates Emulated with Classical Devices}
\label{sec:strategy2}

As an alternative to implementing adiabatic evolution through elementary gates, this section explores the efficacy of using pulse sequences that can produce short-time propagators in a single step, as explained in Section \ref{sec:implementation}. This approach seeks to reduce the depth of the corresponding circuit and, as a result, decrease the time it takes to execute on the device.

To investigate this approach, a two-transmon processor performing adiabatic evolution was emulated with classical computing techniques. Specifically, the Quantum Toolbox in Python (QuTiP)~\cite{johansson_jr2013, johansson_jr2012} was used to solve the Lindblad master equation. This equation is defined as follows:

\begin{equation}
\begin{aligned}
\dot{\rho} =& -\frac{i}{\hbar}
\left[ H_{\rm QPU}, \rho \right]\\
&+ \frac{1}{T_1} \sum_{i=1}^2 \left( a_i\rho a^\dagger_i
- \frac{1}{2}\left\{a_i a^\dagger_i,\rho\right\} \right)\\
&+ \frac{1}{T_2} \sum_{i=1}^2
\left( a^\dagger_i a_i\rho a_i a^\dagger_i - \frac{1}{2}
\left\{a_i a^\dagger_i a^\dagger_i a_i,\rho\right\} \right)\,,
\end{aligned}
\end{equation}

where $\rho$ represents the density of the two-transmon system. In this equation, $H_{\rm QPU}$ denotes the two-transmon Hamiltonian introduced in Equation ~\eqref{eq:HQPU}, with control pulses optimized to achieve the short-time propagators described in Equation ~\eqref{eq:stp}. The master equation accounts for the decoherence mechanisms of relaxation and dephasing, which are represented by the decoherence times $T_1$ and $T_2$. The model utilizes parameter values obtained from calibration data from IBMQ devices that implemented the adiabatic evolution, as seen in Table~\ref{table:IBMQspecs}. This enables the emulation of system interactions with their surrounding environment.

\begin{figure}[htpb]
\centering
\includegraphics[width=0.8\linewidth]{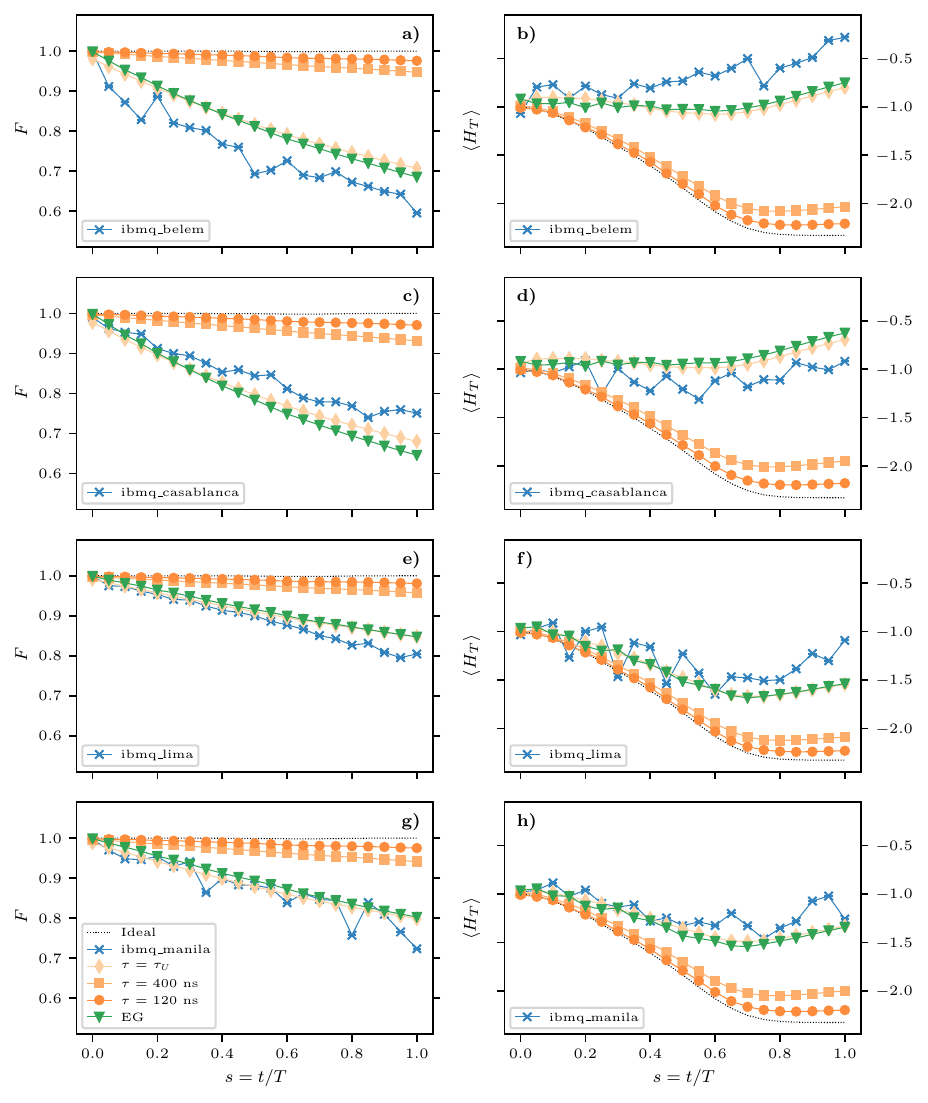}
\caption{(Figure reused from \cite{custom_gate_adiabatic}) A comparison of the emulated outputs of the adiabatic evolution of Eq. \eqref{eq:stp} and the results obtained from \textit{ibmq\_belem}, \textit{ibmq\_casablanca}, \textit{ibmq\_lima}, and \textit{ibmq\_manila} systems. Panels {\textbf{a)}}, {\textbf{c)}}, {\textbf{e)}} and {\textbf{g)}} display the progression of the instantaneous fidelity $F$. Classical emulations implementing the adiabatic evolution through either a circuit of elementary gates or custom gates of length $\tau_U$ (represented by green triangles and orange diamonds, respectively) achieve target state fidelities comparable to those attained by IBMQ systems (indicated by blue crosses). Emulated outputs utilizing shorter gates (denoted by orange squares and circles) reached approximately 95\% fidelity, facilitating enhanced extraction of the target state's spectroscopic information, as demonstrated by the evolution of $\braket{H_T}$ in panels {\textbf{b)}}, {\textbf{d)}}, {\textbf{f)}} and {\textbf{h)}}.}
\label{fig:results}
\end{figure}

We investigated this implementation of quantum adiabatic evolution by simulating various scenarios. First, we emulated the adiabatic evolution by implementing each short-time propagator in Equation ~\eqref{eq:stp} with a single custom gate of pulse length $\tau=120$~ns. This approach yielded target state fidelities of approximately 95\%, a significant enhancement compared to IBMQ system runs, as depicted in Figures ~\ref{fig:results} {\textbf{a)}}, {\textbf{c)}}, {\textbf{e)}}, \& {\textbf{g) } }. 
Although the pulse length of $\tau=120$~ns minimizes gate infidelities below $10^{-4}$, the resulting pulse amplitudes may exceed a threshold, rendering them unsuitable for implementation on quantum devices, represented arbitrarily as the horizontal dotted lines at $|\epsilon_{\rm I,Q}^i|<\alpha/20$, for example.

We also emulated the algorithm implementation using custom gates of length $\tau=400$~ns, which is comparable to the implementation times of the CNOT gates in IBMQ systems. Despite this longer pulse length, our results still surpassed those of IBMQ system runs, achieving target state fidelities of approximately 90\%. This can be explained by the fact that the elementary gate decomposition explored in Section \ref{sec:elemdecomp} requires two CNOT gates per two-qubit unitary evolution, leading to longer overall circuit execution times.

In order to facilitate comparison with IBMQ results, we also simulated the adiabatic evolution using custom gates of length approximately equal to the implementation time of a short-time propagator on IBMQ hardware, denoted $\tau_U$. The specific values for $\tau_U$ are provided in Table~\ref{table:IBMQspecs}. These simulations resulted in gates with fidelity ranging from 65\% to 85\%, which aligned closely with the IBMQ results.  This close match suggests that the total execution time of the circuit may be the most important factor in accurately preparing quantum states on NISQ devices.

\begin{figure}[htpb]
\centering
\includegraphics[width=0.8\linewidth]{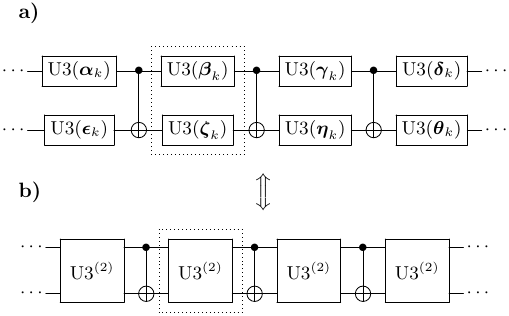}
\caption{(Figure reused from \cite{custom_gate_adiabatic}) Combination of simultaneous U3 gates [circuit {\textbf{a)}}] into two-qubit ${\rm U}^{(2)}$ gates [circuit {\textbf{b)}}]. The CNOT and ${\rm U3}^{(2)}$ gates in circuit {\textbf{b)}} are executed with control pulses, the total durations of which closely correspond to the implementation time of circuit {\textbf{a)}} on IBMQ systems.}
\label{fig:directcomparison}
\end{figure}

For an even more direct comparison, we emulated the output of the IBMQ digital quantum simulation by realizing each elementary gate in the IBMQ quantum circuit with custom control pulses. In this final set of simulations, we combined each pair of simultaneous U3 gates in the short-time propagators into two-qubit ${\rm U3}^{(2)}={\rm U3}_1{\rm U3}_2$ gates, as illustrated in Figure ~\ref{fig:directcomparison}. 
Then, the CNOT and ${\rm U3}^{(2)}$ gates were implemented using control pulses, the cumulative durations of which approximated the execution time of short-time propagators on IBMQ systems, denoted as $\tau_U$.

In Figure \ref{fig:results}, a comparison is made between the output of our emulation and the results of the digital quantum simulations performed on IBMQ systems.

The emulated digital quantum simulation, achieved by implementing the adiabatic evolution through the CNOT and ${\rm U3}^{(2)}$ gates (denoted by green triangles), aligns closely with the IBMQ results (represented by blue crosses). The emulated outputs, obtained by realizing the short-time propagators in Eq. \eqref{eq:stp} via single two-qubit gates with control pulses of lengths $\tau=\tau_U$, $\tau=400$~ns, and $\tau=120$~ns (depicted as orange diamonds, squares, and circles, respectively), highlight the enhancements achievable through the incorporation of custom gates within quantum algorithms. 
The value derived from the shortest classical emulation, $\braket{H_T}(T)=-2.2$, was much closer to the exact result of $E_T=-2.328$ than the energies extracted from IBMQ systems (mean values ranging from $-1.25$ to $-0.25$). Even though this demonstration was conducted on a basic problem involving only two qubits, the enhancement seen over the simple gate composition method indicates that the pursuit of optimal control for custom gates in quantum algorithms is promising for NISQ devices.

\subsection{Discussion of Error Sources}
\begin{figure}[htpb]
\centering
\includegraphics[width=0.8\linewidth]{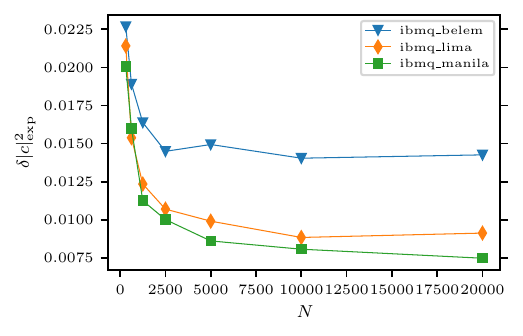}
\caption{(Figure reused from \cite{custom_gate_adiabatic}) Mean error in measured occupation probabilities as a function of the number of shots $N$. A higher number of shots reduces statistical noise and provides an estimation of the measurement error of the quantum computer. The error in occupation probabilities from these simulations is capped at $\delta|c|_{\rm exp}^2=0.016$, which constitutes only a minor portion of the error seen in calculated fidelities and expectation values.}
\label{fig:errorwithN}
\end{figure}

Various error sources impact the performance of this adiabatic evolution method on IBMQ hardware. Our model already considers systematic gate infidelities, stochastic dissipative processes during circuit execution, stochastic measurement errors during readout, and statistical noise. To assess these errors' effects on adiabatic evolution, we compare Qiskit's Aer simulator-based simulations with runs on IBMQ systems. We use Qiskit's Aer simulator to determine the state $H^{(2)}\ket{00}=\left(\ket{00}+\ket{01}+\ket{10}+\ket{11}\right)/2$ via the \textit{AerSimulator.from\_backend} built-in method. This method develops a classical simulator with an approximate noise model for any IBMQ device that includes gate errors, readout errors, and dissipative processes. Figure ~\ref{fig:errorwithN} illustrates the average deviation of the measured probabilities from the ideal value $1/4$, as the number of shots $N$ increases.
As $N$ increases, the statistical error component diminishes and the average error levels off to a device-specific plateau, which we associate with the measurement error for each IBMQ system. These simulation outcomes indicate that measurement and statistical noise are not significant error contributions, as they only cause deviations from the ideal fidelities and expectation values of about $\delta F\approx0.01$ and $\delta\braket{H_T}\approx0.1$ for IBMQ runs with $N=2500$ shots, significantly smaller than those presented in Figure \ref{fig:results}.
\begin{figure}[htpb]
\centering
\includegraphics[width=0.8\linewidth]{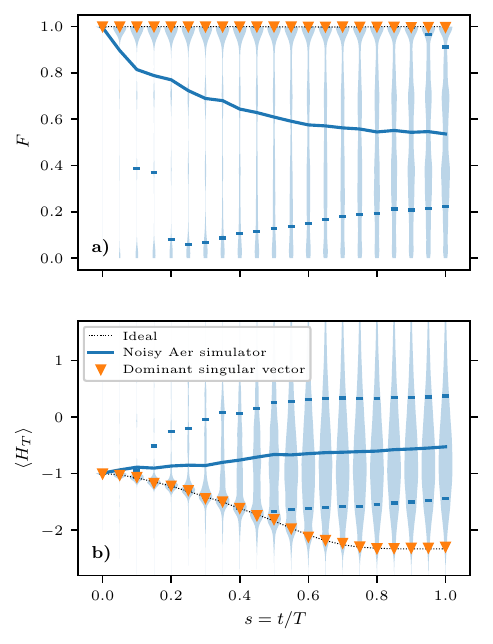}
\caption{(Figure reused from \cite{custom_gate_adiabatic}) Evolution of {\textbf{a)}} fidelity and {\textbf{b)}} expectation value $\braket{H_T}$ using Qiskit's Aer simulator with a noise model based on \textit{ibmq\_belem}. Solid blue lines represent the expected results when incorporating all error sources considered by Aer, while the blue shaded areas show the probable values at each time interval. Dotted black lines indicate the ideal evolution in a noiseless system. Orange triangles come from singular value decomposition filtering of the data supporting the blue distributions. The closeness between the dotted lines and the orange triangles strongly implies that dissipative processes are the primary source of noise in IBMQ simulations, even at the early stages of the evolution.}
\label{fig:noisemodel}
\end{figure}

Next, we repeated the adiabatic evolution simulations using Qiskit's Aer simulator with the \textit{"statevector''} method. This comes with the option to record per-shot amplitudes, although doing so prevents readout error simulation. We ran 100,000 executions of each circuit to determine the instantaneous fidelity and expectation value of $H_T$ for each time value $t_k$. Using the recorded amplitudes, we calculated the energy for each simulated shot, resulting in the distribution of possible energy measurements. We repeated a similar procedure for fidelity. To isolate the effects of dissipative processes from other error sources, we formed the density matrix for each circuit by averaging the shots as follows:
\begin{align}
    \mathbf{\rho} = \frac{1}{N_\textrm{shots}} \sum_{i\in \textrm{shots}} \ket{\psi_i}\bra{\psi_i}\,,
\end{align}

where $\ket{\psi_i}$ is the final state of the simulated circuit.
We then perform the singular value decomposition on the density matrix to identify the most probable state, which is the state with the highest singular value, and then calculate the fidelity and energy based on these amplitudes. If only gate errors and readout errors are present, one should observe a singular value very close to one. However, dissipative processes generate mixed states, causing the highest singular value to be significantly lower than its ideal value, decreasing as the circuit depth increases. In Figure~\ref{fig:noisemodel}, we show the resultant distributions at each step for both {\textbf{a)}} the fidelity and {\textbf{b)}} the expectation value of $H_T$ as blue violin plots, illustrating the mean and $1\sigma$ quantiles. The green lines represent the expected result from an ideal evolution, whereas the orange dashed lines depict the outcomes using the principal singular vector of the density matrices. The discrepancy between the dashed orange and solid blue lines is mainly due to dissipative noise processes. In contrast, the proximity of the dashed orange and solid green lines indicates that other errors are relatively insignificant. The blue markers provide an estimate of the lower bound on the total uncertainty anticipated from simulations on IBMQ systems.


\chapter{Rodeo Algorithm Analysis and Demonstration}\label{cha:rodeo}

Quantum adiabatic evolution, while theoretically an accurate method for preparing eigenstates of quantum Hamiltonians, poses computational challenges when applied to large systems. In this chapter, we discuss the rodeo algorithm as a viable alternative. The Rodeo algorithm (RA) uses stochastic interference to suppress unwanted eigenvectors, offering an exponential speedup over quantum phase estimation and adiabatic evolution \cite{Choi:2020pdg}. 

In Section \ref{sec:ra}, we present a detailed analysis of the rodeo algorithm, and in Section \ref{sec:demorodeo}, we discuss a demonstration of its application on quantum hardware by calculating the eigenvalues of a stochastic one-qubit Hamiltonian using IBM's Casablanca device.

\section{Rodeo algorithm}\label{sec:ra}
Prior to the development of the rodeo algorithm, there had been recent advancements in Hamiltonian evolution on quantum computers using tools such as Lie-Trotter-Suzuki formulas~\cite{Trotter:1959, Suzuki:1976a} and linear combinations of unitary matrices \cite{Childs:2012a}. Hardware limitations presented a significant challenge for all known quantum state preparation approaches. This algorithm was developed in an attempt to mitigate those challenges and enable the preparation of eigenstates on NISQ-era devices.

The rodeo algorithm is an iterative method that prepares a target eigenvector by "shaking off'' all other states like a rodeo horse. It is a dissipative process inspired by the projected cooling algorithm \cite{Lee:2019zze, Gustafson:2020vqg} that can be applied recursively and converges to the desired eigenstate exponentially with the number of repetitions. It can theoretically prepare any target eigenstate of any quantum Hamiltonian given an initial state that has a sufficiently high overlap with that target state. It appears similar to iterative QPE \cite{Kitaev:1995} and fixed-time energy band filtering \cite{Ge:2017,Lu:2020}; however, these methods cannot efficiently prepare individual eigenstates of a general quantum Hamiltonian, making the rodeo algorithm more suitable for quantum simulation applications.

\subsubsection{Algorithm Description}
We designate the Hamiltonian of interest as the "object Hamiltonian'', $H_{obj}$, and the part of the quantum register on which it acts as the "object system''. We then choose an energy interval $[E-\epsilon, E+\epsilon]$. The goal of the algorithm is to prepare this object system in an eigenstate of $H_{obj}$ with an eigenvalue within that energy interval. 

The object system starts in an initial state denoted $\ket{\psi_{I}}$ and evolves into a final state $\ket{\psi_F}$. Auxiliary qubits, called ancilla qubits, allow indirect manipulation of the object system through entangling gate operations and measurements. This set of ancilla qubits is informally termed the 'rodeo arena.' 
When mid-circuit measurements are allowed, as is the case for one-way quantum computers \cite{Raussendorf_2003} and quantum computers that support dynamic circuits \cite{C_rcoles_2021} such as the superconducting quantum computers available on IBMQ, a single ancilla qubit can be reused throughout the recursive process, which allows the rodeo arena to merge into a single qubit. Since mid-circuit measurements are becoming more widely available, subsequent analysis will assume access to them.

One circuit of the RA is implemented as follows:
\begin{enumerate}
    \item \textbf{Choose parameters:} Set $E$ as a guess for the eigenvalue of an eigenstate of $H_{\rm obj}$ and choose a random non-zero value for $t$.
    \item \textbf{Initialization:} The ancilla qubit is prepared in the $\ket0$ state, then a Hadamard gate is applied to place it in a superposition of the computational basis. The object system can be in any state, labeled $\ket{\psi_I}$. For best results, this state should be chosen to have non-negligible overlap with the eigenstate of $H_{\rm obj}$ closest to $E$
    \item \textbf{Controlled time evolution:} Use the ancilla qubit to control the time evolution operation with Hamiltonian $H_{\rm obj}$ and time parameter $t$ on the object system.
    \item \textbf{Phase Shift Gate:} Apply a phase rotation gate $P(Et)$ on the ancilla qubit. This imparts the phase $e^{iEt}$ to the $\ket0$ component while leaving the $\ket1$ component unchanged.
    \item \textbf{Rotate and Measure:} Apply the Hadamard gate to the ancilla qubit once more, then measure only the ancilla qubit.
\end{enumerate}

\begin{figure}
    \centering
    \includegraphics[width=\linewidth]{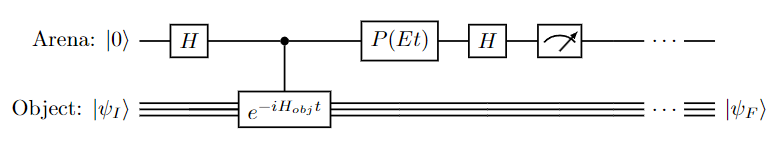}
    \caption{One cycle of the Rodeo algorithm. Additional copies of this circuit can be appended to increase its efficacy. If mid-circuit measurements are allowed, only one ancilla is needed for the arena, otherwise additional ancilla qubits may be used, with one ancilla qubit participating in each cycle.}
    \label{fig:onecycle}
\end{figure}

 The quantum circuit for this algorithm with one ancilla qubit is shown in Figure \ref{fig:onecycle}.

This choice of gates is intuitive because of the similarity between the time evolution operator $e^{-iH_{obj}t_n}$ and the phase rotation $e^{iEt_n}$. As we show in Section \ref{sec:ramath} their combined effect on the quantum register will result in interference in the wave function of the object system, exponentially suppressing its relative overlap with each eigenstate by a factor proportional to the difference between its eigenvalue and $E$. The suppression factor for each eigenstate will also unpredictably depend on $t_n$, but the effect of this choice can be diminished by selecting different values for each $t_n$.

\subsection{Mathematical Analysis of RA}\label{sec:ramath}
By representing gate operations in matrix form, we can exactly predict the state vector output of the RA circuit. The ancilla qubit is initialized in the $\ket 0$ state and the object system is initialized in the $\ket{\psi_I}$ state, so the initial state can be written as $\ket{0}\otimes \ket{\psi_I}$. Writing the gate operations as matrix operations on this initial state shows the effect of one cycle of the RA on this state.

\begin{align}
\rm RA (\ket{0} \otimes \ket{\psi_I}) =  & \begin{bmatrix}
\tfrac{I}{\sqrt{2}} &
\tfrac{I}{\sqrt{2}}\\
\tfrac{I}{\sqrt{2}}& 
\tfrac{-I}{\sqrt{2}}
\end{bmatrix}
\begin{bmatrix}
 I & 0 \\
0 & Ie^{iEt} 
\end{bmatrix}
 \begin{bmatrix}
 I & 0 \\
0 & e^{-iH_{\rm obj}t}
\end{bmatrix} 
\begin{bmatrix}
\tfrac{I}{\sqrt{2}} & 
\tfrac{I}{\sqrt{2}}\\
\tfrac{I}{\sqrt{2}} & 
\tfrac{-I}{\sqrt{2}}
\end{bmatrix}
\begin{bmatrix}
\ket{\psi_I}\\0
\end{bmatrix}\\
&  = \begin{bmatrix}
\left[ \tfrac{I}{2}+\tfrac{1}{2}e^{-i(H_{\rm obj}-E)t} \right] \ket{\psi_I} \\
\left[ \tfrac{I}{2}-\tfrac{1}{2}e^{-i(H_{\rm obj}-E)t} \right] \ket{\psi_I}
\end{bmatrix}\\
& = \ket{\psi_F}
\label{one_rodeo}
\end{align}

Here, $I$ represents the identity operator with the same dimension as the object system. 
 We will represent these eigenvectors of $H_{\rm obj}$ as $\ket{E_k}$ with the corresponding eigenvalues $E_k$. Since these eigenvectors form a basis for the object system, $\ket{\psi_I}$ can be written as a linear combination of these eigenstates as follows:

\begin{equation}
    \ket{\psi_I} = \sum_{k} c_k \ket{E_k}
\end{equation}

$H_{\rm obj}$ commutes with all gates in the circuit, so the total effect of the circuit can be understood in terms of its effect on each of these components of $\ket{\psi_I}$. The probability of measuring the ancilla qubit in the $\ket0$ state can then be extracted from the first entry in the final statevector in Equation \ref{one_rodeo} as follows:

\begin{align}
    P(\ket0) =& \braket{\psi_I|\begin{bmatrix}\frac I2 + \frac12 e^{-i(H_{\rm obj} - E)t}\end{bmatrix}^\dagger\begin{bmatrix}\frac I2 + \frac12 e^{-i(H_{\rm obj} - E)t}\end{bmatrix}|\psi_I}\\
    =& \sum_k c_k^2\left\vert \tfrac{1}{2}+
   \tfrac{1}{2}e^{-i(E_k-E)t} \right\vert^2\\
    =& \sum_k c_k^2\cos^2\left[(E_{k}
   -E)\tfrac{t}{2}\right]
\end{align}

When the RA is repeated for $N$ cycles with different values of the time parameter $t_n$, the probability of measuring the ancilla qubit in the $\ket0$ state each time is the product of these probabilities.

\begin{equation}\label{eq:RAsuccess}
    P_N = \prod^N_{n=1} \sum_k c_k^2\cos ^2\begin{pmatrix}(E_{k} - E)\frac{t_n}{2}\end{pmatrix}
\end{equation}

We call this probability the "success probability'' because when this measurement occurs, it corresponds to a high probability that the final state of the object system is in an eigenstate of $H_{obj}$ with eigenvalue $E$. If $t_n$ are sampled from a uniform distribution, as $N$ becomes large, the spectral weight of any $E_k \neq E$ is suppressed by a factor of $1/4^N$, due to the fact that the geometric mean of the $\cos^2$ function is $\frac14$. If $t_n$ are sampled from a Gaussian distribution where the root mean square value is $\sigma$, then when $\sigma$ scales as $O(1/\epsilon)$, eigenstates that fall outside the interval $[E-\epsilon, E+\epsilon]$ are exponentially expressed. Thus, we can intuitively say that altering $\sigma$ changes the magnification of the energy sensor of the RA

In the case where $E$ is close to an eigenvalue $E_{\rm obj}$ (associated with the eigenvector $\ket{E_{\rm obj}}$) of $H_{obj}$, for large $N$, the probability of consecutive measurement of the ancilla in the $\ket{0}$ state $N$ times converges to the following:

\begin{equation}\label{eq:pnlargen}
    P_N(E) = |\braket{\psi_I|E_{\rm obj}}|^2 \prod_{n=1}^N \cos^2\begin{pmatrix}(E_{\rm obj} - E)\frac{t_n}{2}\end{pmatrix}
\end{equation}

This is a simplified version of Equation\ref{eq:RAsuccess}, where all terms in the sum are excluded except the one involving $\ket{E_j}$, which was the least suppressed, and the coefficient for that term is replaced by the overlap $p = |\braket{\psi_I|E_{\rm obj}}
^2$ of the initial state of the object system. As $E$ approaches the perfect guess of $E=E_{\rm obj}$, this expression simplifies even further to $P_N(E_{\rm obj}) = p$. With the spectral weights of other eigenvectors outside $[E-\epsilon, E+\epsilon]$ reduced by $\delta$, the computational effort for the rodeo algorithm scales as $N\sigma/p = O[|\log\delta|/(p\epsilon)]$. Furthermore, when $t_n$ are sampled from the Gaussian distribution with mean value 0 and root mean square $\sigma$ (that is, $t_n \sim \mathcal{N}(\mu, \sigma^2$)), the product in Equation \ref{eq:pnlargen} can be averaged over the time values $t_n$ to derive a simple expression for the success probability as:

\begin{align}\label{ExpeProbs}
    P_{N}(E) = \left[\frac{1+e^{-(E_{\rm obj}-E)^2\sigma^2/2}}{2}\right]^N.
\end{align}

When a good guess for the parameter $E$ is not available, a simple search algorithm can fine-tune it to approximate the energy eigenvalue $E_{\rm obj}$ within a given confidence interval. Achieving precision $\epsilon$ requires $O(\log \epsilon)$ energy scans, where each scan reduces the energy range by a factor $K$. This can be accomplished by performing each scan at multiple evenly spaced $E$ values with a specified number of rodeo cycles and increasing $\sigma$ by a factor of $K$ each time. The total time evolution scales as $O(1/\epsilon)$, and high-fidelity energy scans add a factor $(\log \epsilon)^2p$. Thus, the computational complexity to identify $E_{\rm obj}$ with error $\epsilon$ scales as $O[(\log \epsilon)^2/(p \epsilon)]$, approaching the theoretical limit of $O(1/\epsilon)$ set by the Heisenberg uncertainty principle. In comparison, phase estimation has a computational complexity of $O[1/(p\epsilon)]$ plus $O[(\log \epsilon)^2]$ for the quantum Fourier transform~\cite{nielsen_ma2010}. Iterative phase estimation, which does not require the Fourier transform, is limited to finding the energies of pure eigenstates and requires $O(1/\epsilon^2)$ measurements to calculate the expectation value of such a state for a Hamiltonian due to statistical errors.

In addition to computing eigenvalues, the RA is also effective as an eigenstate preparation technique. When a good guess for $E_{\rm obj}$ is known and the initial state $\ket{\psi_I}$ has non-zero overlap with the corresponding eigenstate $\ket{E_{\rm obj}}$, the success probability $P_N$ grows rapidly with the number of cycles. Whenever a success occurs (that is, the ancilla qubit was measured in the $\ket0$ state every time), the final state of the object system $\ket{\psi_F}$ has significant overlap with $\ket{E_{\rm obj}}$. This can be explained by the partial collapse of the object system wave function that occurs when the entangled ancilla qubits are measured. The configuration of the entanglement induced by the RA ensures that a measurement of $\ket0$ corresponds to the wave function collapsing toward the target state. Since no measurement operations are necessary on the object system qubits in the RA, these can then be fed forward into another algorithm, such as a quantum simulation.

To prepare an eigenstate $\ket{E_j}$ with a residual orthogonal component $\Delta$, the computational cost is $O(\log\Delta/p)$. We maintain $\sigma$ large and constant to filter the desired eigenstate, which requires $N=O(\log\Delta)$ iterations of the rodeo algorithm, and $1/p$ measurements per iteration. $E$ must be centered on the peak of $E_j$, which introduces a constant factor to the computational cost. For phase estimation, the cost is $O[1/(p\Delta)]$, and for adiabatic evolution, it is $O(1/\Delta)$, adjusted by $p$ based on the adiabatic path. Thus, for eigenstate preparation, the rodeo algorithm is exponentially more efficient as $\Delta$ approaches zero. Useful estimates for the magnitude of the residual orthogonal component $\Delta$ can be written as a function of the number of rodeo cycles $N$ as follows:
 \begin{align}
    F_A \equiv \sqrt{2^{-N}(1-p)/[p+2^{-N}(1-p)]},\nonumber \\
     F_G \equiv \sqrt{4^{-N}(1-p)/[p+4^{-N}(1-p)]}. \label{estimates}
 \end{align}

When $N$ is much smaller than the number of eigenstates that have non-negligible overlap with the initial state, $F_A$ is sufficiently accurate. This has a spectral suppression factor of $1/2$ for undesired eigenstates, equating to the arithmetic mean of $\cos^2 $. On the other hand, $F_G$ is more accurate when $N$ is greater than the number of eigenstates that have non-negligible overlap, with a suppression factor of $1/4$, equivalent to the geometric mean of $\cos^2$.

RA performance largely depends on the overlap between the initial state and the target eigenstate, represented by the overlap probability $p$. Since for most systems of interest it is not trivial to prepare an initial state with a high value of $p$, it may often be helpful to use a preconditioning technique to improve the quality of the initial state and increase the effectiveness of the RA. One method of accomplishing this would be to combine the RA with a variational quantum algorithm; this idea is explored in great detail in Chapter \ref{cha:vra}.


\section{Demonstration of the Rodeo Algorithm}\label{sec:demorodeo}
In this section, we show two examples to demonstrate the application of the RA. In Section \ref{sec:heisra}, we introduce a simple example using the Heisenberg model and emulate the RA circuit on a classical computer. Then, in Section \ref{sec:1qapp}, we use the RA to compute the eigenvalues of a random single-qubit Hamiltonian on the IBM Casablanca quantum computing device through IBMQ. For this single-qubit Hamiltonian, we also fully demonstrate the sequential scanning method discussed in Section \ref{sec:ramath} to compute the full energy spectrum of the Hamiltonian. We also invoke the Hellman-Feynman theorem to compute the expectation values of an arbitrary single-qubit observable. 

\subsection{Heisenberg Model Classical Simulation Example}\label{sec:heisra}
In this section, we consider the spin-$\tfrac{1}{2}$ Heisenberg model in a uniform magnetic field with $10$ interacting sites forming a closed loop in a one-dimensional chain \cite{Heisenberg:1928}. This Hamiltonian can be written in the following way:

\begin{equation}
    H_{\rm obj} = J \sum_{\braket{j,k}} \vec{\sigma}_j\cdot \vec{\sigma}_k + h \sum_{j} \sigma^z_j,
\end{equation}

Where $J$ is the exchange coupling coefficient between the sites, $\vec{\sigma}_j$ are the Pauli matrices on site $j$, $\braket{j,k}$ is shorthand for nearest-neighbor pairs of sites, and $h$ is the coupling to a uniform magnetic field in the $z$ direction. For this example, we will consider the antiferromagnetic case, which requires setting the values $J=1$ and $h=3$. The initial state is chosen to be the alternating tensor product state:
\begin{equation}
    \ket{\psi_I} = \ket{0101010101}.
\end{equation}

\begin{figure}
\centering
\includegraphics[width=\linewidth]{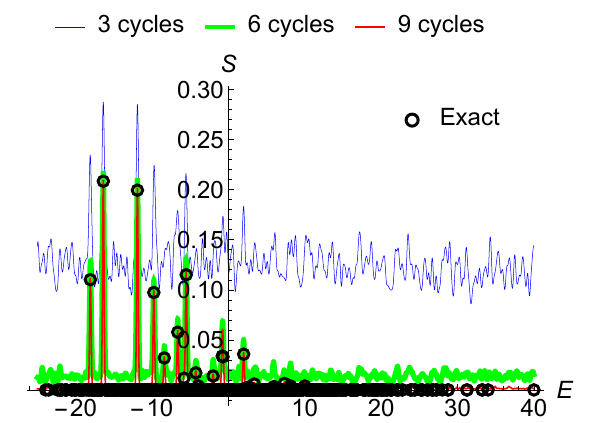}
\caption{(Figure reused from \cite{Choi:2020pdg}) Initial-state spectral function for the Heisenberg model. Using the RA, we depict the initial-state spectral function for the Heisenberg spin chain with $3$ (thin blue line), $6$ (thick green line), and $9$ (medium red line) cycles. The average was taken over $20$ sets of Gaussian random values for $t_n$ with $\sigma=5$. For reference, the exact initial-state spectral function is presented with black open circles.}
\label{fig:Heisenberg_spectrum}
\end{figure}

This choice of initial state is motivated by its high degree of symmetry, which allows us to predict that it will have a nonzero overlap with a relatively small number of energy eigenstates.
The energy eigenstates of \(H_{\rm obj}\) will be labeled \(\ket{E_j}\). The initial-state spectral function \(S(E)\), which describes the energy distribution of \(\ket{\psi_I}\) in terms of these eigenstates, is defined as:
\[
S(E) = 
\begin{cases} 
|\braket{E_j|\psi_I}|^2 & \text{if } E = E_j, \\
0 & \text{otherwise}.
\end{cases}
\]
Here, \(|\braket{E_j|\psi_I}|^2\) is the probability that the initial state \(\ket{\psi_I}\) overlaps with the energy eigenstate \(\ket{E_j}\).
 In cases of degenerate eigenvalues, the total contribution is the sum of contributions from all degenerate states. Figure \ref{fig:Heisenberg_spectrum} shows the initial-state spectral function using the rodeo algorithm for the Heisenberg spin chain with $N=3$ (thin blue line), $N=6$ (thick green line) and $N=9$ (medium red line) cycles. We used $20$ sets of random $t_n$ values sampled from a Gaussian distribution with $\sigma=5$ to reduce stochastic noise and create a constant background distinguishable from the spectral signal. The exact initial-state spectral function is shown with black open circles. Since both $\sigma$ and $N$ are not very large, relatively few gates would be needed to implement this circuit on a real device, which is important on NISQ devices.

In addition to computing the initial-state spectral function, we can generate any energy eigenstate that overlaps the initial state. Table \ref{table:heis_overlaps} shows the overlap of $\ket{\psi_F}$ with the energy eigenvector $\ket{E_j}$ after $N$ cycles of the rodeo algorithm. Gaussian random values are used for $t_n$ with $\sigma = 5$ and $E = E_j$. The table shows that all such energy eigenvectors can be prepared accurately with only a few rodeo cycles.

\begin{table}[h]
\begin{center}
\caption{Overlap of $\ket{\psi_F}$ with energy eigenvector $\ket{E_j}$ after $N$ cycles of the rodeo algorithm using Gaussian random values for $t_n$ with $\sigma = 5$ and $E=E_{j}$.}
\label{table:heis_overlaps}
\begin{tabular}{|c|c|c|c|c|}
\hline
$E_j$ & $N=0$ & $N = 3$ & $N = 6$ & $N= 9$ \\
\hline
$ -18.1 $ & $  0.110 $ & $  0.746 $ & $  0.939 $ & $  0.997  $ \\
$ -16.4 $ & $  0.209 $ & $  0.841 $ & $  0.993 $ & $  1.000  $ \\
$ -11.9 $ & $  0.200 $ & $  0.629 $ & $  0.889 $ & $  0.999  $ \\
$ -9.76 $ & $  0.0974 $ & $  0.488 $ & $  0.903 $ & $  0.999  $ \\
$ -8.38 $ & $  0.0320 $ & $  0.467 $ & $  0.832 $ & $  0.993  $ \\
$ -6.63 $ & $  0.0577 $ & $  0.309 $ & $  0.818 $ & $  0.996  $ \\
$ -5.81 $ & $  0.0118 $ & $  0.179 $ & $  0.637 $ & $  0.817  $ \\
$ -5.52 $ & $  0.115 $ & $  0.456 $ & $  0.766 $ & $  0.997  $ \\
$ -4.26 $ & $  0.0171 $ & $  0.144 $ & $  0.696 $ & $  0.995  $ \\
$ -3.95 $ & $  0.00401 $ & $  0.0430 $ & $  0.343 $ & $  0.952  $ \\
$ -2.00 $ & $  0.0139 $ & $  0.158 $ & $  0.593 $ & $  0.942  $ \\
$ -0.802 $ & $  0.0338 $ & $  0.216 $ & $  0.545 $ & $  0.594  $ \\
$ -0.704 $ & $  0.0331 $ & $  0.286 $ & $  0.540 $ & $  0.585  $ \\
$ 2.00 $ & $  0.0357 $ & $  0.371 $ & $  0.925 $ & $  0.994  $ \\
$ 2.42 $ & $  0.00235 $ & $  0.0122 $ & $  0.0874 $ & $  0.521  $ \\
$ 2.68 $ & $  0.00291 $ & $  0.0845 $ & $  0.639 $ & $  0.929  $ \\
$ 3.39 $ & $  0.00592 $ & $  0.0360 $ & $  0.754 $ & $  0.943  $ \\
$ 5.96 $ & $  0.00336 $ & $  0.0951 $ & $  0.559 $ & $  0.981  $ \\
$ 7.33 $ & $  0.00650 $ & $  0.184 $ & $  0.792 $ & $  0.978  $ \\
$ 8.13 $ & $  0.00393 $ & $  0.0832 $ & $  0.665 $ & $  0.841  $ \\
$ 8.24 $ & $  0.00105 $ & $  0.0275 $ & $  0.142 $ & $  0.289  $ \\
$ 10.0 $ & $  0.00397 $ & $  0.0128 $ & $  0.295 $ & $  0.902 $ \\

\hline
\end{tabular}

\end{center}
\end{table}

\begin{figure}
\centering
\includegraphics[width=\linewidth]{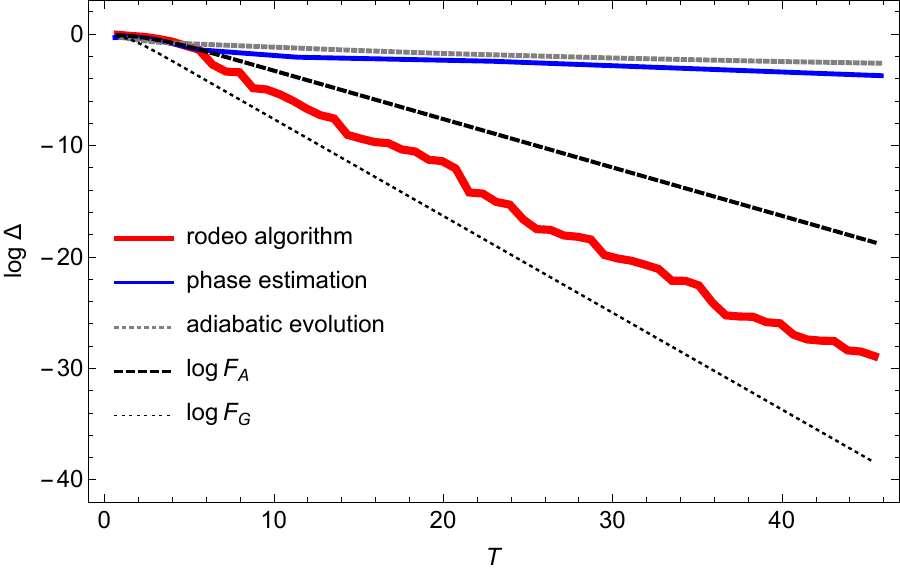}
\caption{ (Figure reused from \cite{Choi:2020pdg}) Logarithm of the wave function error versus the total propagation time for the Heisenberg model. We plot $\log \Delta$, versus the total propagation time, $T$, for the Heisenberg model. We show results for the rodeo algorithm, phase estimation, and adiabatic evolution.  We also show the asymptotic estimates $\log F_A$ and $\log F_G$.}
\label{Heisenberg_error_T}
\end{figure}

In Figure \ref{Heisenberg_error_T}, we show the logarithm of the error of the wave function using the RA to prepare the energy eigenstate $\ket{E_j}$ with $E_j = -18.1$. We plot $\log \Delta$ against the total propagation time $T$ for $\sigma = 1$ to compare with $\log F_A$ and $\log F_G$ from the error estimates in Equation (\ref{estimates}). For small $T$, $\log F_A$ is a good estimate; for large $T$, $\log F_G$ is better. We also compare results using classical simulations of QPE and adiabatic evolution starting from the same initial state for the same total propagation times. For adiabatic evolution, the initial Hamiltonian is $H_I = \sum_{j=1}^{10} (-1)^j \sigma^z_j$ with an interpolating function $H(t) = \cos^2[\pi t/(2T)]H_I + \sin^2[\pi t/(2T)]H_{\rm obj}$. Phase estimation and adiabatic evolution perform similarly, but the rodeo algorithm is exponentially faster.

\subsection{Applications to Single-Qubit Hamiltonians on Quantum Hardware}\label{sec:1qapp}
We consider a quantum register with two qubits: the primary "object'' qubit and the ancillary "arena'' qubit. A single-qubit Hamiltonian is represented in its general form as:

\begin{equation}\label{eq:hobjgeneral}
H_{\rm obj} = c_I I + c_XX + c_YY + c_ZZ,
\end{equation}

where $I$ is the identity operator, $X, Y, Z$ are Pauli operators and $c_I, c_X, c_Y, c_Z$ are real coefficients. The object qubit is initialized in state $\ket{\psi_I}$, and the ancillary qubit is initialized in state $\ket{0}$. We perform $N$ cycles of the RA as defined in Section \ref{sec:ra} with the energy parameter $E$. IBMQ's mid-circuit measurements permit reuse of the ancillary qubit so that only two qubits total are needed. 
The time parameter $t_n$ for each cycle is randomly sampled from a Gaussian distribution with a mean of zero and a standard deviation of $\sigma$. From Equation \ref{ExpeProbs}, it can be inferred that the success probability will decrease exponentially as the difference between $E$ and the nearest eigenstate of $H_{\rm obj}$ increases.

\subsubsection{Gate Implementation For Single Qubit System}
The description of the quantum circuit for the RA in Section \ref{sec:1qapp} includes a controlled time-evolution gate in each cycle. Although such a gate operation could be done on a quantum device using optimal control as in Chapter \ref{cha:adbopt}, in general such an operation is not available, and it must be decomposed into elementary quantum gates in order to be implemented.

The time evolution operator can be written from the single-qubit object Hamiltonian $H_{\rm obj}$ as follows:
\begin{align}
    U(t)=e^{-iH_{\rm obj}t}=e^{-ic_It}e^{-\frac{i\theta}{2}\hat{n}\cdot\vec{\sigma}}\equiv e^{-ic_It}R_{\hat{n}}(\theta),
\end{align}
where $R_{\hat{n}}(\theta)$ represents a rotation matrix around the unit vector $\hat{n}$ in three dimensions by an angle $\theta$. This matrix can parameterize any matrix in the fundamental representation of the $SU(2)$ group and thus can be used as a general representation of any single qubit gate operation. It can be written in terms of the coefficients from Equation \ref{eq:hobjgeneral} as follows:
\begin{align} \label{eq:Rnhat}
R_{\hat{n}}(\theta) &= e^{-\frac{i\theta}{2}\hat{n}\cdot\vec{\sigma}}\nonumber \\
&=\begin{bmatrix}
\cos(\frac{\theta}{2})-i\sin(\frac{\theta}{2})n_Z & -i\sin(\frac{\theta}{2})(n_X-in_Y)\\
-i\sin(\frac{\theta}{2})(n_X+in_Y) & \cos(\frac{\theta}{2})+i\sin(\frac{\theta}{2})n_Z
\end{bmatrix},
\end{align}
where
\begin{align}
\theta = 2t\sqrt{c_X^2+c_Y^2+c_Z^2},
\end{align}
and
\begin{align}
\hat{n}=\frac{1}{\sqrt{c_X^2+c_Y^2+c_Z^2}}\begin{bmatrix}
c_X\\
c_Y\\
c_Z
\end{bmatrix}
=\begin{bmatrix}
n_X\\
n_Y\\
n_Z
\end{bmatrix}.
\end{align}

A generic single-qubit quantum operation $U$ is conventionally parameterized with three Euler angles $\gamma,\beta,\delta$ as follows: 

\begin{align}\label{eq:U3}
U(\gamma,\beta,\delta) &=
\begin{bmatrix} 
\cos\left(\frac{\gamma}{2}\right) & -e^{i\delta} \sin\left(\frac{\gamma}{2}\right) \\
e^{i\beta} \sin\left(\frac{\gamma}{2}\right) & e^{i(\delta + \beta)} \cos\left(\frac{\gamma}{2}\right)
\end{bmatrix}.
\end{align}

Applying the $Z-Y$ decomposition for a single qubit \cite{nielsen_ma2010}, we can rewrite Equation \ref{eq:U3} as:
\begin{align}\label{U3ZY}
U(\gamma,\beta,\delta)
&= e^{i\frac{\delta+\beta}{2}}
\begin{bmatrix}
e^{-i\frac{\delta+\beta}{2}}\cos\left(\frac{\gamma}{2}\right) & -e^{i\frac{\delta-\beta}{2}}\sin\left(\frac{\gamma}{2}\right) \nonumber \\
e^{-i\frac{\delta-\beta}{2}}\sin\left(\frac{\gamma}{2}\right) & e^{i\frac{\delta+\beta}{2}}\cos\left(\frac{\gamma}{2}\right)
\end{bmatrix} \nonumber \\
&= e^{i\frac{\delta+\beta}{2}}R_Z(\beta)R_Y(\gamma)R_Z(\delta) \nonumber \\
&\equiv e^{i\frac{\delta+\beta}{2}}R_{\hat{n}}(\theta).
\end{align}

Equating the upper-left and lower-left entries of the matrices $R_{\hat{n}}(\theta)$ and $R_Z(\beta)R_Y(\gamma)R_Z(\delta)$ yields the following constraints:
\begin{align} \label{ul}
\cos\left(\frac{\delta+\beta}{2}\right)\cos\left(\frac{\gamma}{2}\right) &= \cos\left(\frac{\theta}{2}\right),\\
-\sin\left(\frac{\delta+\beta}{2}\right)\cos\left(\frac{\gamma}{2}\right) &= -n_Z\sin\left(\frac{\theta}{2}\right),\\
\cos\left(\frac{\delta-\beta}{2}\right)\sin\left(\frac{\gamma}{2}\right) &= n_Y\sin\left(\frac{\theta}{2}\right),\\
-\sin\left(\frac{\delta-\beta}{2}\right)\sin\left(\frac{\gamma}{2}\right) &= -n_X\sin\left(\frac{\theta}{2}\right).
\end{align}
The global phase $e^{i\frac{\delta + \beta}{2}}$ in the gate $U$ can be replaced with the overall phase controlled by the term $c_I I$, generating two additional terms $\xi$ in the argument of the phase shift gate. Solving for the parameters $\delta$, $\beta$, $\gamma$, and $\xi$ yields the following:

\begin{align} \label{gensol}
\delta &= \tan^{-1}\left[n_Z\tan\left(\frac{\theta}{2}\right)\right]+\tan^{-1}\left(\frac{n_X}{n_Y}\right), \\
\beta &= \tan^{-1}\left[n_Z\tan\left(\frac{\theta}{2}\right)\right]-\tan^{-1}\left(\frac{n_X}{n_Y}\right), \\
\gamma &= 2\cos^{-1}\left[\frac{\cos(\theta/2)}{\cos\left(\frac{\delta+\beta}{2}\right)}\right], \\
\xi &= -c_I t-\frac{\delta+\beta}{2}.
\end{align}

The parameters $\delta$, $\beta$, and $\gamma$ can be used as arguments in the elementary controlled-$U$ gate operation in order to implement the controlled time evolution in the quantum circuit.

\subsubsection{Determining the Energy Spectrum}

\begin{figure}
\centering
\includegraphics[width=\linewidth]{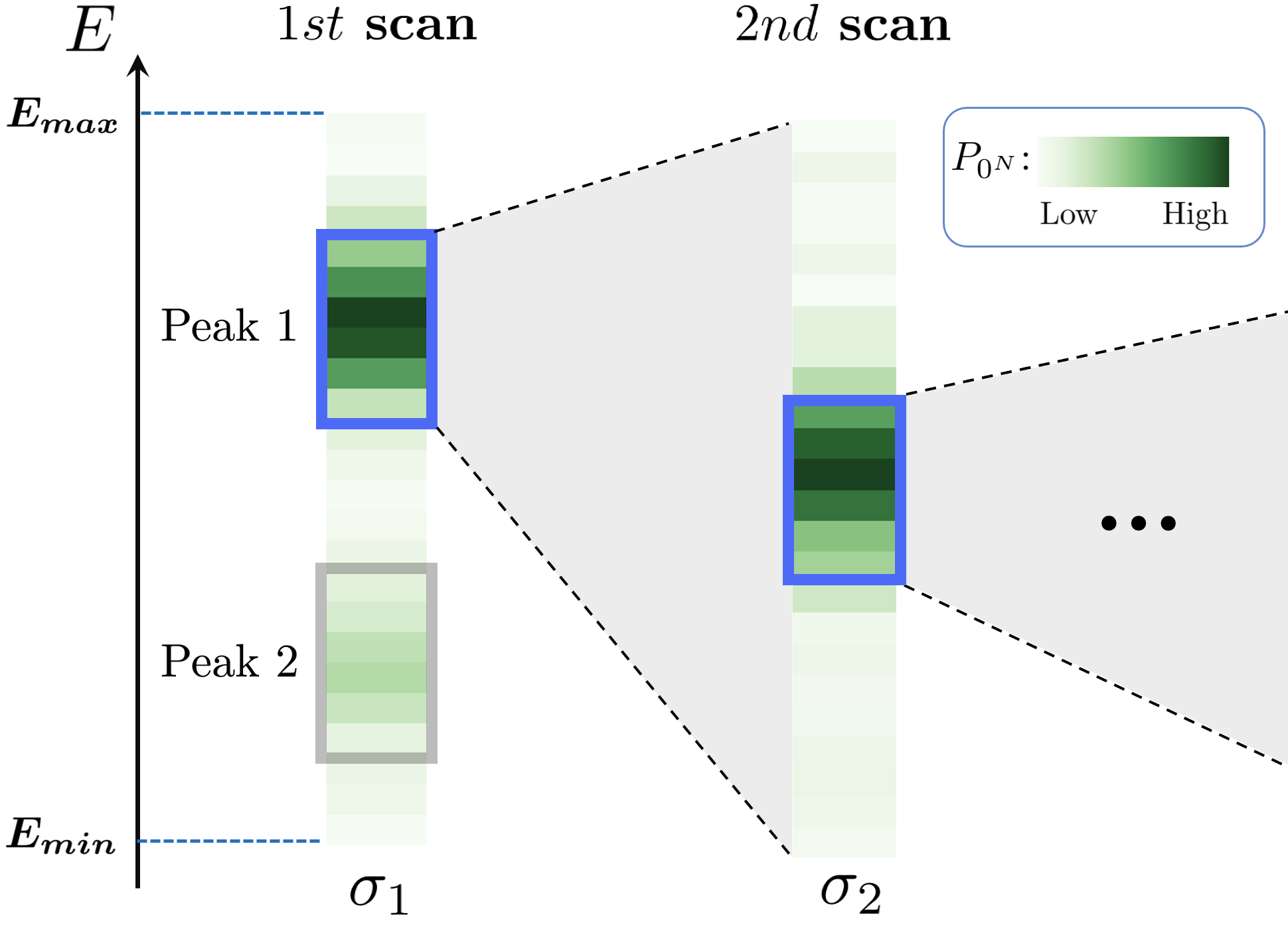}
\caption{(Figure reused from \cite{Qian:2021wya}) Sequential scans of the energy. Each bin represents a distinct RA circuit for target energy $E$ and width parameter $\sigma$. The color and shading indicates the success probability $P_{0^N}(E)$.  Centered around each of the peaks from the first scan, a second scan is performed using with a large value of $\sigma$ and better energy resolution. This is then repeated for the third scan. }
\label{fig:RAScan}
\end{figure}

To find the eigenvalues of $H_{\rm obj}$, we implement RA while scanning the target energy $E$ from $E_{\min}$ to $E_{\max}$, a range that is estimated from the operator norm of $H_{\rm obj}$. The eigenvalues appear as peaks in the success probability distribution. We used multiple scans, each with higher resolution by adjusting the width parameter $\sigma$, inversely proportional to the sharpness of the resolution of the energy. Three scans are used in this example; the second and third scans are centered around the peaks of their previous scans, as illustrated in Figure \ref{fig:RAScan}.

\begin{figure}
\centering
\includegraphics[width=\linewidth]{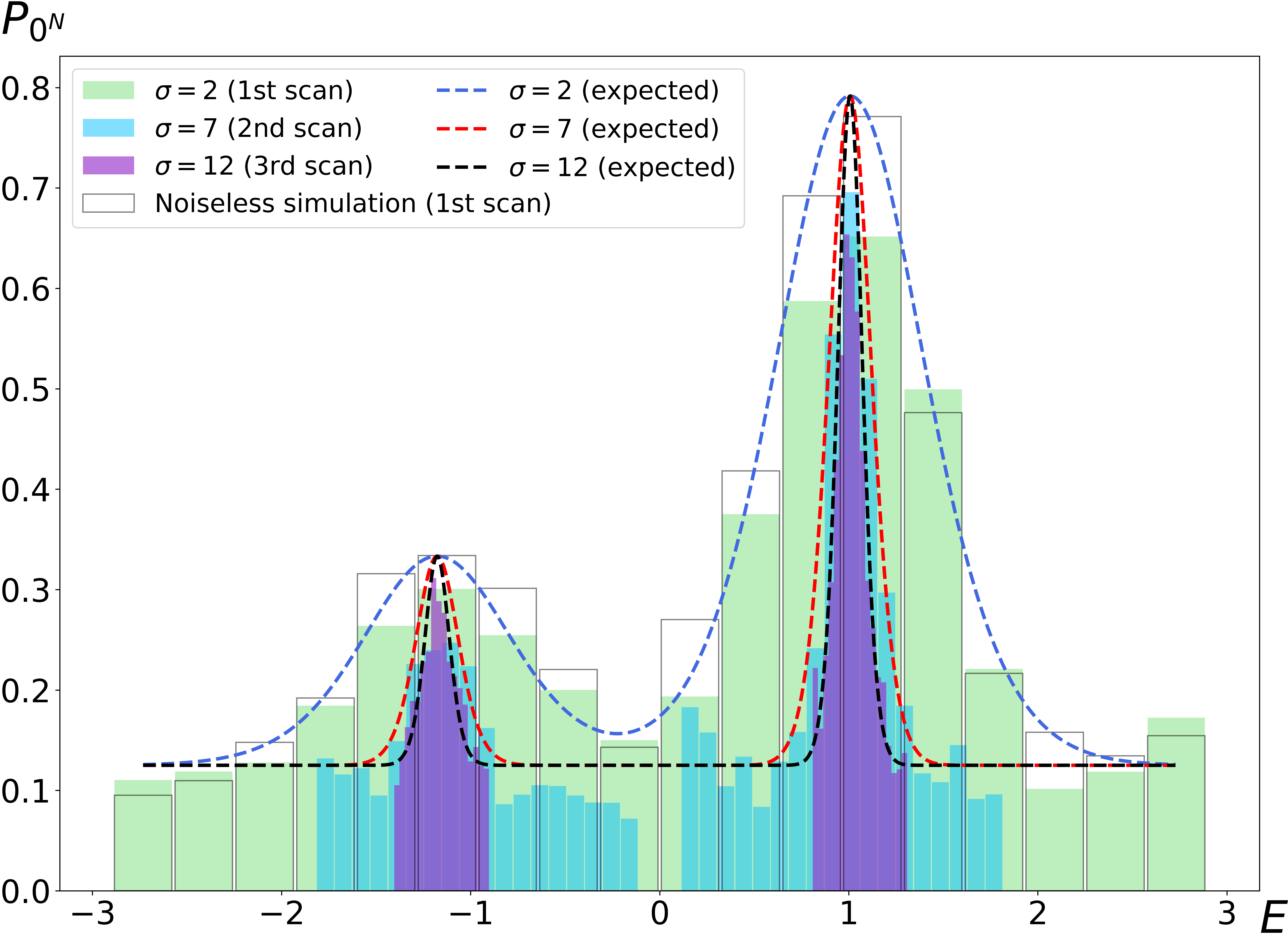}
\caption{(Figure reused from \cite{Qian:2021wya}) Energy scans for $H_{\rm obj}(0)$. The results were obtained from experiments on IBMQ's Casablanca device. The dashed lines show predicted outcomes from classical success probability calculations. The initial scan also includes noiseless quantum device simulation results.}
\label{fig:scanH}
\end{figure}

For the purposes of the following analysis, we consider a one-parameter family of one-qubit Hamiltonians:
\begin{align}& H_{\rm obj}(\phi)=H^{(0)}+\phi H^{(1)},\end{align}
Written in terms of two single-qubit Hamiltonians:
\begin{align}H ^{(0)} = -0.08496 I - 0.89134 X + 0.26536 Y + 0.57205 Z, \end{align}
and
\begin{align}H^{(1)} = -0.84537 I + 0.00673 X - 0.29354 Y + 0.18477 Z. \end{align}
The coefficients in both Hamiltonians were randomly sampled from a uniform distribution over the interval $[-1, 1]$. We also set the number of cycles $N = 3$ for all quantum circuits in this section. This choice minimizes error due to decoherence by keeping the number of gates low, and the results from Section \ref{sec:heisra} suggest that this should be enough to see the effect of RA. The results for $H_{\rm obj}(0)$ are illustrated in Figure \ref{fig:scanH}. The initial state is set to $\ket{0}$, and we perform three separate energy scans with $\sigma$ values set to 2, 7, and 12 to show its inverse effect on the magnification of the energy sensor. The experimental results were obtained using the IBMQ Casablanca device, focusing on two interconnected qubits characterized by minimal real-time error rates. The dashed lines represent the analytic results as derived from classical success probability calculations. Additionally, for the initial scan, we include results from a noiseless quantum device simulator.

Although the heights of the peaks are lower, their locations remain accurate in the presence of noise on the quantum device. Peak positions are estimated by fitting Gaussian functions to the success probability data in the vicinity of each peak. When the difference between $E$ and the target eigenvalue is greater than $1/\sigma$, the probability of measuring $\ket{0}$ is approximately $1/2$ per cycle. With the number of cycles $N=3$, the probability of success is $(1/2)^3 = 0.125$, which is illustrated by the background value in Figure \ref{fig:scanH}.

\subsubsection{Applying the Hellmann-Feynman Theorem}
\begin{figure}[ht]
    \centering
    \includegraphics[width=0.5\linewidth]{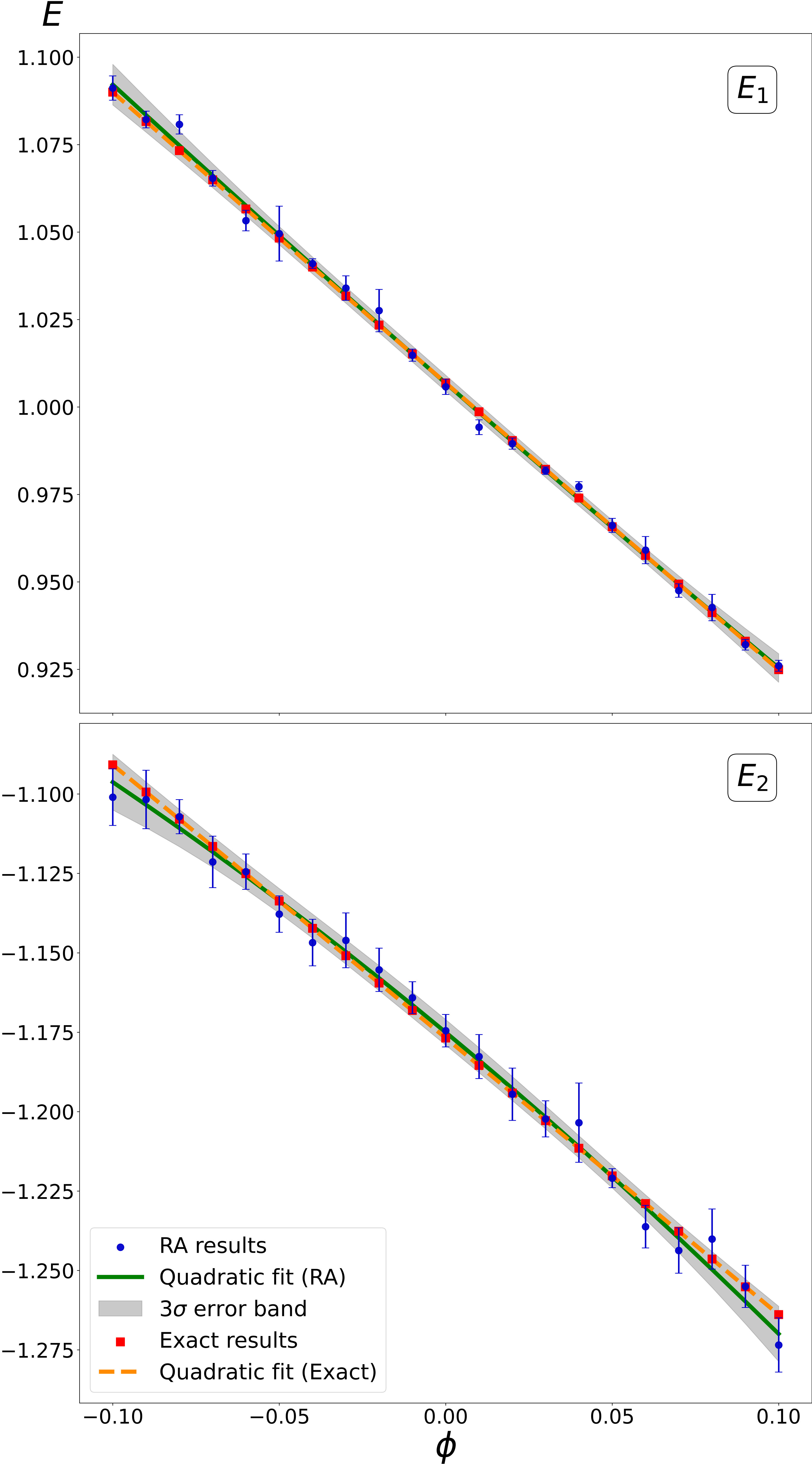}
    \caption{(Figure reused from \cite{Qian:2021wya}) Eigenvalues \(E_1\) (top) and \(E_2\) (bottom) vs. \(\phi\). The plot shows RA data (circles), RA quadratic fit (solid line) with \(3\sigma\) error bands (shaded), exact values (squares), and exact quadratic fit (dashed line).}
    \label{fig:RA_FHResults}
\end{figure}

Using the eigenvalues of $H_{\rm obj}(\phi)$ for small $\phi$, we can apply the Hellmann-Feynman theorem to find the expectation value of $H^{(1)}$ for the eigenstates of $H^{(0)}$ \cite{Feynman:1939}. This theorem essentially represents the first-order perturbation theory for energy. If $E_n(\phi)$ are the energy eigenvalues of $H_{\rm obj}(\phi)$ and $\ket{\psi_n(\phi)}$ are the corresponding eigenstates, then:
\begin{align}
\frac{dE_n(\phi)}{d\phi} = \braket{\psi_n(\phi)|H^{(1)}|\psi_n(\phi)}.
\end{align}
At $\phi = 0$, we obtain the expectation values of $H^{(1)}$ for the eigenstates of $H_{\rm obj}(0)=H^{(0)}$. In Figure \ref{fig:RA_FHResults}, the energy eigenvalues of $H_{\rm obj}(\phi)$ are plotted. The upper panel shows the higher energy eigenvalue $E_1$, and the lower panel shows the lower energy eigenvalue $E_2$. As $\phi$ varies, a scan with $\sigma = 12$ suffices to find each of these eigenvalues. We perform $2500$ measurements for $25$ random sets of $t_n$ for $E_1$ and $5000$ measurements for $50$ random sets of $t_n$ for $E_2$, each set containing 3 numbers due to the chosen value $N=3$. RA results (filled circles) are plotted with a quadratic fit (solid line) and three-standard-deviation error bands (shaded). The exact results (filled squares) and their quadratic fit (dashed line) are also shown. The error bars on the RA data represent errors of one standard deviation from the mean. The smaller error bars for $E_1$ compared to $E_2$ are due to a larger overlap of the initial state with the corresponding eigenstate.

Using quadratic fitting of the RA data points, we determine the energy eigenvalues of $H^{(0)}$ and the expectation values of $H^{(1)}$ for the eigenstates of $H^{(0)}$. The results are in Table \ref{table:fit_results}. The error bars represent the one-sigma uncertainties from statistical noise and Gaussian peak fitting. The exact results are provided for comparison. The relative error in determining the energies of $H^{(0)}$ is 0.08\%, which falls within our error estimates even without the application of error mitigation techniques. The application of the Hellmann-Feynman theorem results in larger error bars due to the error being proportional to the derivative of the energy ; however, the uncertainty in the expectation value of $H^{(1)}$ remains relatively small. The relative error for the eigenvalues of $H^{(1)}$ is 0. 7\%, which is well in agreement with our predicted error estimates.

\begin{table}[h]
\centering
\caption{Results obtained using RA for the energy eigenvalues of $H^{(0)}$ and the expectation values of $H^{(1)}$ corresponding to the eigenstates of $H^{(0)}$. For reference, exact values are also provided.}
\begin{tabular}{|c|c|c|c|c|}
\hline 
 & $\ket{\psi_1(0)}$ & exact & $\ket{\psi_2(0)}$ & exact \\
\hline
$\braket{H^{(0)}}$ & $1.00681(66)$ & $1.00690$ & $-1.1750(12)$ & $-1.1768$  \\
\hline
$\braket{H^{(1)}}$ & $-0.8338(89)$ & $-0.8254$ & $-0.868(14)$ & $-0.8653$   \\
\hline
\end{tabular}
\label{table:fit_results}
\end{table}

\subsubsection{Testing Eigenstate Preparation}
To assess the outcomes of the RA as a method to prepare eigenstates, we used the IBMQ Casablanca device to prepare eigenstates $\ket{\psi_1(0)}$ and $\ket{\psi_2(0)}$ and measure the expectation values of $H^{(0)}$ and $H^{(1)}$. This calculation can be thought of as an upper limit on the accuracy of the variational quantum eigensolver \cite{Peruzzo_2014}. Table \ref{table:no_mitigation} shows the results without using any error mitigation techniques, displaying expectation values of $X, Y, Z$, $H^{(0)}$, and $H^{(1)}$. The error bars represent statistical errors from 10 trials of 5000 measurements for each Pauli operator and eigenstate. The relative error for $H^{(0)}$ is $5\%$ while for $H^{(1)}$ it is $0.6\%$, with the smaller deviation for $H^{(1)}$ attributed to the smaller coefficients of the Pauli matrices. Both deviations significantly exceed statistical error estimates, indicating systematic errors likely due to measurement bias.

\begin{table}[ht]
\centering
\caption{Results for eigenstate preparation using RA without measurement error mitigation.}
\begin{tabular}{|c|c|c|c|c|}
\hline 
 & $\ket{\psi_1(0)}$ & exact & $\ket{\psi_2(0)}$ & exact \\
\hline
\,$\langle X\rangle$\, & -0.7455(44) & -0.8164 & 0.8055(22) & 0.8164  \\
\hline
\,$\langle Y\rangle$\, & 0.2750(36) & 0.2430 & -0.2196(25) & -0.2430  \\
\hline
\,$\langle Z\rangle$\, & 0.5356(46) & 0.5239 & -0.4632(21) & -0.5239  \\
\hline
\,$\langle H^{(0)}\rangle$\, & 0.9589(48)& 1.0069 & -1.1262(24) & -1.1768  \\
\hline
\,$\langle H^{(1)} \rangle$\, &\, -0.8321(14)\, &\, -0.8254\, &\, -0.86109(84)\, &\, -0.8653\,  \\
\hline
\end{tabular}
\label{table:no_mitigation}
\end{table}

As expected, the results without error mitigation on a noisy device were sub-optimal. Thus, we reanalyzed the data while including measurement error mitigation techniques. Before the $10$ trials, we collected data for the confusion matrix using the same method as in Section \ref{sec:confusionmatrix}, showing the probability of measuring $\ket{0}$ or $\ket{1}$ for pure states. We adjusted our measurement statistics by multiplying by the inverse confusion matrix. The results in Table \ref{table:mitigation} show a significant reduction in errors. Using all results for $\ket{\psi_1}$ and $\ket{\psi_2}$, the relative errors for $H^{(0)}$ and $H^{(1)}$ are $0.2\%$ and $0.5\%$, respectively. Residual errors in Pauli operator expectations suggest remaining systematic errors, which shows that further error reduction is limited to techniques beyond increasing the measurement statistics.

\begin{table}[h]
\centering
\caption{Results for eigenstate preparation with measurement error mitigation.}
\begin{tabular}{|c|c|c|c|c|}
\hline 
 & $\ket{\psi_1(0)}$ & exact & $\ket{\psi_2(0)}$ & exact \\
\hline
\,$\langle X\rangle$\, & -0.8119(46) & -0.8164 & 0.8152(27) & 0.8164  \\
\hline
\,$\langle Y\rangle$\, & 0.2569(83) & 0.2430 & -0.2596(79) & -0.2430  \\
\hline
\,$\langle Z\rangle$\, & 0.5297(80) & 0.5239 & -0.5151(89) & -0.5239  \\
\hline
\,$\langle H^{(0)} \rangle$\, & 1.0100(65) & 1.0069 & -1.1751(60) & -1.1768  \\
\hline
\,$\langle H^{(1)}\rangle$\, &\, -0.8283(28)\, &\, -0.8254\, &\, -0.8589(29)\, &\, -0.8653\,  \\
\hline
\end{tabular}
\label{table:mitigation}
\end{table}

\subsection{Discussion}
Our analysis shows that the RA for the one-qubit Hamiltonian $H^{(0)}$ achieves a relative error of $0.08\%$ for energy eigenvalues, outperforming the expectation values of the eigenvector prepared directly from $H^{(0)}$ after applying measurement error mitigation techniques. Despite the Hamiltonian only operating on one qubit, six two-qubit CNOT gates are used, so total errors, which include gate error, measurement error, and qubit decoherence, exceed $0.08\%$. This shows that the RA's resilience and design allow it to achieve precise energy results even without error mitigation and in the presence of substantial noise. Even with noise reducing the spectral weight of the target eigenstate, the signal remains discernible from the background with sufficient statistics.

For single-qubit benchmarks, RA with precise gate calibration can further reduce relative errors. The computational cost to determine energy eigenvalues with relative error $\epsilon$ scales as $O[(\log \epsilon)^2/(p \epsilon)]$ \cite{Choi:2020pdg}, with $p$ being the squared overlap of the initial state and the target eigenvector, contrasted with the scaling $O(1/\epsilon^2)$ for the expectations of the eigenvector prepared directly. This estimate $O(1/\epsilon^2)$ also serves as a lower bound for variational quantum eigensolvers, not including the additional cost of variational search.

The calculations of the Hellmann-Feynman theorem of $H^{(1)}$ using the RA show a relative error of $0.7\%$, an order of magnitude larger, due to the need for numerical derivatives of $E_n(\phi)$. However, the theorem accurately computes observables on a quantum device. To maintain the relative error $\epsilon$ in the operator expectation values, the energies of $H_{\rm obj}(\phi)$ need to be computed with an error tolerance of $O(\epsilon^2)$ for $\phi$ values around $O(\epsilon)$, resulting in a computational complexity scaling of $O[(\log \epsilon)^2/(p \epsilon^2)]$.

\subsubsection{Future Work}
This Section demonstrated RA's performance in both classical simulations for the Heisenberg model Hamiltonian and physical devices executions for a general one-qubit Hamiltonian. An interesting next step would be to test RA on Hamiltonians of more than one qubit. This would introduce systematic errors due to the Suzuki-Trotter decomposition needed for the time evolution operator \cite{Trotter:1959,Suzuki:1976a,Childs:2019b}. However, an effective Hamiltonian can be defined that reproduces the Trotterized time evolution. Based on the results in Section \ref{sec:1qapp}, the eigenvalues of such an effective Hamiltonian can likely be determined with similar accuracy as in the one-qubit case, given a signal strong enough to rise above the random background. Increasing the number of qubits in the objective Hamiltonian will also allow the algorithm to be applied to problems that correspond more closely to real physical systems. 

Although the energy-scanning method has been successful in detecting the energy spectrum, it may encounter difficulties when applied to larger systems that require significantly more scans. One of the main scalability challenges of the RA is ensuring that the initial state has a sufficiently large overlap with the target eigenstate, which may be essential for increasing the success probability enough to discern the energy peaks from the background noise. Achieving this condition becomes especially difficult for larger systems of interest. Therefore, a possible next step, thoroughly investigated in Chapter \ref{cha:vra}, is to combine RA with a variational technique such as QAOA, using classical optimization to help prepare an initial state that ensures a sufficiently high success probability for the RA.


\chapter{Variational Rodeo Algorithm}\label{cha:vra}
This chapter expands on the Rodeo algorithm (RA) studied and evaluated in Chapter \ref{cha:rodeo}, by converting it into a variational technique, which we call the Variational Rodeo Algorithm (VRA). The objective is to address the main shortcoming of the RA, specifically that its success rate is contingent upon the overlap between the initial state and the target eigenstate, which is hard to control for many systems of interest. We test the algorithm by performing classical optimizations using emulated quantum circuits and also compare it with QAOA for the task of ground-state preparation. Our results, combined with those of Chapter \ref{cha:rodeo}, suggest that this method should be effective in preparing high-fidelity eigenstates on quantum computers.

\section{Algorithm}
Like other variational methods, the VRA uses a two-part method that alternates between executing a parameterized quantum circuit and classically optimizing the circuit parameters. The quantum circuit of VRA has two parts: a parameterized section followed by the RA circuit. Any parameterized circuit can be used for the first part, but in this chapter, we consider only the QAOA circuit because of its widespread recognition and because its resemblance to adiabatic evolution may give it an advantage in preparing ground-state eigenvectors. 

The Hamiltonian of interest is labeled as the "object Hamiltonian'', $H_{\rm obj}$. Just like in the RA, a quantum register is divided into an object system, on which $H_{\rm obj}$ is allowed to act, and a set of ancilla qubits called the 'rodeo arena.' If mid-circuit measurements are allowed, only one ancilla qubit is needed for the rodeo arena. First, the QAOA algorithm with mixer Hamiltonian $H_{\rm mix}$ and problem Hamiltonian $H_{\rm obj}$ is applied to the object system (see Section \ref{sec:qaoa}); then the RA with the chosen energy parameter $E$ is applied immediately thereafter. This circuit is executed repeatedly to determine the probability that each ancilla-qubit measurement is in the $\ket{1}$ state, which is the same as the "success probability'' in Section \ref{sec:ramath}. This probability is at a maximum when the final state of the object system after RA, $\ket{\psi_F}$, is an eigenstate of $H_{\rm obj}$. Thus, to prepare eigenstates, VRA aims to maximize this probability, using one minus the probability as the cost function for classical optimization.

\subsection{Comparison to other Methods}
Traditionally, the cost function for the classical optimization part of QAOA is the expectation value of the energy of the state stored in the quantum register with respect to $H_{\rm obj}$, which can be measured from repeated executions of the QAOA circuit without requiring full tomography. Due to the variational principle, the energy expectation value of any state provides an upper bound on the energy of the ground state, and the state with the minimum energy from a sample of states can be used as an approximation of the true ground state~\cite{griffiths2017introduction}. In many but not all cases, this state approaches the true ground state as its energy with respect to $H_{\rm obj}$ tends to its minimum value.
Although it often provides a good approximation for small systems (and is, in fact, universally capable of ground states of specific types of Hamiltonians~\cite{Morales_2020}), QAOA was designed as a heuristic approach and excels at providing approximations for ground state energies rather than preparing eigenstates with high fidelity~\cite{Moussa_2020}. Its limitations become apparent when QAOA is applied to problems where the Hilbert space of $H_{\rm obj}$ is large and the number of parameters in the circuit is low~\cite{Akshay_2020,Pelofske_2023,Streif_2020}.
Due to its inspiration from adiabatic evolution, QAOA can still be useful to converge to eigenstates for many problems due to the adiabatic theorem \cite{Binkowski_2024}.

For simplicity, the mixer Hamiltonian $H_{\rm mix}$ in QAOA is often chosen so that its time evolution operation can be implemented with $O(n)$ quantum gates, but the time evolution of the problem Hamiltonian $H_{\rm obj}$ may be much more complicated, often requiring Trotter approximations and costly 2-qubit gate decompositions~\cite{Willsch_2022}.
Thus, in order to be feasible to compute on NISQ-era devices, QAOA circuits for reasonably-sized problem Hamiltonians are often truncated to a small depth with fewer optimization parameters. Increasing this depth may lead to more accurate results, but would also quickly undermine the effectiveness of the algorithm on physical devices due to decoherence.

Rather than using QAOA alone to estimate the ground state energy, in this novel approach we use its output state as the initial state for the RA. This choice is based on the observation that the states produced by QAOA have significantly higher ground state overlap than randomly generated states, even for large systems~\cite{Pelofske_2023}. This makes it a good candidate as a precursor algorithm for RA, which should be capable of transforming its output state into one with even higher ground state overlap.
Taking this idea one step further, the RA circuit can be appended to the QAOA circuit during optimization so that the cost function depends not on the energy but on the success probability of the RA, which is also easy to estimate from circuit executions.

As we reason in Section \ref{sec:vramath} based on the conclusions of Section \ref{sec:ramath}, the success probability of the RA with the energy parameter $E$ is directly proportional to the overlap of the initial state with the particular problem Hamiltonian eigenstate with the closest eigenvalue to $E$.
In contrast, the relationship between the energy of a quantum state and its ground-state overlap is often unclear. Our results in Section \ref{sec:vraresults} suggest that the state with the minimum energy reachable by a QAOA circuit can differ significantly from the state with the highest overlap. As problem size increases or QAOA steps decrease, the ground-state overlap of the minimum energy state may be no higher than that of a random state in the Hilbert space.
This implies that the RA success probability should provide a more useful cost function for the use case of eigenstate preparation. By avoiding energy minimization, it is also straightforward to prepare excited states with VRA.

\section{Mathematical Comparison of Cost Functions}\label{sec:vramath}
In this section, we explore the effectiveness of three theoretical cost functions for classical optimization in parameterized quantum circuits. First, we investigate energy minimization, the most common approach used in VQE and QAOA methods. We then examine maximizing the overlap with the ground state, which, while ideal in theory, is impractical as it assumes prior knowledge of the system's ground state, an objective typically sought by such algorithms. Lastly, we evaluate the cost function proposed in our method, which maximizes the success probability of the RA. This method requires only an approximate guess for the ground state energy and, as we demonstrate, converges to the maximization of the ground state overlap in the limit where the guessed energy \(E\) approaches the true ground state energy \(E_0\), and the number of RA cycles and the standard deviation \(\sigma\) of the random time parameters become large.

For this section, we will consider a general Hamiltonian $H$ with eigenstates and eigenvalues $\ket{E_n}$ and $E_n$ where $n = \{0,1,\dots,N-1\}$ and $E_0 \leq E_1 \leq \dots \leq E_{N-1}$. Each cost function will be analyzed in the context of the objective of calculating the ground state energy $E_0$. This can be done in the context of a standard gradient descent algorithm by examining the direction of the gradient vector with respect to the optimization parameters to understand the path that an optimization would take.

\subsection{Energy Minimization}

To analyze the energy minimization cost function, we start by decomposing the general state $\ket{\psi}$ into energy eigenstates $\ket{E_n}$:

\begin{equation}
    \ket{\psi} = \sum_{n=0}^{N-1} c_n \ket{E_n}.
\end{equation}

The energy expectation value for the normalized state is given by:

\begin{equation}
    E_\psi = \frac{\braket{\psi|H|\psi}}{\braket{\psi|\psi}} = \frac{\sum_{n=0}^{N-1} |c_n|^2 E_n}{\sum_{n=0}^{N-1} |c_n|^2}.
\end{equation}

The complex phase of each \(c_n\) does not affect the energy expectation value. Thus, we can express \(c_n\) as \( |c_n| e^{i\theta_n} \) and keep each \( e^{i\theta_n} \) fixed for \( n = 0, \cdots, N-1 \). We now consider the squared absolute values \( |c_n|^2 \) as independent variables. The state \(\ket{\psi}\) can be parameterized in terms of the vector \(\{ |c_0|^2, \cdots, |c_{N-1}|^2 \}\). Taking the partial derivative of \( E_\psi \) with respect to \( |c_n|^2 \) yields the following.

\begin{equation}
    \frac{\partial E_\psi}{\partial |c_n|^2} = \frac{\sum_{m=0}^{N-1} |c_m|^2 (E_n - E_m)}{\left[ \sum_{m=0}^{N-1} |c_m|^2 \right]^2}.
\end{equation}

A simple rescaling of the overall normalization of \(\psi\) does not change the energy expectation value, evident from:

\begin{equation}\label{eq:rescalingemin}
    \sum_{n=0}^{N-1} |c_n|^2 \frac{\partial E_\psi}{\partial |c_n|^2} = \frac{\sum_{n=0}^{N-1} \sum_{m=0}^{N-1} |c_n|^2 |c_m|^2 (E_n - E_m)}{\left[ \sum_{m=0}^{N-1} |c_m|^2 \right]^2} = 0.
\end{equation}

Given this rescaling invariance, we normalize \(\ket{\psi}\) such that:
\begin{equation}
    \sum_{m=0}^{N-1} |c_m|^2 = 1.
\end{equation}

Therefore,

\begin{equation}
    \frac{\partial E_\psi}{\partial |c_n|^2} = E_n - E_\psi.
\end{equation}

Following the steepest descent path involves moving in the direction of the infinitesimal vector \(\{ \Delta |c_0|^2, \cdots, \Delta |c_{N-1}|^2 \}\), where:

\begin{equation}
    \Delta |c_n|^2 = - (E_n - E_\psi),
\end{equation}

for \( n = 0, \cdots, N-1 \). This approach shows that \( |c_n|^2 \) decreases for \( E_n > E_\psi \) and increases for \( E_n < E_\psi \),. We note that this optimization path may be inefficient in a large vector space.

\subsection{Ground State Overlap Maximization}

To maximize overlap with the ground state \(\ket{E_0}\), we define:

\begin{equation}
    P^0_\psi = \frac{\braket{\psi|E_0}\braket{E_0|\psi}}{\braket{\psi|\psi}} = \frac{|c_0|^2}{\sum_{m=0}^{N-1} |c_m|^2}.
\end{equation}

Taking the partial derivative of \(P^0_\psi\) with respect to \( |c_n|^2 \) gives:

\begin{equation}
    \frac{\partial P^0_\psi}{\partial |c_n|^2} = \frac{\delta_{n,0} \sum_{m=0}^{N-1} |c_m|^2 - |c_0|^2}{{\left[ \sum_{m=0}^{N-1} |c_m|^2 \right]^2}}.
\end{equation}

Similarly as in Equation \ref{eq:rescalingemin}, a rescaling of the overall normalization of \(\psi\) does not change the overlap:

\begin{equation}
   \sum_{n=0}^{N-1} |c_n|^2 \frac{\partial P^0_\psi}{\partial |c_n|^2} = \frac{\sum_{n=0}^{N-1} |c_n|^2 \delta_{n,0} \sum_{m=0}^{N-1} |c_m|^2 - \sum_{n=0}^{N-1} |c_n|^2 |c_0|^2}{{\left[ \sum_{m=0}^{N-1} |c_m|^2 \right]^2}} = 0.
\end{equation}

The state \(\ket{\psi}\) can be normalized as in Equation \ref{eq:normpsi}.

\begin{equation}\label{eq:normpsi}
    \sum_{m=0}^{N-1} |c_m|^2 = 1,
\end{equation}

Following the steepest ascent path involves moving in the direction of the infinitesimal vector \(\{ \Delta |c_0|^2, \cdots, \Delta |c_{N-1}|^2 \}\):

\begin{equation}
   \Delta |c_n|^2 = \delta_{n,0} - |c_0|^2 = \delta_{n,0} - P^0_\psi,
\end{equation}

for \( n = 0, \cdots, N-1 \). When \( P^0_\psi \) is small, the movement is primarily towards \( |c_0|^2 \) corresponding to the eigenvector of the ground state, facilitating an efficient search in a large vector space due to the low rank of the cost function.

\subsection{Rodeo Algorithm Success Probability Maximization}

To maximize the success probability of the rodeo algorithm for \(M\) cycles, with target energy \(E\) and standard deviation of time values \(\sigma\), we first define the success probability $P_{\psi}$ as follows:

\begin{align}\label{eq:ra_success}
    P_\psi &= \frac{\sum_{n=0}^{N-1}\left[ \frac{1}{2} + \frac{1}{2} e^{-(E_n-E)^2 \sigma^2 / 2} \right]^M \braket{\psi|E_n}\braket{E_n|\psi}}{\braket{\psi|\psi}} \\&= \frac{\sum_{n=0}^{N-1}\left[ \frac{1}{2} + \frac{1}{2} e^{-(E_n-E)^2 \sigma^2 / 2} \right]^M |c_n|^2}{\sum_{m=0}^{N-1} |c_m|^2}.
\end{align}

Taking the partial derivative of \(P_\psi\) with respect to \( |c_n|^2 \) gives:

\begin{equation}
    \frac{\partial P_\psi}{\partial |c_n|^2} = \frac{\sum_{m=0}^{N-1} |c_m|^2 \left\{\left[ \frac{1}{2} + \frac{1}{2} e^{-(E_n-E)^2 \sigma^2 / 2} \right]^M - \left[ 1 + e^{-(E_m-E)^2 \sigma^2 / 2} \right]^M \right\}}{\left[ \sum_{m=0}^{N-1} |c_m|^2 \right]^2}.
\end{equation}

Since normalization of \(\psi\) does not change the success probability of RA, it follows that:

\begin{align}
   &\sum_{n=0}^{N-1} |c_n|^2 \frac{\partial P_\psi}{\partial |c_n|^2} \\&= \frac{\sum_{n=0,m=0}^{N-1} |c_n|^2 |c_m|^2 \left\{\left[ \frac{1}{2} + \frac{1}{2} e^{-(E_n-E)^2 \sigma^2 / 2} \right]^M - \left[ \frac{1}{2} + \frac{1}{2} e^{-(E_m-E)^2 \sigma^2 / 2} \right]^M \right\}}{\left[ \sum_{m=0}^{N-1} |c_m|^2 \right]^2} \\&= 0.
\end{align} 

Following the steepest ascent path, the direction is given by the infinitesimal vector with $N$ components ($n=0,1,\dots,N-1)$ defined as follows:

\begin{equation}
    \Delta |c_n|^2 = \left[ \frac{1}{2} + \frac{1}{2} e^{-(E_n-E)^2 \sigma^2 / 2} \right]^M - P_\psi,
\end{equation}

for \( n = 0, \cdots, N-1 \). When \(P_\psi\) is small, the movement is primarily towards the eigenvectors where \(\left[ \frac{1}{2} + \frac{1}{2} e^{-(E_n-E)^2 \sigma^2 / 2} \right]^M\) is significant. Increasing \(M\) narrows the energy window around \(E\), leading to better search efficiency in a large vector space due to the effective low rank of the rodeo algorithm success probability. As $M$ increases, the gradient tends to increase the magnitude of the component with the smallest value of $|E_n - E|$ and decrease the magnitude of all other components. 

\section{Simulation and Results}\label{sec:vraresults}
\begin{figure}
    \centering
    \includegraphics[width=\linewidth]{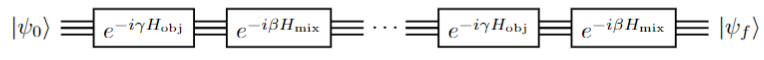}
    \caption{A circuit diagram of a QAOA circuit with objective Hamiltonian $H_f$, mixer Hamiltonian $H_i$, and tunable parameters $\mathbf \gamma$ and $\mathbf{\beta}$. Each time evolution gate has its own adjustable parameter, typically restricted to be in the range $(0,2\pi)$.}
    \label{fig:qaoa_circuit}
\end{figure}

In this section, we investigate the convergence of VRA compared to other variational quantum algorithms by performing classical optimization based on the output of classically simulated quantum circuits. For each of the following methods, we consider a QAOA circuit with $n$ qubits as the parameterized circuit for variational optimization. The objective Hamiltonian $H_{\rm obj}$ is defined by creating a random Hermitian matrix with dimensionality $2^n\times 2^n$. The Hermitian property is enforced by first generating a random triangular matrix of complex values, then summing it with its own conjugate transpose. This choice of Hamiltonian is motivated by its simplicity to implement, as well as the lack of degenerate eigenstates that may arise when choosing a more structured Hamiltonian. For the mixer Hamiltonian $H_{\rm mix}$, we use the so-called $x$-mixer defined as follows:

\begin{equation}
    H_{\rm mix} = \sum_{j=1}^n \mathbf X_j,
\end{equation}
where $\mathbf X_j$ represents an $\mathbf{X}$ gate on the qubit at index $j$. The QAOA circuit is initialized by putting the quantum register in the ground state of $H_{\rm mix}$, denoted $\ket{\psi_0}$. The circuit then consists of time evolution gates acting on the quantum register alternating over $H_{\rm obj}$ and $H_{\rm mix}$ with adjustable time parameters for each. The circuit diagram for this QAOA circuit is shown in figure \ref{fig:qaoa_circuit}. The number of time evolution gates, and therefore parameters, in the circuit is adjusted based on the number of qubits and the desired fidelity of state preparation.

In each of the following demonstrations, the exact final state $\ket{\psi_f}$ of the QAOA circuit is found classically using matrix-vector multiplications so that no circuit executions are necessary. From this state vector, the energy can be calculated as $\braket{\psi_f|H_{\rm obj}| \psi_f}$, and the modulus squared overlap with any particular eigenstate $\ket{E_k}$ of $H_{\rm obj}$ can be calculated as $|\braket{E_k|\psi_f}|^2$. If $\ket{\psi_f}$ is used as the initial state in a Rodeo algorithm circuit with $N$ cycles and time values sampled from a Gaussian distribution with mean 0 and standard deviation $\sigma$, then the probability of success for the Rodeo algorithm is given by $P_N$ in Equation \ref{eq:ra_success}, which we use instead of circuit executions or simulations.

The optimal parameters for each QAOA circuit are estimated using the classical BFGS optimization algorithm (see Section \ref{sec:bfgs}). The optimization is initialized with random parameters in the range $(0,2\pi)$. To avoid local minima or barren plateaus, we repeat the optimization with different random parameters and select the one with the best value of the cost function as the best estimate of the global optimum.  We choose this method because it is simple to implement and each individual optimization run can provide insight into the optimization landscape. Better estimates of global optima can also be obtained through different optimization strategies such as simulated annealing \cite{simanneal} and basin hopping \cite{basinhopping} or QAOA-specific strategies that adjust the depth of the QAOA circuit~\cite{10.1103/physreva.105.032433} or use machine learning to guess better starting parameters~\cite{10.1088/1367-2630/ad4629}.

In Section \ref{sec:vrasim1}, we demonstrate the ability of VRA to prepare excited states of a 6 and 10 qubit system by adjusting the energy parameter $E$ of the Rodeo algorithm. In Section \ref{sec:vrasim2}, we compare VRA with traditional QAOA as well as a two-step optimization approach alternating between both cost functions to prepare the ground state of a 6-qubit system. 

\subsection{Simulation Method 1}\label{sec:vrasim1}
First, we demonstrate the ability of VRA to prepare eigenstates other than the ground state, a feat that is not possible with techniques that use energy minimization. Since VRA aims to maximize the success probability of the Rodeo algorithm, the cost function to minimize is defined as $1-P_N$. We used repeated BFGS optimizations with random initial parameters to estimate the optimal set of parameters, repeating the whole process for different values of the RA energy parameter $E$. The number of cycles is chosen as $N=4$, and the standard deviation of the time values in the definition of $P_N$ given in Equation \ref{eq:ra_success} is set to be constant at $\sigma=3$. We repeat these steps for two random Hamiltonians on 6 and 10 qubits, and we consider 12 parameters and 16 parameters for their QAOA circuits, respectively. 

\begin{figure}
    \centering
    \includegraphics[width=\linewidth]{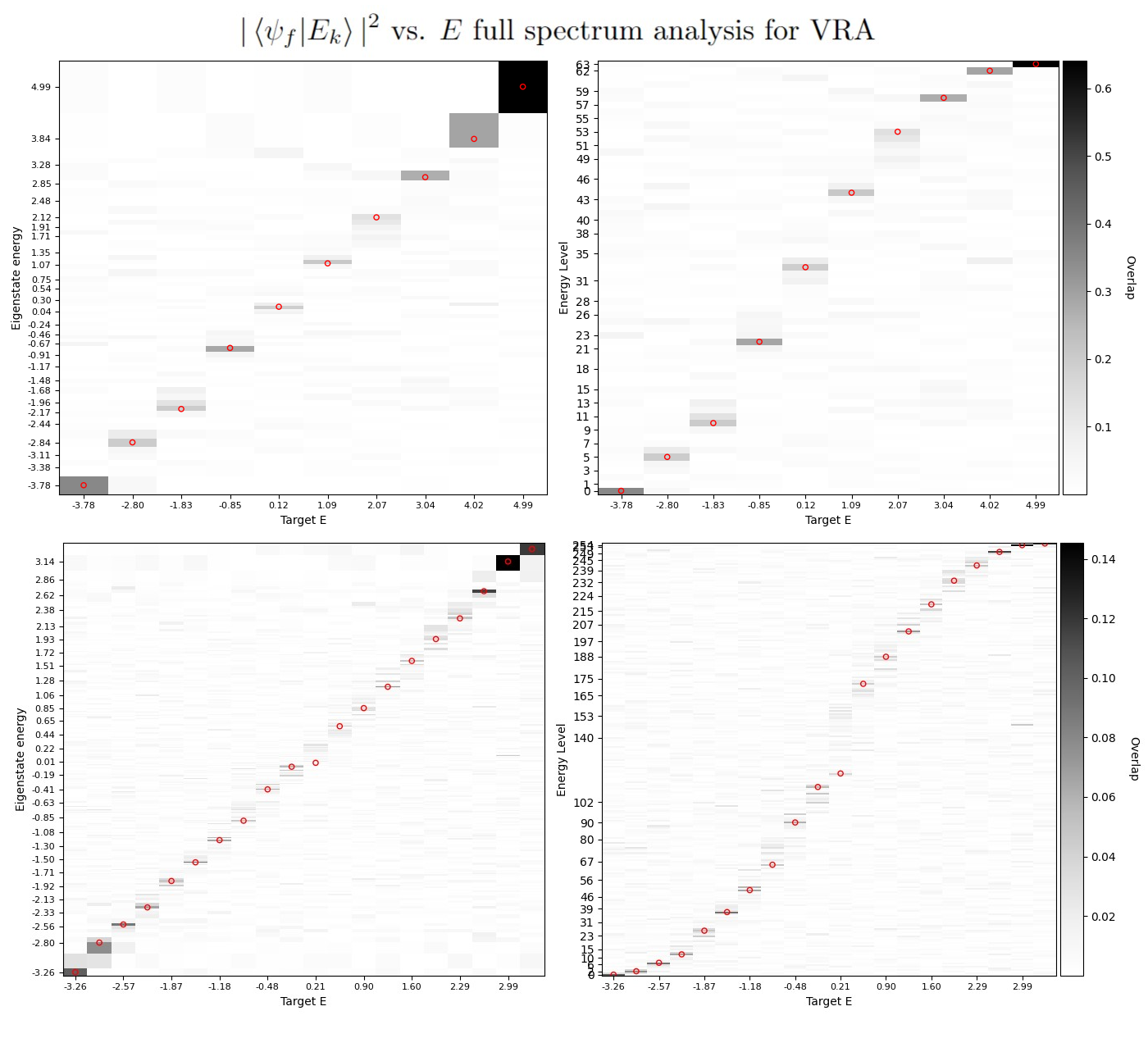}
    \caption{An analysis of the final states  $\ket{\psi_f}$ of VRA simulations on random 6-qubit (top row) and 10-qubit (bottom row) systems, repeated for different values of $E$, increasing linearly along each $x$-axis. The $y$-axes represent the exact eigenstates ordered by their eigenvalues scaled by energy (left column) and level number (right column). The overlap of the final state of the QAOA circuit with each eigenvector is represented by the relative darkness of each shaded rectangle. Note that many rectangles are too light to be seen due to extremely low overlap. For each VRA optimization, a red circle is placed at the eigenstate with the highest overlap  }
    \label{fig:vra1-6q}
\end{figure}

The relative overlap of the output state $\ket{\psi_f}$ of the QAOA circuit with each eigenstate of its corresponding Hamiltonian is visualized with the shaded rectangles in the graphs in Figure \ref{fig:vra1-6q}. In each case, $\ket{\psi_f}$ was dominated by its component with the eigenvector $\ket{E_k}$ with $E_k$ closest to $E$. Many of the shaded squares are therefore too light to see, so for each value of $E$ used in the VRA cost functions, a red dot is placed at the $y$ coordinate corresponding to the eigenstate that had the highest overlap, which was the one with the eigenvalue closest to $E$ in every case. These simulations suggest that VRA may be an effective method for the preparation of excited states on quantum computers.

\subsection{Simulation Method 2}\label{sec:vrasim2}
\begin{figure}
    \centering
    \includegraphics[width=\linewidth]{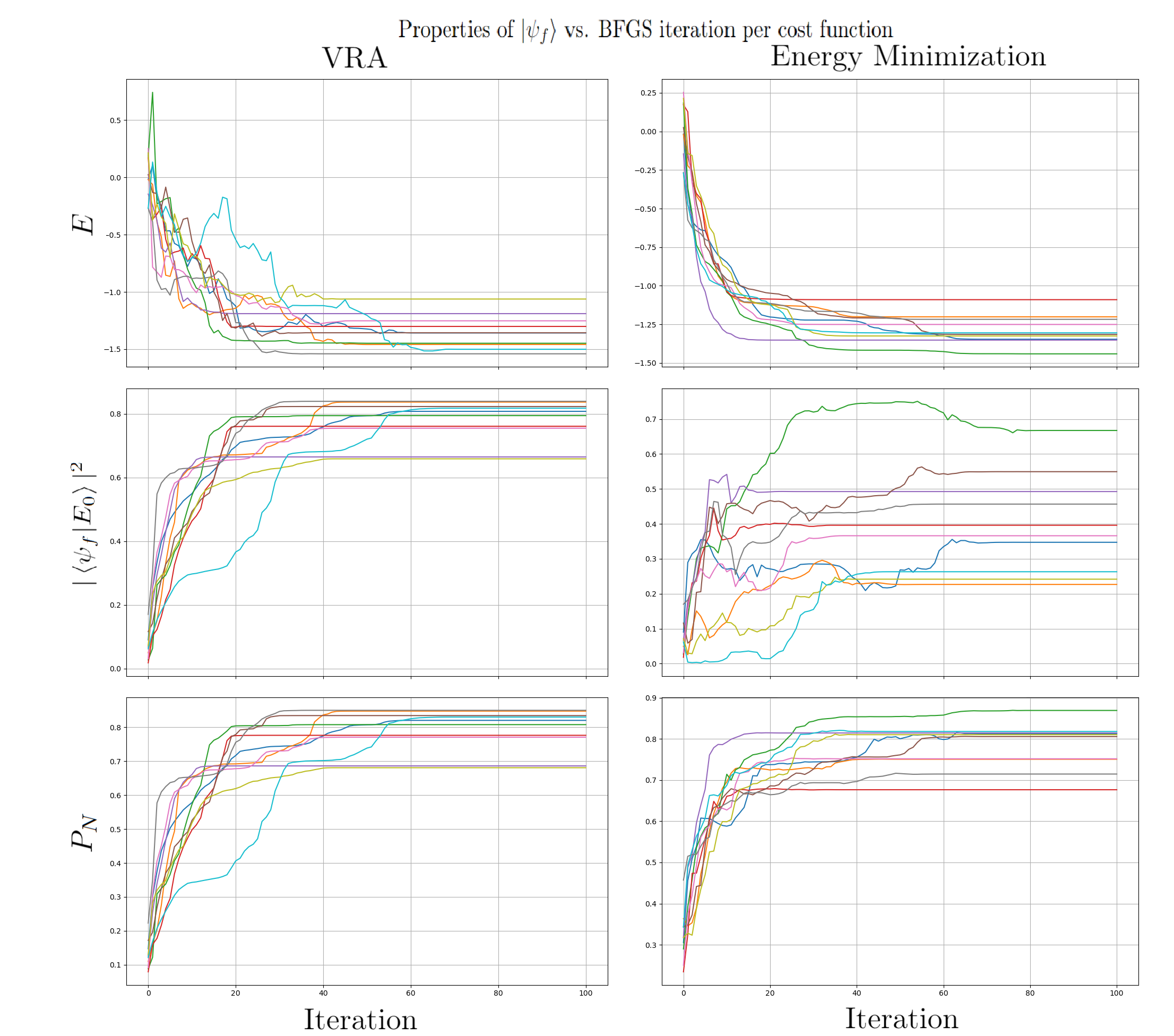}
    \caption{A comparison of the VRA (left) and VQE (right) cost functions for BFGS optimizations of a 20 parameter 6 qubit QAOA circuit. The shared $x$ axes represent the BFGS iteration number from 1 to 100. The $y$ axes in each row represent properties of the output state $\ket{\psi_f}$ of the QAOA circuit, namely energy (top), ground state overlap (middle), and RA success probability (bottom). Note the different scales on the $y$-axes. 10 random sets of 20 parameters each were generated and used as the initial circuit parameters in separate optimizations for both methods, with each set represented by a different color on the plots.}
    \label{fig:vrasim2}
\end{figure}

In this method, we compare VRA with pure QAOA as methods to prepare a ground-state eigenvector of a random 6-qubit Hamiltonian $H_{\rm obj}$. Before performing optimizations, we directly solve for the eigenvectors $\ket{E_k}$ of $H_{\rm obj}$ and use the ground-state energy $E_0$ as the target energy parameter in Equation \ref{eq:ra_success} for $P_N$ along with a $\sigma$ value of 10. We use the traditional VQE cost function (minimizing the energy of the output state $\ket{\psi_f}$) and the cost function $1-P_N$ for VRA as before. The parameterized circuits for both cost functions were QAOA circuits with 20 parameters. As before, we used multiple BFGS optimizations to get better estimates of the global minima. In practice, one could discard the results of all but one optimization that ended with the best final value of the cost function, but here we present all 10 optimizations side-by-side to better illustrate the optimization landscapes. On the $x$ axis is the iteration number of the BFGS optimization, and on the $y$ axes there are several values of interest for $\ket{\psi_f}$ with the circuit parameters in that iteration. Specifically, we tracked energy $\braket{\psi_f|H_{\rm obj}|\psi_f}$, ground state overlap squared $|\braket{E_0|\psi_f}|^2$, and RA success probability $P_N$. The results of each optimization are represented as the different colored lines in the plots in Figure \ref{fig:vrasim2}.

These results show that the VRA cost function is comparable to energy minimization in reaching states with low energies but superior in preparing states with high overlap with the ground state. The difference is noticeable here primarily because 20 QAOA parameters are not sufficient to allow the QAOA circuit to prepare the ground state with perfect fidelity, so the energy minimization technique tends to settle in a mixture of low-lying energy states, while the VRA method reduces the overlap with all states other than the ground state roughly equally. If the depth of the QAOA circuit is increased, the variational principle takes effect and the state produced by energy minimization would theoretically converge to the ground state. However, QAOA is an NISQ-era algorithm intended to be used with a relatively small number of parameters to compensate for the fact that current quantum devices decohere rather quickly, so these results are closer to what should be typical when applying QAOA or VQE to any reasonably sized system. We note that the plots produced by VRA for ground state overlap and $P_N$ are similar because, in this example, the RA parameters $E=E_0$ and $\sigma=10$ make it so that the energy filter of the RA is smaller than the energy gap to the first excited state. The best optimizations for both cost functions resulted in final states with a RA success probability of about 85\%, which makes it clear that the addition of RA to the QAOA circuit after optimization aids in the preparation of the ground state with high fidelity.

To further show the difference between these two cost functions, we combine them in a two-step optimization approach in which the final parameters of a 100 step QAOA optimization with energy minimization are used as initial parameters in a 100 step VRA optimization. This type of approach may be more practical because the VRA circuit depth is higher than that of the QAOA circuit alone. Therefore, it may save time to use energy minimization for the bulk of the optimization and then append the Rodeo algorithm for a few steps of fine-tuning. This approach effectively finds the closest local maximum of the RA success probability in the vicinity of a local energy minimum.

\begin{figure}
    \centering
    \includegraphics[width=0.5\linewidth]{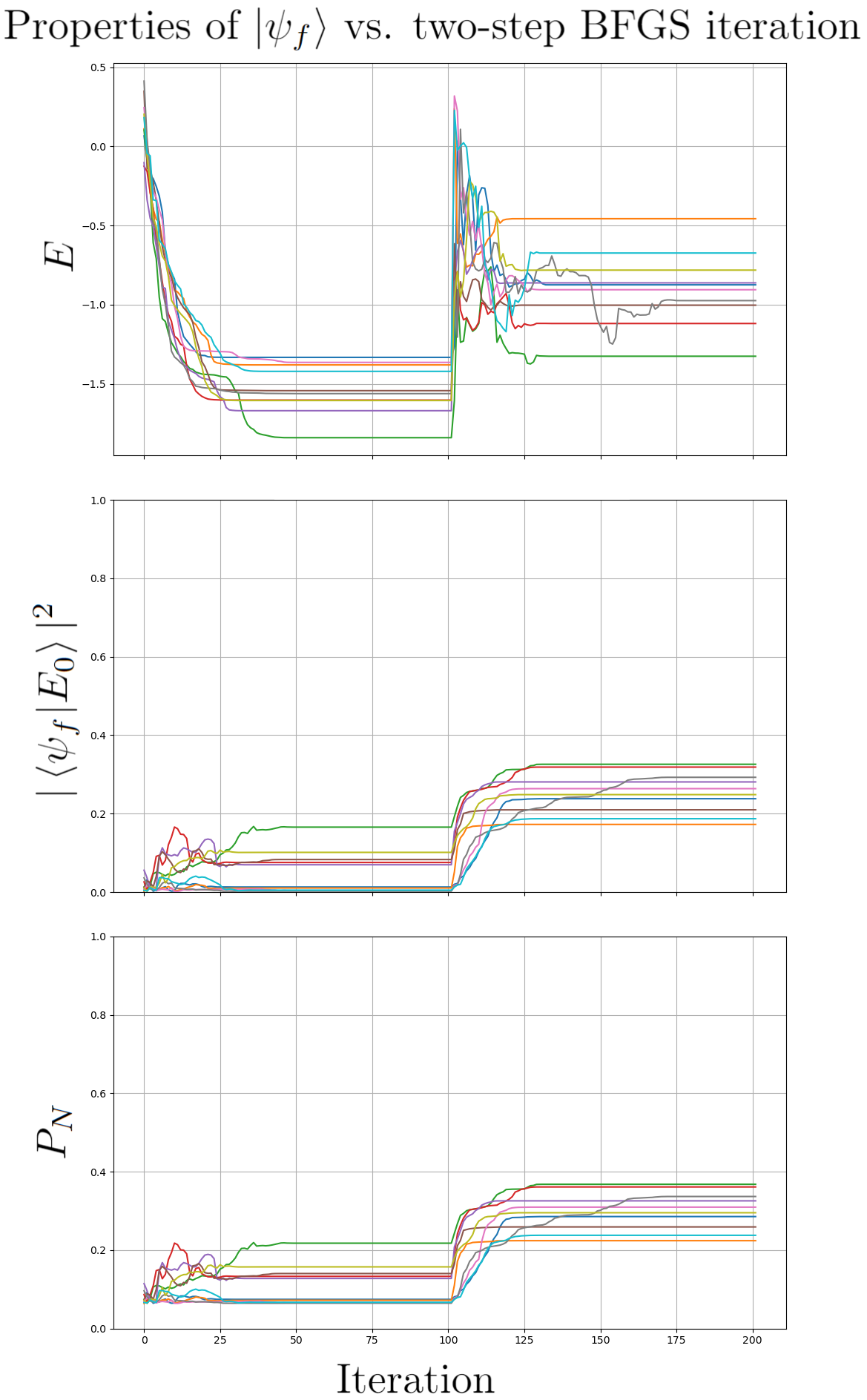}
    \caption{Results for a two-step BFGS optimization of a 12 parameter 6 qubit QAOA circuit with 100 steps of energy minimization followed by 100 steps of VRA. The axes are the same as in Figure \ref{fig:vrasim2} but with 200 total iterations now on the $x$ axis. 10 random sets of 12 parameters each were generated and used as the initial circuit parameters in separate optimizations, with each set represented by a different color on the plots.}
    \label{fig:vrasim3}
\end{figure}

We use a slightly different example from before by choosing a different random 6 qubit Hamiltonian and reducing the number of QAOA parameters to 12. Figure \ref{fig:vrasim3}  shows how the output state $\ket{\psi_f}$ produced by the QAOA circuit evolves throughout the optimizations. For circuits with few parameters like this one where the variational principle is not in effect, these two cost functions do not necessarily share the same local minima. Thus, changing the cost function from minimizing energy to maximizing the RA success probability results in an immediate correction to the output state of the QAOA circuit, giving it higher energy but also a higher RA success probability. This change also results in a larger overlap with the ground state because of its correlation with the probability of success of RA, which can be clearly seen by the similarity of the middle and bottom plots in the figure. In this case, with only 12 parameters, adding VRA to the optimization resulted in a much better overlap with the ground state, increasing the success probability of the best optimizations from roughly 20\% to almost 40\%, with more drastic improvements in the cases where VQE resulted in much lower ground state fidelity. For larger problems as the optimization landscape gets more complicated, we would expect VQE results to have a lower ground state overlap more often, so using VRA for fine-tuning may be critical for state preparation with any reasonable fidelity.

\section{Future Work}\label{sec:future}
This section will describe future avenues of research to improve and test this novel VRA technique, as well as some intermediate results that we have obtained in that direction. 

\subsection{Comparison of Optimization Landscapes}
The results in Section \ref{sec:vrasim2} suggest that the cost functions in QAOA and VRA have different local minima and barren plateaus. A better understanding of these differences can provide insight into when it is worthwhile to use VRA over QAOA. Barren plateaus, which are subspaces in the optimization landscape where the gradient vanishes, typically arise in quantum optimization problems whenever there are more than a few qubits or parameters in the circuit~\cite{McClean_2018}. This makes them inherently difficult to visualize in two- or three-dimensional plots. One possible method to test the density of local minima and barren plateaus is to perform many optimization attempts at different starting sets of parameters and see how frequently they get "stuck'' at a parameter state with zero gradient. However, this would not allow for differentiation between local minima and barren plateaus. 

\begin{figure}
    \centering
    \includegraphics[width=0.8\linewidth]{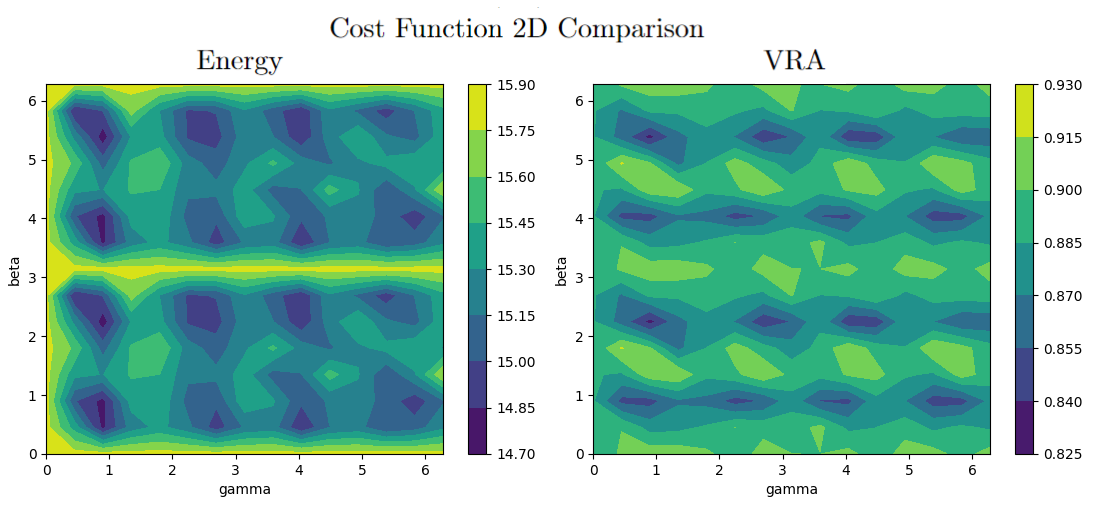}
    \caption{A comparison of values of the two cost functions, $E$ and $1-P_N$ (with $N=4$ and $\sigma =1$), of the output state of a 4-qubit QAOA circuit with two parameters. Darker areas represent areas of lower cost function values. $\beta$ on the $y$ axis and $\gamma$ on the $x$ axis are the time parameters for the time evolutions over $H_{\rm mix}$ and $H_{\rm obj}$, respectively.}
    \label{fig:2dcompare}
\end{figure}

It is also possible to examine the behavior of the cost functions when only a small number of parameters are allowed to vary, which may lead to insights into the optimization landscapes. As an example, we analyze the values of the two cost functions discussed in the above methods for a random four-qubit Hamiltonian with a QAOA circuit with two parameters, $\gamma$ and $\beta$, which are the time parameters for time evolution operations with $H_{\rm obj}$ and $H_{\rm mix}$ respectively. The values of these cost functions sampled on a lattice of values of $\gamma$ and $\beta$ in the range $(0,2\pi)$ are shown in Figure \ref{fig:2dcompare}. We note that our definition of $\beta$ left out a factor of 2 for simplicity, so the landscapes show two periods in the $y$ direction. This intermediate result shows roughly the same density of local minima in the two cost functions, but a more detailed analysis in higher dimensions may be worthwhile.

\subsection{Application to Interesting Problems}
This chapter mainly focused on solving for eigenstates of random Hamiltonians in order to avoid the extra complexity of identifying and justifying the use of a Hamiltonian with implications for solving real problems of interest. However, many such problems exist that may show the practical use cases of VRA. More research in this area may also point towards certain problems where VRA may theoretically outperform classical methods on supercomputers. Many interesting Hamiltonians are available for benchmarking purposes in online databases. For example, the HamLib library~\cite{sawaya2024hamliblibraryhamiltoniansbenchmarking} contains many interesting Hamiltonians in binary-variable optimization, condensed matter physics, and quantum chemistry problems, which are already mapped into qubit representations. Some of these problems may benefit from VRA's enhanced fidelity in preparing eigenstates. It may be especially interesting to apply the VRA to nuclear structure and reaction Hamiltonians, where better solutions may lead to advances in areas such as nuclear medicine and energy. These Hamiltonians typically have more terms than similar Hamiltonians in quantum chemistry and therefore require larger circuit depth to replicate time evolution operations, which is why more focus is currently placed on applying NISQ-era algorithms like QAOA to quantum chemistry applications. However, there may be theoretically interesting small nuclear systems that are good candidates for more accurate approximate solutions on quantum computers.
\chapter{Summary}

\section{Adiabatic Evolution with Optimal Control}
In Chapter \ref{cha:adbopt}, a robust method was presented to execute arbitrary sequences of unitary transformations, focusing on the adiabatic evolution of a two-spin system using Near-term Intermediate Scale Quantum (NISQ) devices. A two-qubit processor model, comprised of two capacitively coupled superconducting transmons, was employed to emulate adiabatic evolution with custom two-qubit gates. The output was then compared with the digital quantum simulations performed on IBMQ systems.

High-fidelity custom gates for short-time propagators could be realized with control pulses of varying duration, from 2500 to 120 ~ns. The lower limit was determined by the validity of the model employed and the specifications of the standard waveform generators. When the implementation time of short-time propagators was similar, the state fidelities of the emulated output and IBMQ were comparable, suggesting that the two-qubit processor model offered a realistic representation of quantum hardware. As the duration of control pulses and the implementation time of adiabatic evolution decreased, the loss of coherence due to the interaction between the quantum computer and its environment was significantly reduced. The fidelity achieved with custom gates exceeded that obtained through elementary gates, reaching fidelity of up to 95\% for control pulses with a length of $\tau=120$~ns. 

\section{Rodeo Algorithm}
In Chapter \ref{cha:rodeo}, we discussed the Rodeo Algorithm for generating quantum eigenstates and analyzing spectral characteristics. In summary, the Rodeo Algorithm uses a tunable energy filter and stochastic methods to prepare eigenstates of a quantum Hamiltonian or observable with exponential efficiency and short gate depth, surpassing the speeds of phase estimation and adiabatic evolution. The latter part of the chapter showcased the RA's performance through a simulation of a basic Heisenberg model and an implementation of a general one-qubit Hamiltonian on an actual device. The findings highlighted the potential of the RA in preparing eigenstates with high accuracy, which subsequently inspired the creation of the Variational Rodeo Algorithm discussed in the following chapter.

\section{Variational Rodeo Algorithm}
In Chapter \ref{cha:vra}, we introduce the Variational Rodeo Algorithm (VRA) as a variational extension of the Rodeo Algorithm (RA) to improve eigenstate preparation by overcoming the RA's dependency on the overlap between the initial and target eigenstates. We evaluated VRA using classical optimizations of emulated quantum circuits and compared its performance with the Quantum Approximate Optimization Algorithm (QAOA) for ground-state preparation. Our preliminary findings indicate that VRA effectively prepares high-fidelity eigenstates on quantum computers.

Unlike QAOA, which relies on energy minimization and often struggles with high-fidelity eigenstate preparation, the VRA maximizes the success probability of the RA, which serves as a more appropriate cost function for optimization, especially for larger systems or those with fewer optimization parameters. 
Simulations demonstrated the ability of VRA to readily prepare excited states by adjusting the RA energy parameter. Using QAOA's output as the initial state for the RA improved ground-state overlap, even when QAOA alone struggled with fidelity.

Overall, VRA presents a promising method for quantum eigenstate preparation, overcoming limitations in other algorithms like QAOA and quantum adiabatic evolution. Future research will focus on gaining a better understanding of VRA's optimization landscapes and applying it to quantum chemistry and nuclear physics problems to demonstrate its potential in solving complex quantum systems.

\section{Outlook and Future Research Directions} This dissertation has demonstrated significant advances in quantum eigenstate preparation through the development and testing of algorithms like the Rodeo Algorithm and its variational counterpart. Although our results provide promising indications of their effectiveness, several open questions and challenges remain, which lay the foundation for future research.

One promising direction is the application of these algorithms to more complex and larger-scale quantum systems, particularly in fields like quantum chemistry, material science, and nuclear physics, where high-fidelity eigenstate preparation is valuable. Moreover, refining these algorithms to function under the noisy conditions of NISQ devices will be essential for their implementation in near-term quantum computing applications. Further work is needed to explore how these methods can be generalized to multi-qubit systems and their performance optimized in the presence of noise and hardware imperfections.

Another area for future research is the development of more efficient optimization techniques for VRA, particularly as the complexity of quantum systems increases. Exploring hybrid quantum-classical approaches, where quantum algorithms are paired with classical machine learning or other advanced optimization methods, could potentially improve both the accuracy and efficiency of eigenstate preparation.

\bibliography{references}

\begin{thebibliography}{134}
\providecommand{\natexlab}[1]{#1}
\providecommand{\url}[1]{\texttt{#1}}
\expandafter\ifx\csname urlstyle\endcsname\relax
  \providecommand{\doi}[1]{doi: #1}\else
  \providecommand{\doi}{doi: \begingroup \urlstyle{rm}\Url}\fi

\bibitem[Aarts and Stamatescu(2008)]{Aarts_2008}
Gert Aarts and Ion-Olimpiu Stamatescu.
\newblock Stochastic quantization at finite chemical potential.
\newblock \emph{Journal of High Energy Physics}, 2008\penalty0 (09):\penalty0 018–018, September 2008.
\newblock ISSN 1029-8479.
\newblock \doi{10.1088/1126-6708/2008/09/018}.
\newblock URL \url{http://dx.doi.org/10.1088/1126-6708/2008/09/018}.

\bibitem[AB(2024)]{nobel2022}
Nobel Prize~Outreach AB.
\newblock The nobel prize in physics 2022, 2024.
\newblock URL \url{https://www.nobelprize.org/prizes/physics/2022/summary/}.
\newblock Accessed: 2024-07-14.

\bibitem[Aharonov(2003)]{toffoliproof}
Dorit Aharonov.
\newblock A simple proof that toffoli and hadamard are quantum universal, 2003.
\newblock URL \url{https://arxiv.org/abs/quant-ph/0301040}.

\bibitem[Akshay et~al.(2020)Akshay, Philathong, Morales, and Biamonte]{Akshay_2020}
V.~Akshay, H.~Philathong, M.E.S. Morales, and J.D. Biamonte.
\newblock Reachability deficits in quantum approximate optimization.
\newblock \emph{Physical Review Letters}, 124\penalty0 (9), March 2020.
\newblock ISSN 1079-7114.
\newblock \doi{10.1103/physrevlett.124.090504}.
\newblock URL \url{http://dx.doi.org/10.1103/PhysRevLett.124.090504}.

\bibitem[Albash and Lidar(2018{\natexlab{a}})]{albash_t2018}
T.~Albash and D.~A. Lidar.
\newblock Adiabatic quantum computation.
\newblock \emph{Rev. Mod. Phys.}, 90:\penalty0 015002, Jan 2018{\natexlab{a}}.
\newblock \doi{10.1103/RevModPhys.90.015002}.
\newblock URL \url{https://link.aps.org/doi/10.1103/RevModPhys.90.015002}.

\bibitem[Albash and Lidar(2018{\natexlab{b}})]{Albash_2018}
Tameem Albash and Daniel~A. Lidar.
\newblock Adiabatic quantum computation.
\newblock \emph{Reviews of Modern Physics}, 90\penalty0 (1), January 2018{\natexlab{b}}.
\newblock ISSN 1539-0756.
\newblock \doi{10.1103/revmodphys.90.015002}.
\newblock URL \url{http://dx.doi.org/10.1103/RevModPhys.90.015002}.

\bibitem[Anand et~al.(2022)Anand, Schleich, Alperin-Lea, Jensen, Sim, Díaz–Tinoco, Kottmann, Degroote, Izmaylov, and Aspuru‐Guzik]{uccvqe}
A.~Anand, P.~Schleich, S.~Alperin-Lea, P.~W.~K. Jensen, S.~Sim, M.~Díaz–Tinoco, J.~S. Kottmann, M.~Degroote, A.~F. Izmaylov, and A.~Aspuru‐Guzik.
\newblock A quantum computing view on unitary coupled cluster theory.
\newblock \emph{Chemical Society Reviews}, 51:\penalty0 1659--1684, 2022.
\newblock \doi{10.1039/d1cs00932j}.

\bibitem[Anis et~al.(2021)]{qiskit_2021}
Md~Sajid Anis et~al.
\newblock {Qiskit: An open-source framework for quantum computing}, 2021.

\bibitem[Aristotle(350 B.C.)]{aristotle_physics}
Aristotle.
\newblock \emph{Physics}.
\newblock Internet Classics Archive, 350 B.C.
\newblock URL \url{http://classics.mit.edu/Aristotle/physics.html}.
\newblock Translated by R.P. Hardie and R.K. Gaye.

\bibitem[Badem et~al.(2018)Badem, Baştürk, Çalışkan, and Yüksel]{beecolony}
H.~Badem, A.~Baştürk, A.~Çalışkan, and M.~E. Yüksel.
\newblock A new hybrid optimization method combining artificial bee colony and limited-memory bfgs algorithms for efficient numerical optimization.
\newblock \emph{Applied Soft Computing}, 70:\penalty0 826--844, 2018.
\newblock \doi{10.1016/j.asoc.2018.06.010}.

\bibitem[Bai and Silverstein(2010)]{bai2010}
Z.~Bai and J.~W. Silverstein.
\newblock Spectral analysis of large dimensional random matrices.
\newblock \emph{Springer Series in Statistics}, 2010.
\newblock \doi{10.1007/978-1-4419-0661-8}.

\bibitem[Banchi and Crooks(2021)]{Banchi_2021}
Leonardo Banchi and Gavin~E. Crooks.
\newblock Measuring analytic gradients of general quantum evolution with the stochastic parameter shift rule.
\newblock \emph{Quantum}, 5:\penalty0 386, January 2021.
\newblock ISSN 2521-327X.
\newblock \doi{10.22331/q-2021-01-25-386}.
\newblock URL \url{http://dx.doi.org/10.22331/q-2021-01-25-386}.

\bibitem[{Barenco, A. and Bennett, C. H. and Cleve, R. and DiVincenzo, D. P. and Margolus, N. and Shor, P. and Sleator, T. and Smolin, J. A. and Weinfurter, H.}(1995)]{barenco_a1995}
{Barenco, A. and Bennett, C. H. and Cleve, R. and DiVincenzo, D. P. and Margolus, N. and Shor, P. and Sleator, T. and Smolin, J. A. and Weinfurter, H.}
\newblock Elementary gates for quantum computation.
\newblock \emph{Phys. Rev. A}, 52:\penalty0 3457, Nov 1995.
\newblock \doi{10.1103/PhysRevA.52.3457}.
\newblock URL \url{https://journals.aps.org/pra/abstract/10.1103/PhysRevA.52.3457}.

\bibitem[Baxter(1972)]{BAXTER1972323}
Rodney~J Baxter.
\newblock One-dimensional anisotropic heisenberg chain.
\newblock \emph{Annals of Physics}, 70\penalty0 (2):\penalty0 323--337, 1972.
\newblock ISSN 0003-4916.
\newblock \doi{https://doi.org/10.1016/0003-4916(72)90270-9}.
\newblock URL \url{https://www.sciencedirect.com/science/article/pii/0003491672902709}.

\bibitem[Benhar and Fantoni(2020)]{benhar2020nuclear}
O.~Benhar and S.~Fantoni.
\newblock \emph{Nuclear Matter Theory}.
\newblock CRC Press, Taylor \& Francis Group, 2020.
\newblock ISBN 9780815386667.
\newblock URL \url{https://books.google.com/books?id=OI_wygEACAAJ}.

\bibitem[Bernien et~al.(2017)Bernien, Schwartz, Keesling, Levine, Omran, Pichler, Choi, Zibrov, Endres, Greiner, Vuletić, and Lukin]{rydbergatoms}
H.~Bernien, S.~Schwartz, A.~Keesling, H.~Levine, A.~Omran, H.~Pichler, S.~Choi, A.~Zibrov, M.~Endres, M.~Greiner, V.~Vuletić, and M.~Lukin.
\newblock Probing many-body dynamics on a 51-atom quantum simulator.
\newblock \emph{Nature}, 551:\penalty0 579--584, 2017.
\newblock \doi{10.1038/nature24622}.

\bibitem[Binkowski et~al.(2024)Binkowski, Koßmann, Ziegler, and Schwonnek]{Binkowski_2024}
Lennart Binkowski, Gereon Koßmann, Timo Ziegler, and René Schwonnek.
\newblock Elementary proof of qaoa convergence.
\newblock \emph{New Journal of Physics}, 26\penalty0 (7):\penalty0 073001, July 2024.
\newblock ISSN 1367-2630.
\newblock \doi{10.1088/1367-2630/ad59bb}.
\newblock URL \url{http://dx.doi.org/10.1088/1367-2630/ad59bb}.

\bibitem[Blatt et~al.(2004)Blatt, Häffner, Roos, Becher, and Schmidt‐Kaler]{trapion}
R.~Blatt, H.~Häffner, C.~F. Roos, C.~Becher, and F.~Schmidt‐Kaler.
\newblock Ion trap quantum computing with ca+ ions.
\newblock \emph{Quantum Information Processing}, 3:\penalty0 61--73, 2004.
\newblock \doi{10.1007/s11128-004-3105-1}.

\bibitem[Boixo and Somma(2010)]{boixo_s2010}
S.~Boixo and R.~D. Somma.
\newblock Necessary condition for the quantum adiabatic approximation.
\newblock \emph{Phys. Rev. A}, 81:\penalty0 032308, Mar 2010.
\newblock \doi{10.1103/PhysRevA.81.032308}.
\newblock URL \url{https://link.aps.org/doi/10.1103/PhysRevA.81.032308}.

\bibitem[Boixo et~al.(2009)Boixo, Knill, and Somma]{boixo_s2009}
S.~Boixo, E.~Knill, and R.~D. Somma.
\newblock Eigenpath traversal by phase randomization.
\newblock \emph{Quantum Inf. Comput.}, 9\penalty0 (9{\&}10):\penalty0 833--855, 2009.
\newblock \doi{10.26421/QIC9.9-10-7}.
\newblock URL \url{https://doi.org/10.26421/QIC9.9-10-7}.

\bibitem[Born and Fock(1928)]{born_m1928}
M.~Born and V.~Fock.
\newblock {Beweis des Adiabatensatzes}.
\newblock \emph{{Zeitschrift f\"ur Physik}}, 51:\penalty0 165, March 1928.
\newblock \doi{10.1007/BF01343193}.
\newblock URL \url{https://link.springer.com/article/10.1007/BF01343193}.

\bibitem[Bowman et~al.(1979)Bowman, Christoffel, and Tobin]{bowman1979application}
Joel~M Bowman, Kurt Christoffel, and Frank Tobin.
\newblock Application of scf-si theory to vibrational motion in polyatomic molecules.
\newblock \emph{Journal of Physical Chemistry}, 83\penalty0 (8):\penalty0 905--912, 1979.

\bibitem[Brassard et~al.(2002)Brassard, Høyer, Mosca, and Tapp]{Brassard_2002}
Gilles Brassard, Peter Høyer, Michele Mosca, and Alain Tapp.
\newblock Quantum amplitude amplification and estimation, 2002.
\newblock ISSN 0271-4132.
\newblock URL \url{http://dx.doi.org/10.1090/conm/305/05215}.

\bibitem[Byrd et~al.(1995)Byrd, Lu, Nocedal, and Zhu]{bfgs}
R.~H. Byrd, P.~Lu, J.~Nocedal, and C.~Zhu.
\newblock A limited memory algorithm for bound constrained optimization.
\newblock \emph{SIAM Journal on Scientific Computing}, 16:\penalty0 1190--1208, 1995.
\newblock \doi{10.1137/0916069}.

\bibitem[Cao et~al.(2019)Cao, Romero, Olson, Degroote, Johnson, Kieferov{\'a}, Kivlichan, Menke, Peropadre, Sawaya, et~al.]{cao2019quantum}
Yudong Cao, Jonathan Romero, Jonathan~P Olson, Matthias Degroote, Peter~D Johnson, M{\'a}ria Kieferov{\'a}, Ian~D Kivlichan, Tim Menke, Borja Peropadre, Nicolas~PD Sawaya, et~al.
\newblock Quantum chemistry in the age of quantum computing.
\newblock \emph{Chemical reviews}, 119\penalty0 (19):\penalty0 10856--10915, 2019.

\bibitem[Carlson et~al.(2018)Carlson, Dean, Hjorth-Jensen, Kaplan, Preskill, Roche, Savage, and Troyer]{osti_1631143}
Joseph Carlson, David~J. Dean, Morten Hjorth-Jensen, David Kaplan, John Preskill, Kenneth Roche, Martin~J. Savage, and Matthias Troyer.
\newblock Quantum computing for theoretical nuclear physics, a white paper prepared for the u.s. department of energy, office of science, office of nuclear physics, 1 2018.
\newblock URL \url{https://www.osti.gov/biblio/1631143}.

\bibitem[Cheng et~al.(2020)Cheng, Deumens, Freericks, Li, and Sanders]{osti_10291929}
Hai-Ping Cheng, Erik Deumens, James~K. Freericks, Chenglong Li, and Beverly~A. Sanders.
\newblock Application of quantum computing to biochemical systems: A look to the future.
\newblock \emph{Frontiers in Chemistry}, 8, 2020.
\newblock \doi{10.3389/fchem.2020.587143}.
\newblock URL \url{https://par.nsf.gov/biblio/10291929}.

\bibitem[Childs and Wiebe(2012)]{Childs:2012a}
Andrew~M. Childs and Nathan Wiebe.
\newblock Hamiltonian simulation using linear combinations of unitary operations.
\newblock \emph{Quantum Inf. Comput.}, 12\penalty0 (11-12):\penalty0 901--924, 2012.
\newblock ISSN 1533-7146.

\bibitem[Childs et~al.(2021)Childs, Su, Tran, Wiebe, and Zhu]{Childs:2019b}
Andrew~M. Childs, Yuan Su, Minh~C. Tran, Nathan Wiebe, and Shuchen Zhu.
\newblock Theory of trotter error with commutator scaling.
\newblock \emph{Physical Review X}, 11\penalty0 (1), February 2021.
\newblock ISSN 2160-3308.
\newblock \doi{10.1103/physrevx.11.011020}.
\newblock URL \url{http://dx.doi.org/10.1103/PhysRevX.11.011020}.

\bibitem[Choi et~al.(2021)Choi, Lee, Bonitati, Qian, and Watkins]{Choi:2020pdg}
Kenneth Choi, Dean Lee, Joey Bonitati, Zhengrong Qian, and Jacob Watkins.
\newblock Rodeo algorithm for quantum computing.
\newblock \emph{Phys. Rev. Lett.}, 127\penalty0 (4):\penalty0 040505, 2021.
\newblock \doi{10.1103/PhysRevLett.127.040505}.
\newblock URL \url{https://journals.aps.org/prl/abstract/10.1103/PhysRevLett.127.040505}.

\bibitem[Coello~P\'erez et~al.(2022)Coello~P\'erez, Bonitati, Lee, Quaglioni, and Wendt]{custom_gate_adiabatic}
Eduardo~A. Coello~P\'erez, Joey Bonitati, Dean Lee, Sofia Quaglioni, and Kyle~A. Wendt.
\newblock Quantum state preparation by adiabatic evolution with custom gates.
\newblock \emph{Phys. Rev. A}, 105:\penalty0 032403, Mar 2022.
\newblock \doi{10.1103/PhysRevA.105.032403}.
\newblock URL \url{https://link.aps.org/doi/10.1103/PhysRevA.105.032403}.

\bibitem[Cox et~al.(2000)Cox, Gattringer, Holland, Scarlet, and Wiese]{Cox_2000}
J.~Cox, C.~Gattringer, K.~Holland, B.~Scarlet, and U.-J. Wiese.
\newblock Meron-cluster solution of fermion and other sign problems.
\newblock \emph{Nuclear Physics B - Proceedings Supplements}, 83–84:\penalty0 777–791, April 2000.
\newblock ISSN 0920-5632.
\newblock \doi{10.1016/s0920-5632(00)91804-8}.
\newblock URL \url{http://dx.doi.org/10.1016/S0920-5632(00)91804-8}.

\bibitem[Córcoles et~al.(2021)Córcoles, Takita, Inoue, Lekuch, Minev, Chow, and Gambetta]{C_rcoles_2021}
A.D. Córcoles, Maika Takita, Ken Inoue, Scott Lekuch, Zlatko~K. Minev, Jerry~M. Chow, and Jay~M. Gambetta.
\newblock Exploiting dynamic quantum circuits in a quantum algorithm with superconducting qubits.
\newblock \emph{Physical Review Letters}, 127\penalty0 (10), August 2021.
\newblock ISSN 1079-7114.
\newblock \doi{10.1103/physrevlett.127.100501}.
\newblock URL \url{http://dx.doi.org/10.1103/PhysRevLett.127.100501}.

\bibitem[Dibner and Snow(2016)]{national2016science}
K.A. Dibner and C.E. Snow.
\newblock \emph{Science Literacy: Concepts, Contexts, and Consequences}.
\newblock National Academies Press, 2016.
\newblock ISBN 9780309447591.
\newblock URL \url{https://books.google.com/books?id=57fODQAAQBAJ}.

\bibitem[Eisert et~al.(1999)Eisert, Wilkens, and Lewenstein]{Eisert_1999}
Jens Eisert, Martin Wilkens, and Maciej Lewenstein.
\newblock Quantum games and quantum strategies.
\newblock \emph{Physical Review Letters}, 83\penalty0 (15):\penalty0 3077–3080, October 1999.
\newblock ISSN 1079-7114.
\newblock \doi{10.1103/physrevlett.83.3077}.
\newblock URL \url{http://dx.doi.org/10.1103/PhysRevLett.83.3077}.

\bibitem[Epelbaum et~al.(2009)Epelbaum, Hammer, and Meißner]{Epelbaum_2009}
E.~Epelbaum, H.-W. Hammer, and Ulf-G. Meißner.
\newblock Modern theory of nuclear forces.
\newblock \emph{Reviews of Modern Physics}, 81\penalty0 (4):\penalty0 1773–1825, December 2009.
\newblock ISSN 1539-0756.
\newblock \doi{10.1103/revmodphys.81.1773}.
\newblock URL \url{http://dx.doi.org/10.1103/RevModPhys.81.1773}.

\bibitem[Evans(2020)]{su2so3}
Jonny Evans.
\newblock Example: Su(2) to so(3), 2020.
\newblock URL \url{https://jde27.uk/lgla/17_example_homo.html}.
\newblock Licensed under CC-BY-SA 4.0.

\bibitem[Farhi et~al.(2000)Farhi, Goldstone, Gutmann, and Sipser]{Farhi:2000a}
Edward Farhi, Jeffrey Goldstone, Sam Gutmann, and Michael Sipser.
\newblock Quantum computation by adiabatic evolution, 2000.
\newblock URL \url{https://arxiv.org/abs/quant-ph/0001106}.

\bibitem[Feynman(1939)]{Feynman:1939}
R.~P. Feynman.
\newblock Forces in molecules.
\newblock \emph{Phys. Rev.}, 56:\penalty0 340--343, Aug 1939.
\newblock \doi{10.1103/PhysRev.56.340}.
\newblock URL \url{https://link.aps.org/doi/10.1103/PhysRev.56.340}.

\bibitem[Feynman(1982)]{feynman_rp1982}
R.~P. Feynman.
\newblock Simulating physics with computers.
\newblock \emph{Int. J. Theor. Phys.}, 21:\penalty0 467, 1982.
\newblock \doi{10.1007/BF02650179}.
\newblock URL \url{https://link.springer.com/article/10.1007/BF02650179}.

\bibitem[Fradkin(2013)]{fieldthoeries}
E.~Fradkin.
\newblock \emph{Field theories of condensed matter physics}.
\newblock Cambridge University Press, 2013.
\newblock \doi{10.1017/cbo9781139015509}.

\bibitem[Frame et~al.(2018)Frame, He, Ipsen, Lee, Lee, and Rrapaj]{ec}
D.~Frame, R.~He, I.~C.~F. Ipsen, D.~Lee, D.~Lee, and E.~Rrapaj.
\newblock Eigenvector continuation with subspace learning.
\newblock \emph{Physical Review Letters}, 121, 2018.
\newblock \doi{10.1103/physrevlett.121.032501}.

\bibitem[Fuchs et~al.(2020)Fuchs, Øie Kolden, Aase, and Sartor]{fuchs2020efficientencodingweightedmax}
Franz~Georg Fuchs, Herman Øie Kolden, Niels~Henrik Aase, and Giorgio Sartor.
\newblock Efficient encoding of the weighted max k-cut on a quantum computer using qaoa, 2020.
\newblock URL \url{https://arxiv.org/abs/2009.01095}.

\bibitem[Fucito et~al.(1981)Fucito, Marinari, Parisi, and Rebbi]{Fucito:1980fh}
F.~Fucito, E.~Marinari, G.~Parisi, and C.~Rebbi.
\newblock {A Proposal for Monte Carlo Simulations of Fermionic Systems}.
\newblock \emph{Nucl. Phys. B}, 180:\penalty0 369, 1981.
\newblock \doi{10.1016/0550-3213(81)90055-9}.

\bibitem[Gallager(2014)]{gallager2014information}
R.~Gallager.
\newblock \emph{Information Theory and Reliable Communication: Course held at the Department for Automation and Information July 1970}.
\newblock CISM International Centre for Mechanical Sciences. Springer Vienna, 2014.
\newblock ISBN 9783709129456.
\newblock URL \url{https://books.google.com/books?id=KOkkBAAAQBAJ}.

\bibitem[Ge et~al.(2018)Ge, Tura, and Cirac]{Ge:2017}
Yimin Ge, Jordi Tura, and J.~Ignacio Cirac.
\newblock Faster ground state preparation and high-precision ground energy estimation with fewer qubits, 2018.
\newblock URL \url{https://arxiv.org/abs/1712.03193}.

\bibitem[Gersch and Knollman(1963)]{PhysRev.129.959}
H.~A. Gersch and G.~C. Knollman.
\newblock Quantum cell model for bosons.
\newblock \emph{Phys. Rev.}, 129:\penalty0 959--967, Jan 1963.
\newblock \doi{10.1103/PhysRev.129.959}.
\newblock URL \url{https://link.aps.org/doi/10.1103/PhysRev.129.959}.

\bibitem[Granade et~al.(2017)Granade, Ferrie, and Flammia]{qctomography}
C.~Granade, C.~Ferrie, and S.~Flammia.
\newblock Practical adaptive quantum tomography.
\newblock \emph{New Journal of Physics}, 19:\penalty0 113017, 2017.
\newblock \doi{10.1088/1367-2630/aa8fe6}.

\bibitem[Griffiths(2017)]{griffiths2017introduction}
D.J. Griffiths.
\newblock \emph{Introduction to Quantum Mechanics}.
\newblock Cambridge University Press, 2017.
\newblock ISBN 9781107179868.
\newblock URL \url{https://books.google.com/books?id=0h-nDAAAQBAJ}.

\bibitem[Grover(1996)]{grover1996fastquantummechanicalalgorithm}
Lov~K. Grover.
\newblock A fast quantum mechanical algorithm for database search, 1996.
\newblock URL \url{https://arxiv.org/abs/quant-ph/9605043}.

\bibitem[Guerreschi and Smelyanskiy(2017)]{guerreschi2017practicaloptimizationhybridquantumclassical}
Gian~Giacomo Guerreschi and Mikhail Smelyanskiy.
\newblock Practical optimization for hybrid quantum-classical algorithms, 2017.
\newblock URL \url{https://arxiv.org/abs/1701.01450}.

\bibitem[Gustafson(2020)]{Gustafson:2020vqg}
Erik~J. Gustafson.
\newblock Projective cooling for the transverse {I}sing model.
\newblock \emph{Phys. Rev. D}, 101\penalty0 (7):\penalty0 071504(R), 5, 2020.
\newblock ISSN 2470-0010,2470-0029.
\newblock \doi{10.1103/physrevd.101.071504}.
\newblock URL \url{https://doi.org/10.1103/physrevd.101.071504}.

\bibitem[Hadfield et~al.(2019)Hadfield, Wang, O’Gorman, Rieffel, Venturelli, and Biswas]{Hadfield_2019}
Stuart Hadfield, Zhihui Wang, Bryan O’Gorman, Eleanor~G. Rieffel, Davide Venturelli, and Rupak Biswas.
\newblock From the quantum approximate optimization algorithm to a quantum alternating operator ansatz.
\newblock \emph{Algorithms}, 12\penalty0 (2):\penalty0 34, February 2019.
\newblock ISSN 1999-4893.
\newblock \doi{10.3390/a12020034}.
\newblock URL \url{http://dx.doi.org/10.3390/a12020034}.

\bibitem[Hagen et~al.(2014)Hagen, Papenbrock, Hjorth-Jensen, and Dean]{Hagen_2014}
G~Hagen, T~Papenbrock, M~Hjorth-Jensen, and D~J Dean.
\newblock Coupled-cluster computations of atomic nuclei.
\newblock \emph{Reports on Progress in Physics}, 77\penalty0 (9):\penalty0 096302, September 2014.
\newblock ISSN 1361-6633.
\newblock \doi{10.1088/0034-4885/77/9/096302}.
\newblock URL \url{http://dx.doi.org/10.1088/0034-4885/77/9/096302}.

\bibitem[Hager(2021)]{linalgch7}
William~W. Hager.
\newblock \emph{Applied Numerical Linear Algebra}.
\newblock SIAM, 2021.
\newblock \doi{10.1137/1.9781611976861.ch7}.
\newblock URL \url{https://epubs.siam.org/doi/abs/10.1137/1.9781611976861.ch7}.

\bibitem[Hall(2003)]{hall2003lie}
B.C. Hall.
\newblock \emph{Lie Groups, Lie Algebras, and Representations: An Elementary Introduction}.
\newblock Graduate Texts in Mathematics. Springer, 2003.
\newblock ISBN 9780387401225.
\newblock URL \url{https://books.google.com/books?id=m1VQi8HmEwcC}.

\bibitem[Heisenberg(1928)]{Heisenberg:1928}
W.~Heisenberg.
\newblock Zur theorie des ferromagnetismus.
\newblock \emph{Zeitschrift Fr Physik}, 49:\penalty0 619--636, 1928.
\newblock \doi{10.1007/bf01328601}.

\bibitem[Holland et~al.(2020)Holland, Wendt, Kravvaris, Wu, Ormand, DuBois, Quaglioni, and Pederiva]{holland_et2020}
Eric~T. Holland, Kyle~A. Wendt, Konstantinos Kravvaris, Xian Wu, W.~Erich Ormand, Jonathan~L DuBois, Sofia Quaglioni, and Francesco Pederiva.
\newblock Optimal control for the quantum simulation of nuclear dynamics.
\newblock \emph{Phys. Rev. A}, 101:\penalty0 062307, Jun 2020.
\newblock \doi{10.1103/PhysRevA.101.062307}.
\newblock URL \url{https://journals.aps.org/pra/abstract/10.1103/PhysRevA.101.062307}.

\bibitem[Hume(2006)]{hume}
David Hume.
\newblock \emph{An Enquiry Concerning Human Understanding}.
\newblock Project Gutenberg, 2006.
\newblock URL \url{https://www.gutenberg.org/ebooks/9662}.
\newblock eBook \#9662, Project Gutenberg.

\bibitem[{IBM Quantum}(2021)]{ibmq_2021}
{IBM Quantum}.
\newblock https://quantum-computing.ibm.com, 2021.
\newblock URL \url{https://quantum-computing.ibm.com}.

\bibitem[Ibrahim et~al.(2014)Ibrahim, Mamat, and Leong]{ibrahim2014}
M.~A.~H. Ibrahim, M.~Mamat, and W.~J. Leong.
\newblock The hybrid bfgs-cg method in solving unconstrained optimization problems.
\newblock \emph{Abstract and Applied Analysis}, 2014:\penalty0 1--6, 2014.
\newblock \doi{10.1155/2014/507102}.

\bibitem[Johansson et~al.(2012)Johansson, Nation, and Nori]{johansson_jr2012}
J.~R. Johansson, P.~D. Nation, and F.~Nori.
\newblock {QuTiP: An open-source Python framework for the dynamics of open quantum systems}.
\newblock \emph{Comput. Phys. Commun.}, 183\penalty0 (8):\penalty0 1760, 2012.
\newblock ISSN 0010-4655.
\newblock \doi{doi.org/10.1016/j.cpc.2012.02.021}.
\newblock URL \url{https://www.sciencedirect.com/science/article/pii/S0010465512000835}.

\bibitem[Johansson et~al.(2013)Johansson, Nation, and Nori]{johansson_jr2013}
J.~R. Johansson, P.~D. Nation, and F.~Nori.
\newblock {QuTiP 2: A Python framework for the dynamics of open quantum systems}.
\newblock \emph{Comput. Phys. Commun.}, 184\penalty0 (4):\penalty0 1234, 2013.
\newblock ISSN 0010-4655.
\newblock \doi{doi.org/10.1016/j.cpc.2012.11.019}.
\newblock URL \url{https://www.sciencedirect.com/science/article/pii/S0010465512003955}.

\bibitem[{Jones} et~al.(2021){Jones}, {Steven}, {Poncini}, {Rose}, and {Fedorov}]{jones_t2021}
Tyler {Jones}, Kaiah {Steven}, Xavier {Poncini}, Matthew {Rose}, and Arkady {Fedorov}.
\newblock {Approximations in transmon simulation}.
\newblock \emph{arXiv e-prints}, art. arXiv:2102.09721, February 2021.
\newblock URL \url{https://arxiv.org/abs/2102.09721}.

\bibitem[Kandala et~al.(2017)Kandala, Mezzacapo, Temme, Takita, Brink, Chow, and Gambetta]{Kandala_2017}
Abhinav Kandala, Antonio Mezzacapo, Kristan Temme, Maika Takita, Markus Brink, Jerry~M. Chow, and Jay~M. Gambetta.
\newblock Hardware-efficient variational quantum eigensolver for small molecules and quantum magnets.
\newblock \emph{Nature}, 549\penalty0 (7671):\penalty0 242–246, September 2017.
\newblock ISSN 1476-4687.
\newblock \doi{10.1038/nature23879}.
\newblock URL \url{http://dx.doi.org/10.1038/nature23879}.

\bibitem[Kingma and Ba(2017)]{kingma2017adammethodstochasticoptimization}
Diederik~P. Kingma and Jimmy Ba.
\newblock Adam: A method for stochastic optimization, 2017.
\newblock URL \url{https://arxiv.org/abs/1412.6980}.

\bibitem[Kirkpatrick et~al.(1983)Kirkpatrick, Gelatt, and Vecchi]{siman}
S.~Kirkpatrick, C.~D. Gelatt, and M.~Vecchi.
\newblock Optimization by simulated annealing.
\newblock \emph{Science}, 220:\penalty0 671--680, 1983.
\newblock \doi{10.1126/science.220.4598.671}.

\bibitem[Kitaev(1995{\natexlab{a}})]{Kitaev:1995}
A.~Yu. Kitaev.
\newblock Quantum measurements and the abelian stabilizer problem, 1995{\natexlab{a}}.
\newblock URL \url{https://arxiv.org/abs/quant-ph/9511026}.

\bibitem[Kitaev(1995{\natexlab{b}})]{Kitaev:1995qy}
A.~Yu. Kitaev.
\newblock Quantum measurements and the abelian stabilizer problem, 1995{\natexlab{b}}.
\newblock URL \url{https://arxiv.org/abs/quant-ph/9511026}.

\bibitem[Kitaev et~al.(2002)Kitaev, Shen, and Vyalyi]{kitaev2002classical}
A.Y. Kitaev, A.~Shen, and M.N. Vyalyi.
\newblock \emph{Classical and Quantum Computation}.
\newblock Graduate studies in mathematics. American Mathematical Society, 2002.
\newblock ISBN 9780821832295.
\newblock URL \url{https://books.google.com/books?id=qYHTvHPvmG8C}.

\bibitem[Koch et~al.(2007)Koch, Yu, Gambetta, Houck, Schuster, Majer, Blais, Devoret, Girvin, and Schoelkopf]{koch_j2007}
Jens Koch, Terri~M. Yu, Jay Gambetta, A.~A. Houck, D.~I. Schuster, J.~Majer, Alexandre Blais, M.~H. Devoret, S.~M. Girvin, and R.~J. Schoelkopf.
\newblock Charge-insensitive qubit design derived from the cooper pair box.
\newblock \emph{Phys. Rev. A}, 76:\penalty0 042319, Oct 2007.
\newblock \doi{10.1103/PhysRevA.76.042319}.
\newblock URL \url{https://journals.aps.org/pra/abstract/10.1103/PhysRevA.76.042319}.

\bibitem[Kockum and Nori(2019)]{jjunction}
A.~F. Kockum and F.~Nori.
\newblock Quantum bits with josephson junctions.
\newblock \emph{Fundamentals and Frontiers of the Josephson Effect}, pages 703--741, 2019.
\newblock \doi{10.1007/978-3-030-20726-7_17}.

\bibitem[Korotkov(2009)]{Korotkov2009}
Alexander Korotkov.
\newblock Special issue on quantum computing with superconducting qubits.
\newblock \emph{Quantum Information Processing - QUANTUM INF PROCESS}, 8:\penalty0 51--54, 06 2009.
\newblock \doi{10.1007/s11128-009-0104-2}.

\bibitem[Krantz et~al.(2019)Krantz, Kjaergaard, Yan, Orlando, Gustavsson, and Oliver]{krantz_p2019}
P.~Krantz, M.~Kjaergaard, F.~Yan, T.~P. Orlando, S.~Gustavsson, and W.~D. Oliver.
\newblock A quantum engineer's guide to superconducting qubits.
\newblock \emph{Appl. Phys. Rev.}, 6\penalty0 (2):\penalty0 021318, 2019.
\newblock \doi{10.1063/1.5089550}.
\newblock URL \url{https://doi.org/10.1063/1.5089550}.

\bibitem[Langfeld and Lucini(2016)]{langfeld2016formdensityofstatesmethodfinite}
Kurt Langfeld and Biagio Lucini.
\newblock Form the density-of-states method to finite density quantum field theory, 2016.
\newblock URL \url{https://arxiv.org/abs/1606.03879}.

\bibitem[Lawler(1985)]{lawler1985travelling}
E.L. Lawler.
\newblock \emph{The Travelling Salesman Problem: A Guided Tour of Combinatorial Optimization}.
\newblock Wiley-Interscience series in discrete mathematics and optimization. John Wiley \& Sons, 1985.
\newblock URL \url{https://books.google.com/books?id=qbFlMwEACAAJ}.

\bibitem[Lee et~al.(2004)Lee, Borasoy, and Schaefer]{Lee_2004}
Dean Lee, Buḡra Borasoy, and Thomas Schaefer.
\newblock Nuclear lattice simulations with chiral effective field theory.
\newblock \emph{Physical Review C}, 70\penalty0 (1), July 2004.
\newblock ISSN 1089-490X.
\newblock \doi{10.1103/physrevc.70.014007}.
\newblock URL \url{http://dx.doi.org/10.1103/PhysRevC.70.014007}.

\bibitem[Lee et~al.(2020)Lee, Bonitati, Given, Hicks, Li, Lu, Rai, Sarkar, and Watkins]{Lee:2019zze}
Dean Lee, Joey Bonitati, Gabriel Given, Caleb Hicks, Ning Li, Bing-Nan Lu, Abudit Rai, Avik Sarkar, and Jacob Watkins.
\newblock Projected cooling algorithm for quantum computation.
\newblock \emph{Physics Letters B}, 807:\penalty0 135536, 2020.
\newblock ISSN 0370-2693.
\newblock \doi{https://doi.org/10.1016/j.physletb.2020.135536}.
\newblock URL \url{https://www.sciencedirect.com/science/article/pii/S0370269320303403}.

\bibitem[Lloyd(1996)]{lloyd_s1996}
S.~Lloyd.
\newblock {Universal Quantum Simulators}.
\newblock \emph{Science}, 273\penalty0 (5278):\penalty0 1073, 1996.
\newblock ISSN 0036-8075.
\newblock \doi{10.1126/science.273.5278.1073}.
\newblock URL \url{https://science.sciencemag.org/content/273/5278/1073}.

\bibitem[Lu et~al.(2021)Lu, Bañuls, and Cirac]{Lu:2020}
Sirui Lu, Mari~Carmen Bañuls, and J.~Ignacio Cirac.
\newblock Algorithms for quantum simulation at finite energies.
\newblock \emph{PRX Quantum}, 2\penalty0 (2), May 2021.
\newblock ISSN 2691-3399.
\newblock \doi{10.1103/prxquantum.2.020321}.
\newblock URL \url{http://dx.doi.org/10.1103/PRXQuantum.2.020321}.

\bibitem[Lubasch et~al.(2020)Lubasch, Joo, Moinier, Kiffner, and Jaksch]{vqenl}
M.~Lubasch, J.~Joo, P.~Moinier, M.~Kiffner, and D.~Jaksch.
\newblock Variational quantum algorithms for nonlinear problems.
\newblock \emph{Physical Review A}, 101, 2020.
\newblock \doi{10.1103/physreva.101.010301}.

\bibitem[Lucas(2014)]{Lucas_2014}
Andrew Lucas.
\newblock Ising formulations of many np problems.
\newblock \emph{Frontiers in Physics}, 2, 2014.
\newblock ISSN 2296-424X.
\newblock \doi{10.3389/fphy.2014.00005}.
\newblock URL \url{http://dx.doi.org/10.3389/fphy.2014.00005}.

\bibitem[Maksymov et~al.(2021)Maksymov, Nguyen, Chaplin, Nam, and Markov]{maksymov2021detectingqubitcouplingfaultsiontrap}
Andrii~O. Maksymov, Jason Nguyen, Vandiver Chaplin, Yunseong Nam, and Igor~L. Markov.
\newblock Detecting qubit-coupling faults in ion-trap quantum computers, 2021.
\newblock URL \url{https://arxiv.org/abs/2108.03708}.

\bibitem[McClean et~al.(2016)McClean, Romero, Babbush, and Aspuru‐Guzik]{mccleanvqe}
J.~R. McClean, J.~Romero, R.~Babbush, and A.~Aspuru‐Guzik.
\newblock The theory of variational hybrid quantum-classical algorithms.
\newblock \emph{New Journal of Physics}, 18:\penalty0 023023, 2016.
\newblock \doi{10.1088/1367-2630/18/2/023023}.

\bibitem[McClean et~al.(2018)McClean, Boixo, Smelyanskiy, Babbush, and Neven]{McClean_2018}
Jarrod~R. McClean, Sergio Boixo, Vadim~N. Smelyanskiy, Ryan Babbush, and Hartmut Neven.
\newblock Barren plateaus in quantum neural network training landscapes.
\newblock \emph{Nature Communications}, 9\penalty0 (1), November 2018.
\newblock ISSN 2041-1723.
\newblock \doi{10.1038/s41467-018-07090-4}.
\newblock URL \url{http://dx.doi.org/10.1038/s41467-018-07090-4}.

\bibitem[Morales et~al.(2020)Morales, Biamonte, and Zimborás]{Morales_2020}
M.~E.~S. Morales, J.~D. Biamonte, and Z.~Zimborás.
\newblock On the universality of the quantum approximate optimization algorithm.
\newblock \emph{Quantum Information Processing}, 19\penalty0 (9), August 2020.
\newblock ISSN 1573-1332.
\newblock \doi{10.1007/s11128-020-02748-9}.
\newblock URL \url{http://dx.doi.org/10.1007/s11128-020-02748-9}.

\bibitem[Motzoi et~al.(2009)Motzoi, Gambetta, Rebentrost, and Wilhelm]{motzoi_f2009}
F.~Motzoi, J.~M. Gambetta, P.~Rebentrost, and F.~K. Wilhelm.
\newblock Simple pulses for elimination of leakage in weakly nonlinear qubits.
\newblock \emph{Phys. Rev. Lett.}, 103:\penalty0 110501, Sep 2009.
\newblock \doi{10.1103/PhysRevLett.103.110501}.
\newblock URL \url{https://journals.aps.org/prl/abstract/10.1103/PhysRevLett.103.110501}.

\bibitem[Moussa et~al.(2020)Moussa, Calandra, and Dunjko]{Moussa_2020}
Charles Moussa, Henri Calandra, and Vedran Dunjko.
\newblock To quantum or not to quantum: towards algorithm selection in near-term quantum optimization.
\newblock \emph{Quantum Science and Technology}, 5\penalty0 (4):\penalty0 044009, October 2020.
\newblock ISSN 2058-9565.
\newblock \doi{10.1088/2058-9565/abb8e5}.
\newblock URL \url{http://dx.doi.org/10.1088/2058-9565/abb8e5}.

\bibitem[{N. Khaneja and T. Reiss and C. Kehlet and T. Schulte-Herbr\"uggen and S. J. Glaser}(2005)]{khaneja_n2005}
{N. Khaneja and T. Reiss and C. Kehlet and T. Schulte-Herbr\"uggen and S. J. Glaser}.
\newblock {Optimal control of coupled spin dynamics: design of NMR pulse sequences by gradient ascent algorithms}.
\newblock \emph{J. Magn. Reson.}, 172\penalty0 (2):\penalty0 296, 2005.
\newblock ISSN 1090-7807.
\newblock \doi{10.1016/j.jmr.2004.11.004}.
\newblock URL \url{https://www.sciencedirect.com/science/article/pii/S1090780704003696}.

\bibitem[Nakanishi et~al.(2019)Nakanishi, Mitarai, and Fujii]{Nakanishi_2019}
Ken~M. Nakanishi, Kosuke Mitarai, and Keisuke Fujii.
\newblock Subspace-search variational quantum eigensolver for excited states.
\newblock \emph{Physical Review Research}, 1\penalty0 (3), October 2019.
\newblock ISSN 2643-1564.
\newblock \doi{10.1103/physrevresearch.1.033062}.
\newblock URL \url{http://dx.doi.org/10.1103/PhysRevResearch.1.033062}.

\bibitem[Nelder and Mead(1965)]{neldermead}
J.~A. Nelder and R.~Mead.
\newblock {A Simplex Method for Function Minimization}.
\newblock \emph{The Computer Journal}, 7\penalty0 (4):\penalty0 308--313, 01 1965.
\newblock ISSN 0010-4620.
\newblock \doi{10.1093/comjnl/7.4.308}.
\newblock URL \url{https://doi.org/10.1093/comjnl/7.4.308}.

\bibitem[Nielsen and Chuang(2010)]{nielsen_ma2010}
M.~A. Nielsen and I.~L. Chuang.
\newblock \emph{{Quantum Computation and Quantum Information}}.
\newblock Cambridge University Press, 2010.
\newblock \doi{10.1017/CBO9780511976667}.

\bibitem[NIST(2024)]{NIST2024}
NIST.
\newblock 2022 codata value: Newtonian constant of gravitation.
\newblock The NIST Reference on Constants, Units, and Uncertainty, May 2024.
\newblock URL \url{https://physics.nist.gov/cgi-bin/cuu/Value?bg}.
\newblock \url{https://physics.nist.gov/cgi-bin/cuu/Value?bg}.

\bibitem[Nogga et~al.(2006)Nogga, Navrátil, Barrett, and Vary]{Nogga_2006}
A.~Nogga, P.~Navrátil, B.~R. Barrett, and J.~P. Vary.
\newblock Spectra and binding energy predictions of chiral interactions for<mml:math xmlns:mml="http://www.w3.org/1998/math/mathml" display="inline"><mml:mmultiscripts><mml:mi mathvariant="normal">li</mml:mi><mml:mprescripts /><mml:none /><mml:mrow><mml:mn>7</mml:mn></mml:mrow></mml:mmultiscripts></mml:math>.
\newblock \emph{Physical Review C}, 73\penalty0 (6), June 2006.
\newblock ISSN 1089-490X.
\newblock \doi{10.1103/physrevc.73.064002}.
\newblock URL \url{http://dx.doi.org/10.1103/PhysRevC.73.064002}.

\bibitem[Pan and Meng(2024)]{Pan_2024}
Gaopei Pan and Zi~Yang Meng.
\newblock \emph{The sign problem in quantum Monte Carlo simulations}, page 879–893.
\newblock Elsevier, 2024.
\newblock ISBN 9780323914086.
\newblock \doi{10.1016/b978-0-323-90800-9.00095-0}.
\newblock URL \url{http://dx.doi.org/10.1016/B978-0-323-90800-9.00095-0}.

\bibitem[Pan et~al.(2022)Pan, Tong, and Yang]{10.1103/physreva.105.032433}
Y.~Pan, Y.~Tong, and Y.~Yang.
\newblock Automatic depth optimization for a quantum approximate optimization algorithm.
\newblock \emph{Physical Review A}, 105, 2022.
\newblock \doi{10.1103/physreva.105.032433}.

\bibitem[Parra-Rodriguez et~al.(2020)Parra-Rodriguez, Lougovski, Lamata, Solano, and Sanz]{Parra_Rodriguez_2020}
Adrian Parra-Rodriguez, Pavel Lougovski, Lucas Lamata, Enrique Solano, and Mikel Sanz.
\newblock Digital-analog quantum computation.
\newblock \emph{Physical Review A}, 101\penalty0 (2), February 2020.
\newblock ISSN 2469-9934.
\newblock \doi{10.1103/physreva.101.022305}.
\newblock URL \url{http://dx.doi.org/10.1103/PhysRevA.101.022305}.

\bibitem[Pelofske et~al.(2023)Pelofske, Bärtschi, and Eidenbenz]{Pelofske_2023}
Elijah Pelofske, Andreas Bärtschi, and Stephan Eidenbenz.
\newblock \emph{Quantum Annealing vs. QAOA: 127 Qubit Higher-Order Ising Problems on NISQ Computers}, page 240–258.
\newblock Springer Nature Switzerland, 2023.
\newblock ISBN 9783031320415.
\newblock \doi{10.1007/978-3-031-32041-5_13}.
\newblock URL \url{http://dx.doi.org/10.1007/978-3-031-32041-5_13}.

\bibitem[Peruzzo et~al.(2014)Peruzzo, McClean, Shadbolt, Yung, Zhou, Love, Aspuru-Guzik, and O’Brien]{Peruzzo_2014}
Alberto Peruzzo, Jarrod McClean, Peter Shadbolt, Man-Hong Yung, Xiao-Qi Zhou, Peter~J. Love, Alán Aspuru-Guzik, and Jeremy~L. O’Brien.
\newblock A variational eigenvalue solver on a photonic quantum processor.
\newblock \emph{Nature Communications}, 5\penalty0 (1), July 2014.
\newblock ISSN 2041-1723.
\newblock \doi{10.1038/ncomms5213}.
\newblock URL \url{http://dx.doi.org/10.1038/ncomms5213}.

\bibitem[Pieper and Wiringa(2001)]{gfmc}
S.~C. Pieper and R.~B. Wiringa.
\newblock Quantum monte carlo calculations of light nuclei.
\newblock \emph{Annual Review of Nuclear and Particle Science}, 51:\penalty0 53--90, 2001.
\newblock \doi{10.1146/annurev.nucl.51.101701.132506}.

\bibitem[{Poulin, D. and Qarry, A. and Somma, R. and Verstraete, F.}(2011)]{poulin_d2011}
{Poulin, D. and Qarry, A. and Somma, R. and Verstraete, F.}
\newblock {Quantum Simulation of Time-Dependent Hamiltonians and the Convenient Illusion of Hilbert Space}.
\newblock \emph{Phys. Rev. Lett.}, 106:\penalty0 170501, Apr 2011.
\newblock \doi{10.1103/PhysRevLett.106.170501}.
\newblock URL \url{https://journals.aps.org/prl/abstract/10.1103/PhysRevLett.106.170501}.

\bibitem[Poundstone(1992)]{poundstone1992prisoner}
W.~Poundstone.
\newblock \emph{Prisoner's Dilemma}.
\newblock Doubleday, 1992.
\newblock ISBN 9780385415675.
\newblock URL \url{https://books.google.com/books?id=9uruAAAAMAAJ}.

\bibitem[Preskill(2018)]{preskillnisq}
J.~Preskill.
\newblock Quantum computing in the nisq era and beyond.
\newblock \emph{Quantum}, 2:\penalty0 79, 2018.
\newblock \doi{10.22331/q-2018-08-06-79}.

\bibitem[Pérez-Obiol et~al.(2023)Pérez-Obiol, Romero, Menéndez, Rios, García-Sáez, and Juliá-Díaz]{P_rez_Obiol_2023}
A.~Pérez-Obiol, A.~M. Romero, J.~Menéndez, A.~Rios, A.~García-Sáez, and B.~Juliá-Díaz.
\newblock Nuclear shell-model simulation in digital quantum computers.
\newblock \emph{Scientific Reports}, 13\penalty0 (1), July 2023.
\newblock ISSN 2045-2322.
\newblock \doi{10.1038/s41598-023-39263-7}.
\newblock URL \url{http://dx.doi.org/10.1038/s41598-023-39263-7}.

\bibitem[Qian et~al.(2024)Qian, Watkins, Given, Bonitati, Choi, and Lee]{Qian:2021wya}
Zhengrong Qian, Jacob Watkins, Gabriel Given, Joey Bonitati, Kenneth Choi, and Dean Lee.
\newblock Demonstration of the rodeo algorithm on a quantum computer, 2024.
\newblock URL \url{https://arxiv.org/abs/2110.07747}.

\bibitem[Raussendorf et~al.(2003)Raussendorf, Browne, and Briegel]{Raussendorf_2003}
Robert Raussendorf, Daniel~E. Browne, and Hans~J. Briegel.
\newblock Measurement-based quantum computation on cluster states.
\newblock \emph{Physical Review A}, 68\penalty0 (2), August 2003.
\newblock ISSN 1094-1622.
\newblock \doi{10.1103/physreva.68.022312}.
\newblock URL \url{http://dx.doi.org/10.1103/PhysRevA.68.022312}.

\bibitem[Roggero et~al.(2020)Roggero, Gu, Baroni, and Papenbrock]{roggero_a2020}
A.~Roggero, C.~Gu, A.~Baroni, and T.~Papenbrock.
\newblock Preparation of excited states for nuclear dynamics on a quantum computer.
\newblock \emph{Phys. Rev. C}, 102:\penalty0 064624, Dec 2020.
\newblock \doi{10.1103/PhysRevC.102.064624}.
\newblock URL \url{https://journals.aps.org/prc/abstract/10.1103/PhysRevC.102.064624}.

\bibitem[Saad(2003)]{cgiterative}
Y.~Saad.
\newblock \emph{Iterative methods for sparse linear systems}.
\newblock Philadelphia, PA, 2003.
\newblock \doi{10.1137/1.9780898718003}.

\bibitem[Sadiku et~al.(2017)Sadiku, Tembely, and Musa]{largenum}
M.~Sadiku, M.~Tembely, and S.~Musa.
\newblock quantum computing: a primer.
\newblock \emph{International Journal of Advanced Research in Computer Science and Software Engineering}, 7:\penalty0 129, 2017.
\newblock \doi{10.23956/ijarcsse.v7i11.501}.

\bibitem[Santagati et~al.(2024)Santagati, Aspuru-Guzik, Babbush, Degroote, González, Kyoseva, Moll, Oppel, Parrish, Rubin, Streif, Tautermann, Weiss, Wiebe, and Utschig-Utschig]{Santagati_2024}
Raffaele Santagati, Alan Aspuru-Guzik, Ryan Babbush, Matthias Degroote, Leticia González, Elica Kyoseva, Nikolaj Moll, Markus Oppel, Robert~M. Parrish, Nicholas~C. Rubin, Michael Streif, Christofer~S. Tautermann, Horst Weiss, Nathan Wiebe, and Clemens Utschig-Utschig.
\newblock Drug design on quantum computers.
\newblock \emph{Nature Physics}, 20\penalty0 (4):\penalty0 549–557, March 2024.
\newblock ISSN 1745-2481.
\newblock \doi{10.1038/s41567-024-02411-5}.
\newblock URL \url{http://dx.doi.org/10.1038/s41567-024-02411-5}.

\bibitem[Sarty et~al.(1993)]{Sarty:1993zz}
A.~J. Sarty et~al.
\newblock {Measurement of the reaction He-3 (gamma, pp) n and its relation to three-body forces}.
\newblock \emph{Phys. Rev. C}, 47:\penalty0 459--467, 1993.
\newblock \doi{10.1103/PhysRevC.47.459}.

\bibitem[Sawaya et~al.(2024)Sawaya, Marti-Dafcik, Ho, Tabor, Neira, Magann, Premaratne, Dubey, Matsuura, Bishop, de~Jong, Benjamin, Parekh, Tubman, Klymko, and Camps]{sawaya2024hamliblibraryhamiltoniansbenchmarking}
Nicolas~PD Sawaya, Daniel Marti-Dafcik, Yang Ho, Daniel~P Tabor, David E~Bernal Neira, Alicia~B Magann, Shavindra Premaratne, Pradeep Dubey, Anne Matsuura, Nathan Bishop, Wibe~A de~Jong, Simon Benjamin, Ojas~D Parekh, Norm Tubman, Katherine Klymko, and Daan Camps.
\newblock Hamlib: A library of hamiltonians for benchmarking quantum algorithms and hardware, 2024.
\newblock URL \url{https://arxiv.org/abs/2306.13126}.

\bibitem[Scalettar et~al.(1995)Scalettar, Batrouni, Kampf, and Zimányi]{hubbardmodel}
R.~Scalettar, G.~G. Batrouni, A.~P. Kampf, and G.~T. Zimányi.
\newblock Simultaneous diagonal and off-diagonal order in the bose-hubbard hamiltonian.
\newblock \emph{Physical Review B}, 51:\penalty0 8467--8480, 1995.
\newblock \doi{10.1103/physrevb.51.8467}.

\bibitem[Shende et~al.(2004)Shende, Markov, and Bullock]{shende_vv2004}
Vivek~V. Shende, Igor~L. Markov, and Stephen~S. Bullock.
\newblock Minimal universal two-qubit controlled-not-based circuits.
\newblock \emph{Phys. Rev. A}, 69:\penalty0 062321, Jun 2004.
\newblock \doi{10.1103/PhysRevA.69.062321}.
\newblock URL \url{https://link.aps.org/doi/10.1103/PhysRevA.69.062321}.

\bibitem[Smite-Meister(2024)]{blochsphere}
Smite-Meister.
\newblock Image from wikimedia commons.
\newblock \url{https://commons.wikimedia.org/w/index.php?curid=5829358}, 2024.
\newblock CC BY-SA 3.0.

\bibitem[Streif and Leib(2020)]{Streif_2020}
Michael Streif and Martin Leib.
\newblock Forbidden subspaces for level-1 quantum approximate optimization algorithm and instantaneous quantum polynomial circuits.
\newblock \emph{Physical Review A}, 102\penalty0 (4), October 2020.
\newblock ISSN 2469-9934.
\newblock \doi{10.1103/physreva.102.042416}.
\newblock URL \url{http://dx.doi.org/10.1103/PhysRevA.102.042416}.

\bibitem[Suzuki(1976)]{Suzuki:1976a}
Masuo Suzuki.
\newblock Generalized {T}rotter's formula and systematic approximants of exponential operators and inner derivations with applications to many-body problems.
\newblock \emph{Comm. Math. Phys.}, 51\penalty0 (2):\penalty0 183--190, 1976.
\newblock ISSN 0010-3616,1432-0916.
\newblock URL \url{http://projecteuclid.org/euclid.cmp/1103900351}.

\bibitem[Suzuki et~al.(2020)Suzuki, Uno, Raymond, Tanaka, Onodera, and Yamamoto]{Suzuki_2020}
Yohichi Suzuki, Shumpei Uno, Rudy Raymond, Tomoki Tanaka, Tamiya Onodera, and Naoki Yamamoto.
\newblock Amplitude estimation without phase estimation.
\newblock \emph{Quantum Information Processing}, 19\penalty0 (2), January 2020.
\newblock ISSN 1573-1332.
\newblock \doi{10.1007/s11128-019-2565-2}.
\newblock URL \url{http://dx.doi.org/10.1007/s11128-019-2565-2}.

\bibitem[Szabo and Ostlund(1996)]{szabo1996modern}
A.~Szabo and N.S. Ostlund.
\newblock \emph{Modern Quantum Chemistry: Introduction to Advanced Electronic Structure Theory}.
\newblock Dover Books on Chemistry. Dover Publications, 1996.
\newblock ISBN 9780486691862.
\newblock URL \url{https://books.google.com/books?id=6mV9gYzEkgIC}.

\bibitem[Trabesinger(2012)]{Trabesinger:2012icv}
Andreas Trabesinger.
\newblock {Quantum simulation}.
\newblock \emph{Nature Phys.}, 8\penalty0 (4):\penalty0 263, 2012.
\newblock \doi{10.1038/nphys2258}.

\bibitem[Trotter(1959)]{Trotter:1959}
H.~F. Trotter.
\newblock On the product of semi-groups of operators.
\newblock \emph{Proc. Amer. Math. Soc.}, 10:\penalty0 545--551, 1959.
\newblock ISSN 0002-9939,1088-6826.
\newblock \doi{10.2307/2033649}.
\newblock URL \url{https://doi.org/10.2307/2033649}.

\bibitem[Vatan and Williams(2004)]{vatan_f2004}
Farrokh Vatan and Colin Williams.
\newblock Optimal quantum circuits for general two-qubit gates.
\newblock \emph{Phys. Rev. A}, 69:\penalty0 032315, Mar 2004.
\newblock \doi{10.1103/PhysRevA.69.032315}.
\newblock URL \url{https://link.aps.org/doi/10.1103/PhysRevA.69.032315}.

\bibitem[Venkateswaran et~al.(2022)Venkateswaran, Ramachandran, Kurinjimalar, Vidhya, and Mathivanan]{simanneal}
C.~Venkateswaran, M.~Ramachandran, R.~Kurinjimalar, P.~Vidhya, and G.~Mathivanan.
\newblock Application of simulated annealing in various field.
\newblock \emph{Materials and Its Characterization}, 1:\penalty0 01--08, 2022.
\newblock \doi{10.46632/mc/1/1/1}.

\bibitem[Verresen(2023)]{verresen2023quantumisingmodel}
Ruben Verresen.
\newblock Everything is a quantum ising model, 2023.
\newblock URL \url{https://arxiv.org/abs/2301.11917}.

\bibitem[Vidal and Dawson(2004)]{vidal_g2004}
G.~Vidal and C.~M. Dawson.
\newblock Universal quantum circuit for two-qubit transformations with three controlled-not gates.
\newblock \emph{Phys. Rev. A}, 69:\penalty0 010301, Jan 2004.
\newblock \doi{10.1103/PhysRevA.69.010301}.
\newblock URL \url{https://link.aps.org/doi/10.1103/PhysRevA.69.010301}.

\bibitem[Wainwright and Jordan(2008)]{wainwright2008graphical}
M.J. Wainwright and M.I. Jordan.
\newblock \emph{Graphical Models, Exponential Families, and Variational Inference}.
\newblock Foundations and trends in machine learning. Now Publishers, 2008.
\newblock ISBN 9781601981844.
\newblock URL \url{https://books.google.com/books?id=zp5Mo3VsJbgC}.

\bibitem[Wales and Doye(1997)]{basinhopping}
David~J. Wales and Jonathan P.~K. Doye.
\newblock Global optimization by basin-hopping and the lowest energy structures of lennard-jones clusters containing up to 110 atoms.
\newblock \emph{The Journal of Physical Chemistry A}, 101\penalty0 (28):\penalty0 5111--5116, 1997.
\newblock \doi{10.1021/jp970984n}.
\newblock URL \url{https://doi.org/10.1021/jp970984n}.

\bibitem[Williams(2011)]{williams2011}
C.~P. Williams.
\newblock Quantum gates.
\newblock \emph{Texts in Computer Science}, pages 51--122, 2011.
\newblock \doi{10.1007/978-1-84628-887-6_2}.

\bibitem[Willsch et~al.(2022)Willsch, Willsch, Jin, Michielsen, and De~Raedt]{Willsch_2022}
Dennis Willsch, Madita Willsch, Fengping Jin, Kristel Michielsen, and Hans De~Raedt.
\newblock Gpu-accelerated simulations of quantum annealing and the quantum approximate optimization algorithm.
\newblock \emph{Computer Physics Communications}, 278:\penalty0 108411, September 2022.
\newblock ISSN 0010-4655.
\newblock \doi{10.1016/j.cpc.2022.108411}.
\newblock URL \url{http://dx.doi.org/10.1016/j.cpc.2022.108411}.

\bibitem[Wu et~al.(2020)Wu, Tomarken, Petersson, Martinez, Rosen, and DuBois]{qudits_xian}
Xian Wu, S.~L. Tomarken, N.~Anders Petersson, L.~A. Martinez, Yaniv~J. Rosen, and Jonathan~L. DuBois.
\newblock High-fidelity software-defined quantum logic on a superconducting qudit.
\newblock \emph{Phys. Rev. Lett.}, 125:\penalty0 170502, Oct 2020.
\newblock \doi{10.1103/PhysRevLett.125.170502}.
\newblock URL \url{https://link.aps.org/doi/10.1103/PhysRevLett.125.170502}.

\bibitem[Yarkoni et~al.(2022)Yarkoni, Raponi, Bäck, and Schmitt]{qannealing}
S.~Yarkoni, E.~Raponi, T.~Bäck, and S.~Schmitt.
\newblock quantum annealing for industry applications: introduction and review.
\newblock \emph{Reports on Progress in Physics}, 85:\penalty0 104001, 2022.
\newblock \doi{10.1088/1361-6633/ac8c54}.

\bibitem[Zahidi(2024)]{WEF}
Saadia Zahidi.
\newblock The global risks report 2024, 2024.
\newblock URL \url{https://www3.weforum.org/docs/WEF_The_Global_Risks_Report_2024.pdf}.
\newblock ISBN: 978-2-940631-64-3.

\bibitem[Zhang et~al.(2020)Zhang, Yuan, and Yin]{zhang2020variationalquantumeigensolversvariance}
Dan-Bo Zhang, Zhan-Hao Yuan, and Tao Yin.
\newblock Variational quantum eigensolvers by variance minimization, 2020.
\newblock URL \url{https://arxiv.org/abs/2006.15781}.

\bibitem[Zhao et~al.(2024)Zhao, Cheng, Wang, Fan, and Ma]{10.1088/1367-2630/ad4629}
R.~Zhao, T.~Cheng, R.~Wang, X.~Fan, and H.~Ma.
\newblock Artificial intelligence warm-start approach: optimizing the generalization capability of qaoa in complex energy landscapes.
\newblock \emph{New Journal of Physics}, 26:\penalty0 053016, 2024.
\newblock \doi{10.1088/1367-2630/ad4629}.

\end{thebibliography}

\end{document}